\documentclass[a4paper,11pt]{article}

\usepackage{jheppub} 

\usepackage{graphicx}

\usepackage{amsmath}
\usepackage[utf8]{inputenc}

\usepackage{amssymb}
\usepackage{amsfonts}
\usepackage{amsbsy}
\usepackage{url}
\usepackage{enumerate}
\usepackage{caption}
\usepackage{color}
\usepackage{ulem}
\usepackage{bm}
\usepackage{multirow,bigdelim}

\newcommand{\be}{\begin{eqnarray}}
\newcommand{\ee}{\end{eqnarray}}
\newcommand{\p}{\partial}

\makeatletter
\@addtoreset{equation}{section}
\makeatother

\newcommand\rsout{\bgroup\markoverwith{\textcolor{red}{\rule[0.5ex]{2pt}{0.4pt}}}\ULon}

\title{
Dynamics of global and local vortices with orientational moduli
}

\author{Minoru Eto$^{a,b}$, Adam Peterson$^c$, Fidel I. Schaposnik Massolo$^d$, Gianni Tallarita$^e$}

\affiliation{$^a$Department of Physics, Yamagata University, Kojirakawa-machi 1-4-12, Yamagata, Yamagata 990-8560, Japan,\\
\\
$^b$Research and Education Center for Natural Sciences, Keio University, 4-1-1 Hiyoshi, Yokohama, Kanagawa 223-8521, Japan,\\
\\
 $^c$Lawrence Berkeley National Laboratory, Center for Computational Science and Engineering, 1 Cyclotron Rd, Berkeley, CA 94720, USA,\\
\\
$^d$Institut des Hautes \'{E}tudes Scientifiques, 35 route de Chartres, 91440 Bures-sur-Yvette, France,\\
\\
$^e$Departamento de Ciencias, Facultad de Artes Liberales, Universidad Adolfo Ib\'a\~nez, Santiago 7941169, Chile.}

\emailAdd{$^{a,b}$meto@sci.kj.yamagata-u.ac.jp, $^c$ajpeterson@lbl.gov, $^d$fidels@ihes.fr, $^e$gianni.tallarita@uai.cl  }

\abstract{
The dynamics of both global and local vortices with non-Abelian orientational moduli 
is investigated in detail. Head-on collisions of these vortices are numerically simulated for parallel, anti-parallel and orthogonal internal orientations where we find interesting dynamics of the orientational moduli. A detailed study of the inter-vortex force is provided and a phase diagram separating Abelian and non-Abelian vortex types is constructed. 
Some results on scatterings with non-zero impact parameter and multi-vortex collisions are included.
}
\preprint{YGHP-20-08}

\begin{document}
\maketitle


\section{Introduction}
\label{sec:intro} 

Abrikosov realized long ago that vortex solutions in superconductors confine magnetic ``charges"  \cite{abrikosov}, a result that was used to formulate the idea that a similar mechanism should occur for the confinement of quarks \cite{Nambu:1974zg,thooft,mandel}, despite the physics of the problem being dramatically different. These differences are fundamental, quarks are firstly electric sources and not magnetic and furthermore they are also charged under the colour group, absent from all previous superconductor-related discussions. Therefore, if this idea is to be successful in describing real quarks some analogous and yet dual mechanism for vortex formation must be responsible, spiced up with some new ingredients. Rather than condensation of electric charges, as per the well-known Cooper pair mechanism, what should occur is condensation of magnetic monopoles, leading to a similar Meissner effect only now for electric rather than magnetic fields. In addition, the resulting vortices must possess some additional internal degrees of freedom, responsible for the transmission of the colour flux. The problem is further complicated by the fact that the physics of quarks is inherently strongly coupled, meaning all effective classical descriptions are at best inaccurate.

Despite the overwhelming complications, a major breakthrough in the realization of this idea was the work of Seiberg and Witten \cite{Seiberg:1994rs}, who discovered that magnetic monopoles do indeed condense in a softly broken $\mathcal{N}=2$ Super Yang-Mills theory. Their condensation does in fact yield electric flux tubes, one of the main desired ingredients, however the mechanism behind this remains purely Abelian and fails to describe any of the actual physics of Nature's quarks. In particular, these flux tubes predict extra multiplicities in the hadron spectrum \cite{Vainshtein:2000hu,shifman2} not observed in any experiments, not mentioning the need for supersymmetry. It therefore becomes important to study alternative types of vortex solutions. Non-abelian vortices \cite{Hanany:2003hp,Auzzi:2003fs} are standard solitonic vortex solutions with additional orientational moduli localized at their cores. These moduli are the Goldstone bosons resulting from the breaking of a non-Abelian global symmetry in the vortex core. They can be realized in some vacua of $\mathcal{N}=2$ SQCD in which they do in fact yield interesting confinement phenomena \cite{Carlino:2000uk}, as well as several more relations between two and four dimensional theories \cite{Dorey:1999zk,Hanany:2004ea,Shifman:2004dr}. Since their initial discovery, an extensive literature on this subject has been developed \cite{Eto:2006dx,Tong:2005un,Eto:2006pg,Konishi:2008vj,shifbook1,shifbook2} which mainly involves far from straightforward models usually including some degree of supersymmetry. In great part because of these complications, the dynamics of such non-Abelian vortices remains an open and important question. 

There are however plenty of analytical and/or numerical studies on dynamics of the global/local Abelian vortices in the literature, see for example \cite{Vilenkin:2000jqa,Samols:1991ne,Shellard:1988zx} and references therein. In comparison, studies on the dynamics of non-Abelian vortices are indeed very few. The static interaction (semi-dynamical problem) of the non-Abelian global vortices with $\mathbb{C}P^{N-1}$ orientational moduli was studied in Ref.~\cite{Nakano:2007dq}. That of the non-Abelian local vortices with $\mathbb{C}P^{N-1}$ orientational moduli was studied in Refs.~\cite{Auzzi:2007iv,Auzzi:2007wj}, and that for the non-Abelian local vortices with $S^{N-1}$ orientational moduli in Ref.~\cite{Tallarita:2017opp}.
 Studies which have attempted to investigate such dynamics \cite{Hashimoto:2005hi,Eto:2006db,Eto:2011pj} have resorted to moduli approximations for slowly moving BPS vortices, avoiding the full numerical bulk calculations necessary for an in depth study of this rich physics. This is not surprising, the bulk supersymmetric models even when truncated to their bosonic subsectors involve a plethora of fields, complicating the calculations dramatically.

Interestingly, non-Abelian vortex cousins have also been found in different contexts without supersymmetries.
One of them is high-density QCD where a mixture of non-Abelian global and local vortices exists \cite{Balachandran:2005ev,Nakano:2007dr,Nakano:2008dc,Eto:2009kg,Eto:2009bh,Eto:2009tr}. 
There are also intensive studies on these vortices from various perspectives \cite{Gorsky:2011hd,Eto:2011mk,Hirono:2010gq,Vinci:2012mc,Eto:2013hoa,Chatterjee:2015lbf,Alford:2016dco,Chatterjee:2016ykq,Alford:2018mqj,Chatterjee:2018nxe,Hirono:2018fjr,Chatterjee:2019tbz,Hidaka:2019jtv}. The other recent topic is related to non-Abelian vortices in the so-called two Higgs doublet model,
which is one of the simplest extensions of the Standard Model (SM) with an additional Higgs doublet \cite{Lee:1973iz}.
A topologically stable vortex was first found in Ref.~\cite{Dvali:1993sg}, and has been recently revived 
in Ref.~\cite{Eto:2018hhg,Eto:2018tnk}. A topological monopole attached to the non-Abelian strings was found in \cite{Eto:2019hhf,Eto:2020hjb,Eto:2020opf}. Similar configurations are also studied in an axion model \cite{Abe:2020ure}.\newline

Recently, inspired by Witten's superconducting string model \cite{Witten:1984eb}, Shifman \cite{Shifman:2012vv} has proposed a particulary simple model which also contains non-Abelian vortices, free of any supersymmetry. This model was easily generalized to many non-Abelian soliton setups. In \cite{Peterson:2014nma,Peterson:2015tpa} the model was used to investigate cholesteric and spin vortices in an analogous model to liquid crystals in condensed matter, and in \cite{Shifman:2015ama} the system was generalized to include isospin degrees of freedom on t'Hooft-Polyakov monopole solutions. These moduli were also added to Skyrmions in \cite{Canfora:2016spb}, and a dual non-Abelian monopole-vortex complex was constructed in \cite{Tallarita:2017xhh}. A preliminary study of non-Abelian vortex lattices was performed in \cite{Tallarita:2017opp}, relying upon the simplicity of the original model and its close analogue to the Abrikosov lattice solution. These works are all in the classical regime of the theory, ignoring all strong coupling effects. Attempts to include such effects in this model were first initiated in \cite{Tallarita:2015mca}, where non-Abelian vortices were first built in a holographic context assuming no backreaction of the matter fields in the gravitational sector. This was extended in \cite{Tallarita:2019czh}, where the fully backreacted non-Abelian vortex solution was built.

The simplicity of this model allows one to keep the numerical calculations of non-Abelian vortex dynamics simple enough for a bona-fide bulk simulation. This paper is devoted to precisely such a study. By taking advantage of recent technical advances in the field of neural networks we were able to implement all necessary numerical simulations on graphical processing units, thus speeding up the complex two-dimensional relaxation procedures by two orders of magnitude. This dramatic reduction in computing time allowed us to delve into the dynamics of non-Abelian vortices in fine detail, elucidating the phase diagram, scattering physics and additional aspects of this fascinating subject. Our work covers global and local, BPS and non-BPS, and slow and fast moving vortices.

This paper is structured as follows. In Section~\ref{sec:model} we introduce the model, which is a slight variation of Shifman's original model corresponding to a specific combination of coupling constants in the potential; in Section~\ref{sec:GNAV} we begin our scattering investigations in the global vortex context, which serves as a warmup for the gauged case. In Section~ \ref{sec:static_int_lnA} we provide a detailed analytic study of the interaction between non-Abelian vortices, in particular extending and refining the results of \cite{Tallarita:2017opp} to higher accuracy in the mass parameters. In Section~\ref{sec:scattering_lnA} we finally investigate the full scattering dynamics of local non-Abelian vortices, restricting to the scattering of parallel, anti-parallel and orthogonal internal orientations. Some exotic results on anti-vortex and multi-vortex scatterings are included in Section~\ref{exotic} to illustrate the full generality and applicability of our methods. Finally, Section~\ref{concs} is devoted to summarizing the results of the paper.


\section{The model}
\label{sec:model}

The model we study in this work is an Abelian-Higgs model with neutral scalar fields.
The matter contents are the $U(1)$ gauge field $A_\mu$, the charged Higgs field $\phi$,
and the neutral real scalar fields $\chi_a$ ($a=1,\cdots, N$). We consider one of
the simplest models which admits a nontrivial vortex with internal orientational moduli.
The Lagrangian is
\be
{\cal L} &=& - \frac{1}{4}F_{\mu\nu}F^{\mu\nu} + D_\mu \phi (D^\mu \phi)^* 
+ \p_\mu \chi_a\p^\mu\chi_a - V,\label{eq:lag}\\
V &=& \Omega^2 \chi_a^2 + \lambda \left(|\phi|^2 + \chi_a^2 - v^2\right)^2.\label{eq:pot}
\ee
The covariant derivative is defined as $D_\mu \phi = (\p_\mu + ie A_\mu)\phi$ where
$e$ is a $U(1)$ charge.
In addition to the $U(1)$ gauge symmetry, this model has $O(N)$ global symmetry 
$\vec \chi \to U \vec \chi$ with $U \in O(N)$. In what follows, we assume $\Omega^2 > 0$,
$v^2 > 0$, and $\lambda > 0$. Then the potential minimum reads
\be
|\phi| = v,\qquad \vec \chi = 0.
\ee
This is the Higgs branch in the sense that the $U(1)$ gauge symmetry is broken.
On the other hand, the global $O(N)$ symmetry is unbroken.
To be more precise, let us examine the potential minima by considering the Hessian 
\be
H = \left(
\begin{array}{cc}
\frac{\p^2 V}{\p x^2} & \frac{\p^2 V}{\p x\p y}\\
\frac{\p^2 V}{\p y\p x} & \frac{\p^2 V}{\p y^2}
\end{array}
\right),\qquad x \equiv |\phi| \ge 0,\quad y = \sqrt{\chi_i^2} \ge 0.
\ee
There are three candidates for local minima: 
$(x,y) = (0,0)$, $(v,0)$ and $(0,\tilde v)$ with $\tilde v \equiv \sqrt{\frac{2v^2\lambda -\Omega^2}{2\lambda}}$, the latter existing only when $2v^2\lambda \ge \Omega^2$.
\begin{table}
\begin{center}
\begin{tabular}{c|ccc}
\hline
$(x,y)$  &  $(0,0)$ & $(v,0)$ & $(0,\tilde v)|_{2v^2\lambda > \Omega^2}$ \\
\hline
$\det H$ & $8v^2\lambda(2v^2\lambda-\Omega^2)$ & $16v^2\lambda \Omega^2$ & $-8\Omega^2(2v^2\lambda - \Omega^2)$\\
${\rm tr} H$ & $-2(4v^2\lambda - \Omega^2)$ & $2(4v^2\lambda +\Omega^2)$ & $2(4v^2\lambda^2-3\Omega^2)$\\
\hline
\end{tabular}
\caption{Determinant and trace of the Hessian $H$.}
\label{tab:hessian}
\end{center}
\end{table}
The determinant and trace of the Hessian $H$ are shown in Table~\ref{tab:hessian}. For a point to be local minimum, both
$\det H$ and ${\rm tr} H$ are required to be positive. Therefore, under the conditions $\Omega^2 > 0$,
$v^2 > 0$ and $\lambda > 0$, $(x,y) = (0,0)$ and $(0,\tilde v)$ cannot be local minima. Instead, $(x,y) = (v,0)$ is
always a local minimum where all  bosonic excitations are massive and
\be
m_\phi^2 = 4v^2\lambda ,\quad  m_\gamma^2 = 2 e^2v^2,\quad  m_\chi^2 = \Omega^2.
\ee

\begin{figure}[ht]
\begin{center}
\includegraphics[width=15cm]{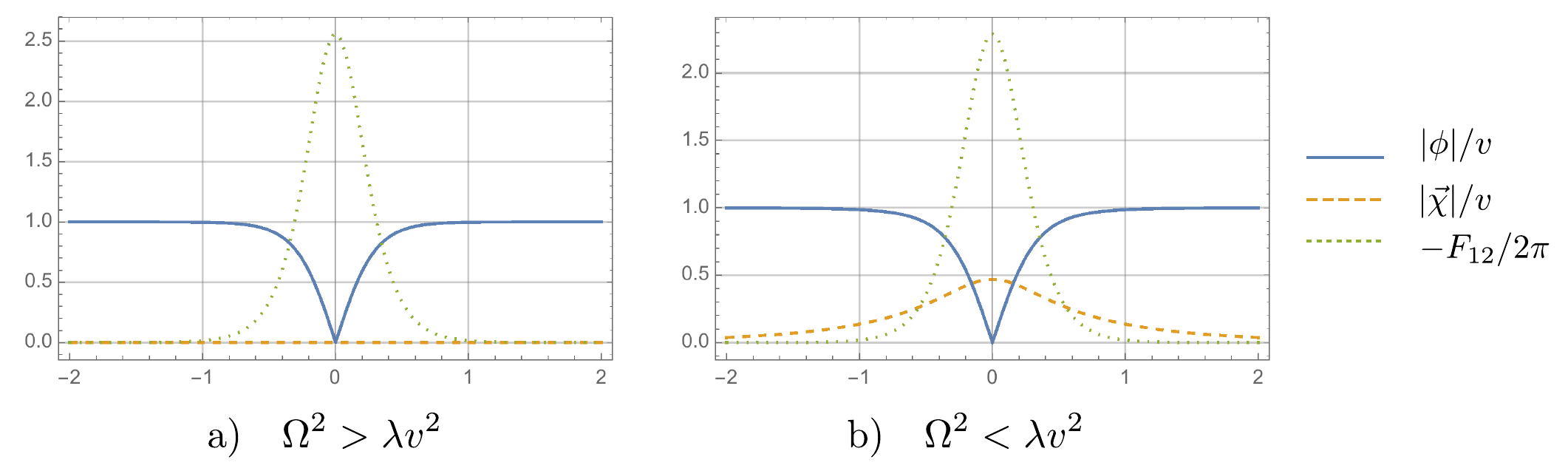}
\caption{The non-superconducting vortex (a) and the superconducting vortex (b).
The parameter choice is $(e,\lambda,v)=(2,2,2)$ with $\Omega = 3$ for (a) and $\Omega = 1$ for (b).}
\label{fig:vortex_profile}
\end{center}
\end{figure}

The spontaneously broken $U(1)$ symmetry gives rise to the well-known Abrikosov-Nielsen-Olesen
vortices which are topological solitons associated with the first homotopy group
\be
\pi_1(U(1)) = \mathbb{Z}.
\ee
The topological winding number is related to the asymptotic behavior of the phase of $\phi$ by
\be
\phi\Big|_{r\to \infty} \to v e^{ik\theta},
\ee
with $\{r,\theta\}$ the usual polar coordinates. The single-valuedness condition on $\phi$
forces $k$ to be an integer, which is nothing but the topological winding number $k \in \mathbb{Z}$.
Then, the finiteness of energy requires
\be
D_\theta \phi \to 0
\quad \Rightarrow \quad
A_\theta \to \frac{i}{e}\phi^{-1}\p_\theta \phi = \frac{k}{e},
\ee
therefore the magnetic flux is quantized as
\be
\Phi = \int d^2x\, F_{12} = \int^{2\pi}_0 d\theta~  A_\theta\big|_{r\to\infty} 
= \frac{2\pi}{e}k.
\ee
Additionally, the finiteness of energy requires $\phi$ to vanish at the center of vortex as
\be
\phi\Big|_{r\to0} \to 0.
\ee
While the asymptotic behavior of $\phi$ is determined by the topology, that of $\vec\chi$ depends
on dynamical details. To see this, let us consider an effective mass of $\vec\chi$ under
the vortex background for $\phi$:
\be
\tilde m_\chi^2 \equiv \Omega^2 + 2\lambda\left(|\phi|^2-v^2\right)
\to \left\{
\begin{array}{ccl}
\Omega^2 & \quad & \text{as}\quad r\to \infty\\
\Omega^2 - 2\lambda v^2  & & \text{as}\quad r\to 0
\end{array}
\right.\,.
\label{eq:condition_condensation}
\ee
This implies that condensation of $\vec\chi$ tends to be suppressed if $\Omega^2 \gg 2\lambda v^2$.
On the contrary, when  $\Omega^2 \ll 2\lambda v^2$, $\vec\chi$ tends to condense only in the vicinity
of the vortex core. Namely, the vortex becomes superconducting.\footnote{Strictly speaking this is an abuse of terminology, borrowed from Witten's superconducting string setup, since in this case the string is not superconducting as the internal condensate is global. We will however use this terminology here to describe those vortices which possess a non-zero condensate in the core.
There are many papers studying conditions for a scalar field to condense on the Abelian vortex \cite{Witten:1984eb,Hill:1987ye,MacKenzie:1987ye,Haws:1988ax,Amsterdamski:1988zp,Babul:1987me,Babul:1988qt,Davis:1988jp,Davis:1988jq,Hill:1987qx}.}
Figs.~\ref{fig:vortex_profile}a and \ref{fig:vortex_profile}b show examples of non-superconducting and superconducting 
vortices, respectively.

In what follows, we will be mainly interested in the superconducting vortex because
it has the peculiar feature of being accompanied by the so-called non-Abelian moduli.
This can be understood as follows. As shown in Fig.~\ref{fig:vortex_profile}b, the $\vec\chi$ field
condenses at the center of the vortex. There, the global $SO(N)$ symmetry is spontaneously
broken to $SO(N-1)$. Hence, the Nambu-Goldstone bosons of $SO(N)/SO(N-1)$ appear but
they are localized around the vortex core since $SO(N)$ is unbroken in the bulk. In addition, 
the presence of the vortex also breaks the translational symmetry, so that the moduli
space of the single vortex is
\be
{\cal M}_{\text{1-vortex}} = \mathbb{R}^2 \times \frac{SO(N)}{SO(N-1)}
\simeq \mathbb{R}^2 \times S^{N-1}.
\ee
The orientational moduli $S^{N-1}$ are the so-called non-Abelian moduli.

\begin{figure}[ht]
\begin{center}
\includegraphics[width=12cm]{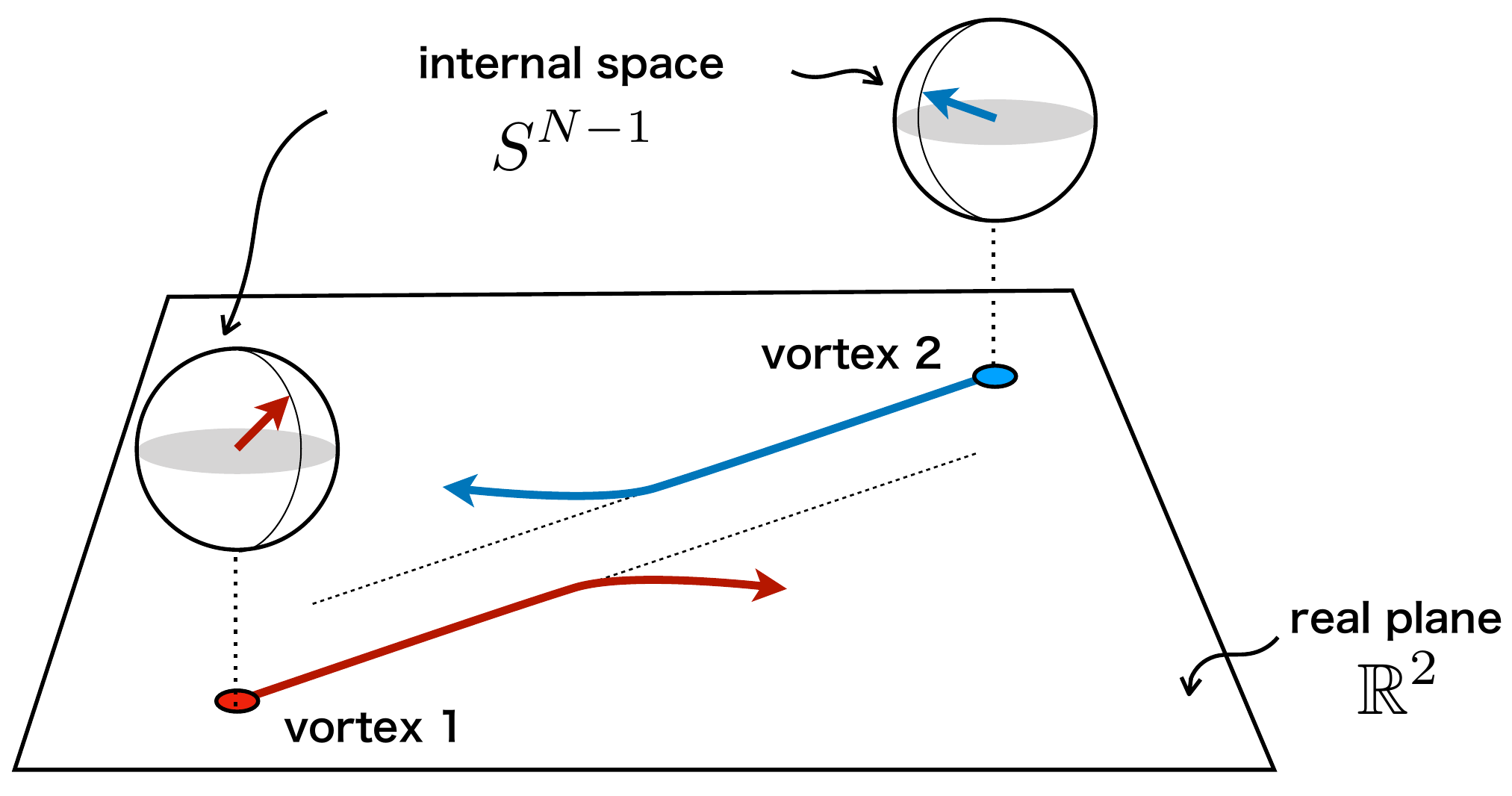}
\caption{Schematic picture of scattering of two non-Abelian vortices on the plane. Each vortex
has its own internal orientation $S^{N-1}$.}
\label{fig:schematic_scattering}
\end{center}
\end{figure}

In the subsequent sections, we will study scattering of two non-Abelian vortices.
The vortices can be regarded as particles of a finite size moving on the real plane $\mathbb{R}^2$,
since their individualities are obviously distinguishable as long as they are well separated.
In addition, the vortices should scatter inside the internal $S^{N-1}$ space.
The situation somehow resembles the scattering of two particles with spin
as shown in Fig.~\ref{fig:schematic_scattering}.


\section{Scattering of global non-Abelian vortices: A warm up}
\label{sec:GNAV}

\subsection{Set up}

In this section we will study the non-Abelian vortices in the simplified model where
the $U(1)$ symmetry is a global symmetry. Namely, we consider the Lagrangian
\be
{\cal L} &=& \p_\mu \phi \p^\mu \phi^* 
+ \p_\mu \chi_a\p^\mu\chi_a - V,
\ee
where the scalar potential is same as the one in Eq.~(\ref{eq:pot}).
This simplified model also admits vortices associated with the spontaneously broken
global $U(1)$ symmetry. However they are not local but global solitons.
The biggest difference is in their masses. The mass of the local vortex is finite,
whereas that of the global vortex diverges as $\log \Lambda$ where $\Lambda$
is an IR cutoff scale. 
On the other hand, concerning the condensation of $\vec \chi$ the global vortex is
quite similar to the local vortex. According to the ratio between $\Omega^2$ and $\lambda v^2$,
the global vortex induces the local condensation of $\vec\chi$ as the local vortex
shown in Fig.~\ref{fig:vortex_profile}.

Even though the dynamics of the global vortices with non-Abelian orientations is itself interesting, we only study it here as a first step before the more complicated numerical simulation
of local vortices in subsequent sections.

\begin{figure}[t]
\begin{center}
\includegraphics[width=14cm]{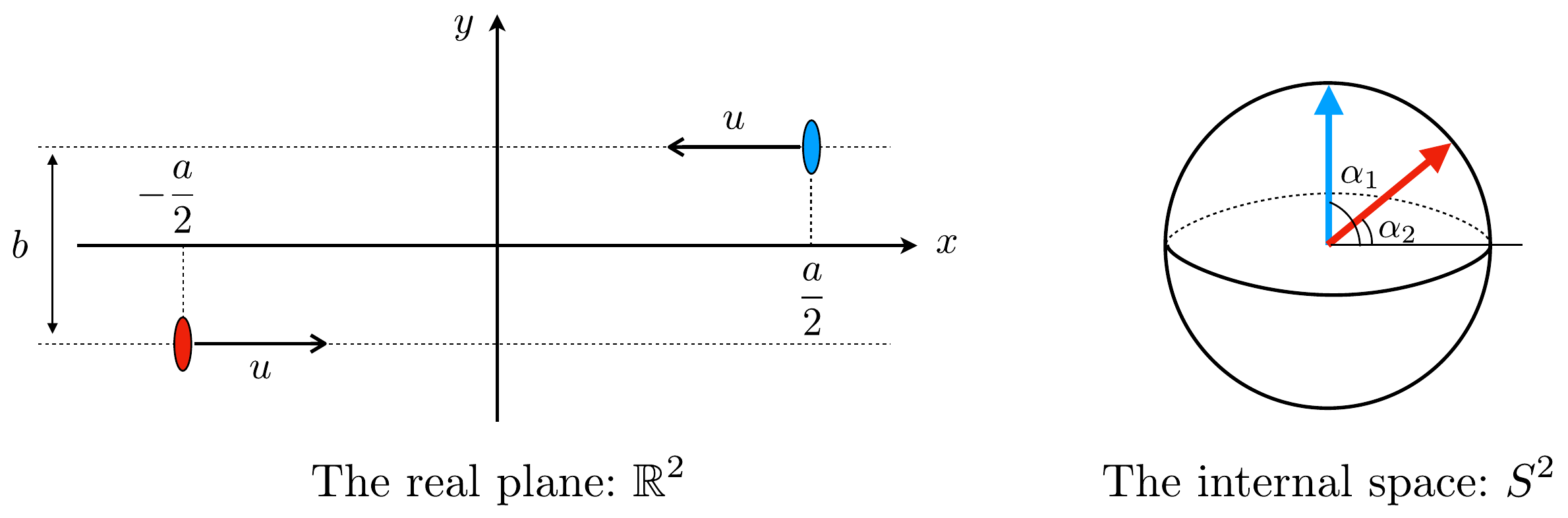}
\caption{Schematic picture of the initial state of two non-Abelian vortices.
The left diagram shows the vortices in the real plane and the right the vortices in the
internal plane. The painted ovals stand for the boosted vortices which are squashed
because of the Lorentz contraction effect.}
\label{fig:initial}
\end{center}
\end{figure}

In what follows, we will consider the case of $N=3$. Namely, the internal orientation manifold is $S^2$.
We will numerically simulate the scattering problem of two superconducting vortices 
$(\Omega^2 < \lambda v^2)$. We prepare the initial configurations as follows:
\begin{enumerate}[1)]
\item First, numerically solve the equations of motion for the static single non-Abelian vortex at the origin.
Let $\phi^{(0)}$ and $\chi^{(0)}$ be the solution
\be
\phi = \phi^{(0)}(x,y),\quad
\bm{\chi} = \chi^{(0)}(x,y) \bm{n}.
\ee
Here $\bm{n}$ is a constant three vector satisfying $|\bm{n}| = 1$ which determines a point on the internal $S^2$ moduli space.
In what follows, we will set $\bm{n}$ to be oriented along the third axis denoting $\bm{n}_0 = (0,0,1)$, which is always possible
without loss of generality. 
\item We shift the static solution by $\delta x =\pm a/2$ and $\delta y = \pm b/2$, 
and boost it with velocity $u$ along the $x$ axis. We also act with an $SO(3)$ rotation 
on $\bm{\chi}$ as
\be
\phi^{(1)} = \phi^{(0)}\!\left(\gamma \left(x-\frac{a}{2} + ut\right), y-\frac{b}{2}\right),\ \ 
\bm{\chi}^{(1)} = 
\chi^{(0)}\!\left(\gamma \left(x-\frac{a}{2} + ut\right), y-\frac{b}{2}\right) \bm{n}_1,
\label{eq:ini_phi_chi_1}\\
\phi^{(2)} = \phi^{(0)}\!\left(\gamma \left(x+\frac{a}{2} - ut\right), y+\frac{b}{2}\right),\ \ 
\bm{\chi}^{(2)} = 
\chi^{(0)}\!\left(\gamma \left(x+\frac{a}{2} - ut\right), y+\frac{b}{2}\right) \bm{n}_2,\nonumber
\label{eq:ini_phi_chi_2}
\ee
where the rapidity is $\gamma = 1/\sqrt{1-u^2}$, and $\bm{n}_i = R_i\bm{n}_0$ is a rotated $S^2$ vector with $SO(3)$ rotation
matrix $R_i$
\be
R_i = \left(
\begin{array}{ccc}
1 & 0 & 0\\
0 & \sin\alpha_i & \cos\alpha_i\\
0 & -\cos\alpha_i & \sin \alpha_i
\end{array}
\right),
\ee
with $\alpha_i \in [-\pi/2,\pi/2]$. We refer to $\alpha_1 - \alpha_2$ as the relative internal angle at the initial time.
\item We prepare the initial configuration for integrating the equations of motion
in the real time \textit{\`a la} Abrikosov as
\be
\phi(0,x,y) &=& \frac{1}{v} \phi^{(1)}(0,x,y)\phi^{(2)}(0,x,y),\\
\chi_a(0,x,y) &=& \chi^{(1)}_a(0,x,y) + \chi^{(2)}_a(0,x,y),
\ee
and
\be
\dot\phi(0,x,y) &=& \frac{1}{v} \left(\dot \phi^{(1)}(0,x,y)\phi^{(2)}(0,x,y) + 
\phi^{(1)}(0,x,y)\dot\phi^{(2)}(0,x,y)\right),\\
\dot\chi_a(0,x,y) &=& \dot\chi^{(1)}_a(0,x,y) + \dot\chi^{(2)}_a(0,x,y).
\ee
\end{enumerate}
The initial configurations prepared through these steps are illustrated in Fig.~\ref{fig:initial}.

\subsection{Head on collisions}

We now show results of the head on collisions of two non-Abelian vortices
with three different $\alpha_2 = \frac{\pi}{2}, 0, -\frac{\pi}{2}$ whereas we fix $\alpha_1 = \frac{\pi}{2}$. 
We call the two vortices parallel if
$\alpha_1 - \alpha_2 = 0$, anti-parallel if $\alpha_1 - \alpha_2 = \pi$, and 
orthogonal if $\alpha_1 - \alpha_2 = \frac{\pi}{2}$.
We will set our parameters $(\lambda,v,\Omega) = (2,2,1)$, so that the vortices are in the
superconducting regime ($|\bm{\chi}|\neq 0$ in the vortex core). 
Moreover, we always set the initial distance as $a=10$ 
in the unit of $\Omega$. The impact parameter is set to zero, $b=0$.

\paragraph{Parallel vortices $(\alpha_1,\alpha_2) = \left(\frac{\pi}{2},\frac{\pi}{2}\right)$:}

First we consider the scattering of two vortices with parallel orientations.
We repeatedly perform numerical simulations by changing the initial velocity $u$ as
$u = 0.6, 0.65, 0.7, 0.75, 0.8, 0.85$. 
\begin{figure}[b]
\begin{center}
\includegraphics[width=13cm]{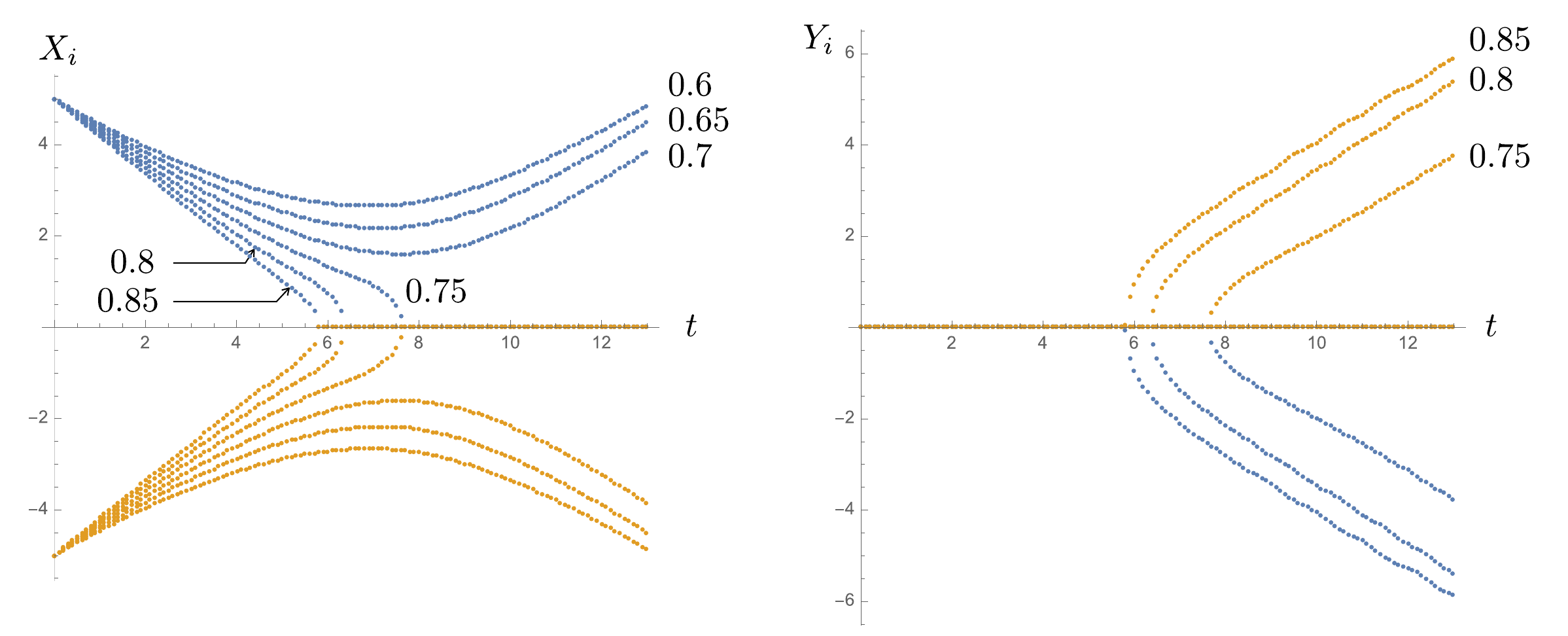}
\caption{The trajectories $(X_i(t),Y_i(t))$ of the two parallel vortices for $u=0.6, 0.65, 0.7,
0.75, 0.8, 0.85$.}
\label{fig:D01}
\end{center}
\end{figure}
Since the two vortices are global vortices with positive topological charges $k=1$, we expect that they interact by a long range repulsive force
as usual global vortices without the orientational degrees of freedom. Our results are consistent
with this expectation. Indeed, for $a=10$ a threshold is found at $u_c \in (0.7, 0.75)$ above which
the vortices scatter at right angles while below they backscatter along the incoming
directions. 
Fig.~\ref{fig:D01} shows the positions $(X_i(t),Y_i(t))$ of the vortices 
determined by $\phi(X_i(t),Y_i(t)) = 0$. As is seen from Fig.~\ref{fig:D01}, the vortices
bounce off on the $x$ axis for $u=0.6, 0.65, 0.7$. On the other hand, the vortices go beyond
the potential barrier and collide head on and scatter at right angles (the motion transitions from
 the $x$ axis to the $y$ axis) for $u=0.75, 0.8, 0.85$.

\begin{figure}[h]
\begin{center}
\includegraphics[width=15cm]{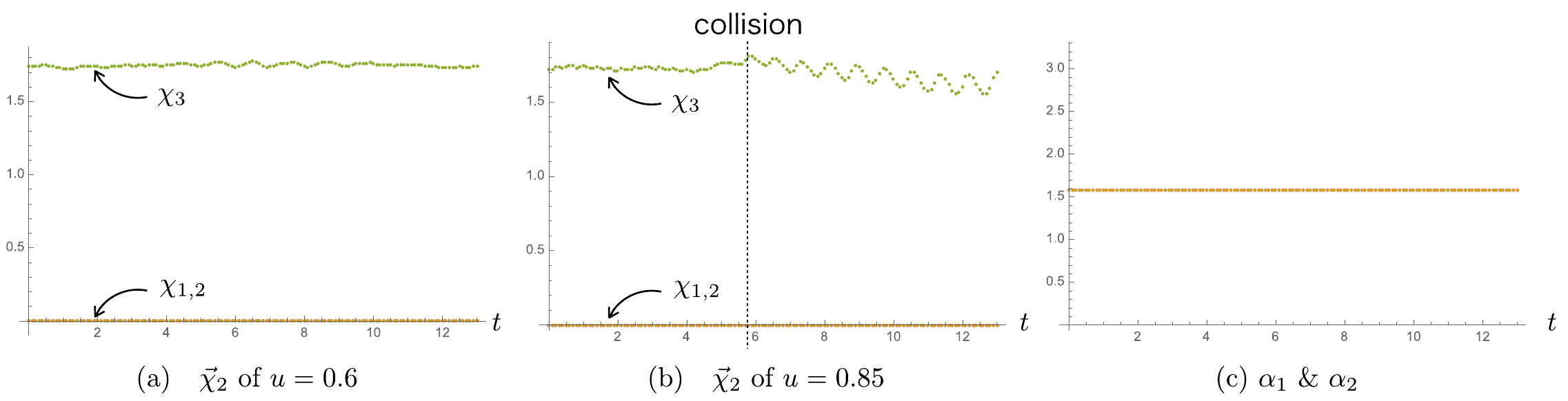}
\caption{Evolution of $\vec \chi_2 = \vec \chi(X_2(t),Y_2(t))$ for $u=0.6$ and $u=0.85$.}
\label{fig:D01_alpha}
\end{center}
\end{figure}
\begin{figure}[h]
\begin{center}
\includegraphics[width=15cm]{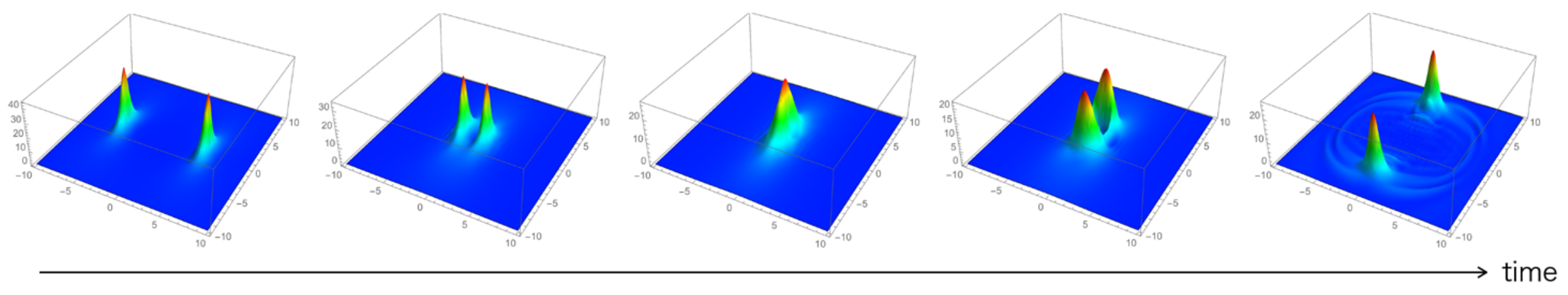}
\caption{Evolution of energy density for two parallel vortices scattering with velocities $u=0.85$.}
\label{fig:D01_u085}
\end{center}
\end{figure}

We also observed how the internal orientations evolve. 
We find that they are preserved for all cases,
namely, $\alpha_1= \alpha_2 = \frac{\pi}{2}$ throughout the whole simulations. 
However, the amplitude $\left|\bm{\chi}\right|$ is not a constant of motion. 
A massive (Higgs) mode is excited during the scattering as shown in Fig.~\ref{fig:D01_alpha}, the excitation being particularly intense after the vortices scatter at right angles.
This is because the fields $\phi$ and $\bm{\chi}$ are strongly deformed during the collision, where the individual presence of two vortices is not obvious at all,
see Fig.~\ref{fig:D01_u085}.

\begin{figure}[h]
\begin{center}
\includegraphics[width=7cm]{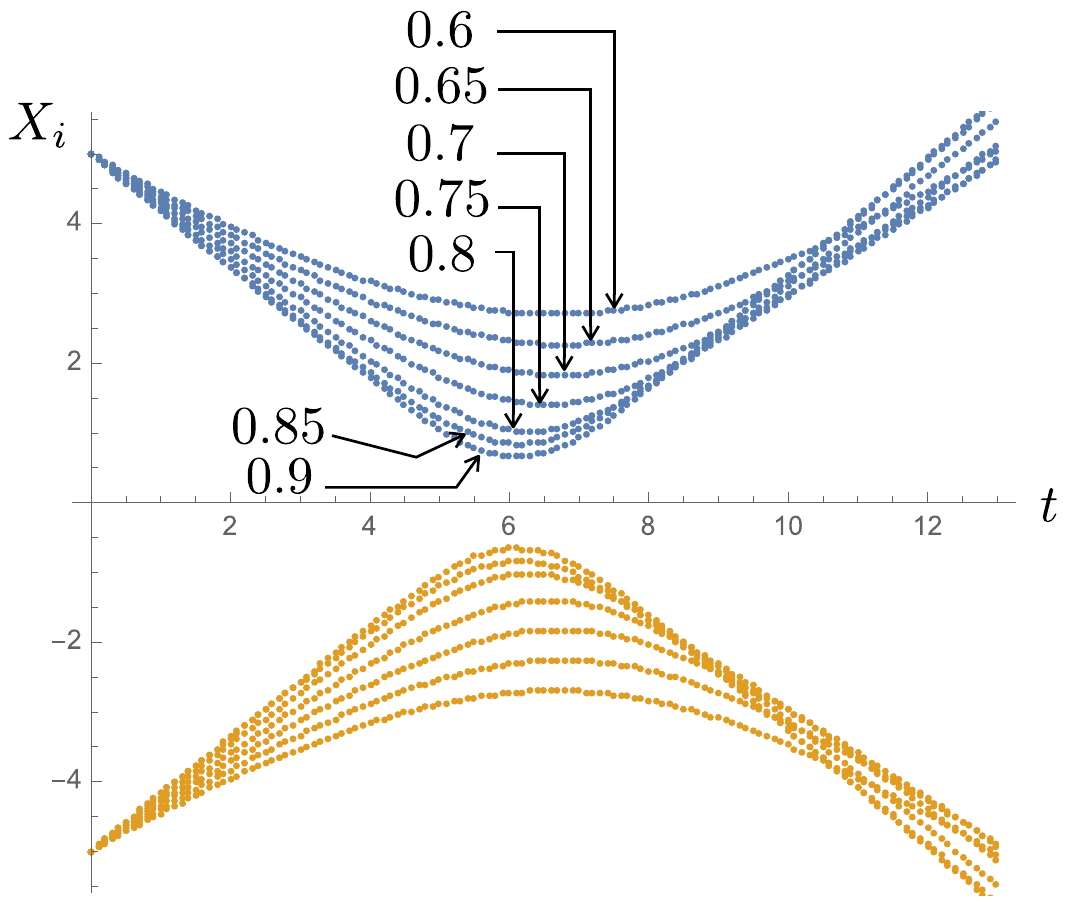}
\caption{The trajectories $X_i(t)$ of two anti-parallel vortices for $u=0.6, 0.65, 0.7,
0.75, 0.8, 0.85, 0.9$. We do not show $Y_i(t)$ because it always vanishes.}
\label{fig:E01}
\end{center}
\end{figure}

\begin{figure}[h]
\begin{center}
\includegraphics[width=15cm]{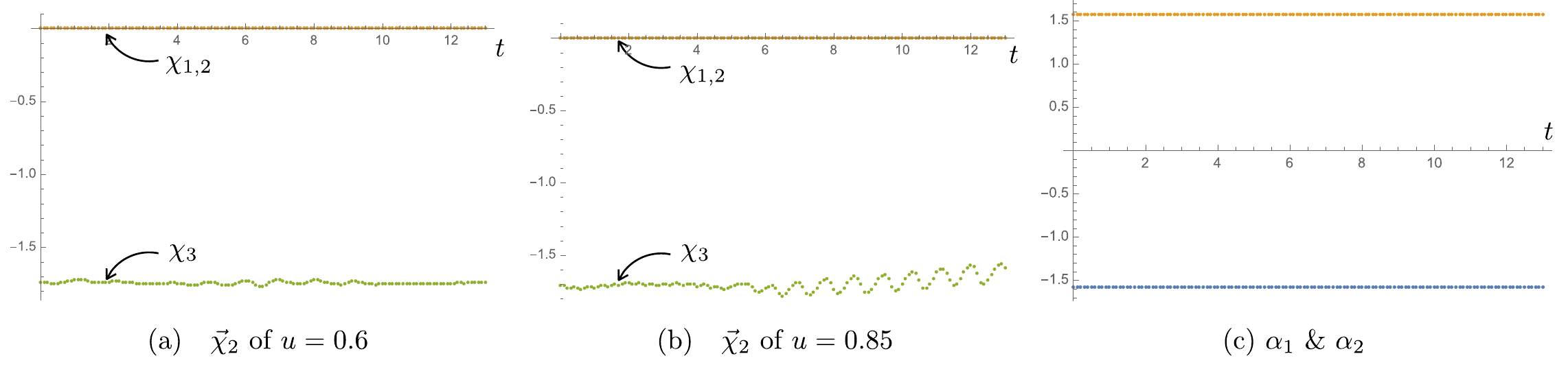}
\caption{Evolution of $\vec \chi_2 = \vec \chi(X_2(t),Y_2(t))$ for $u=0.6$ and $u=0.85$.}
\label{fig:E01_alpha}
\end{center}
\end{figure}

\begin{figure}[h]
\begin{center}
\includegraphics[width=15cm]{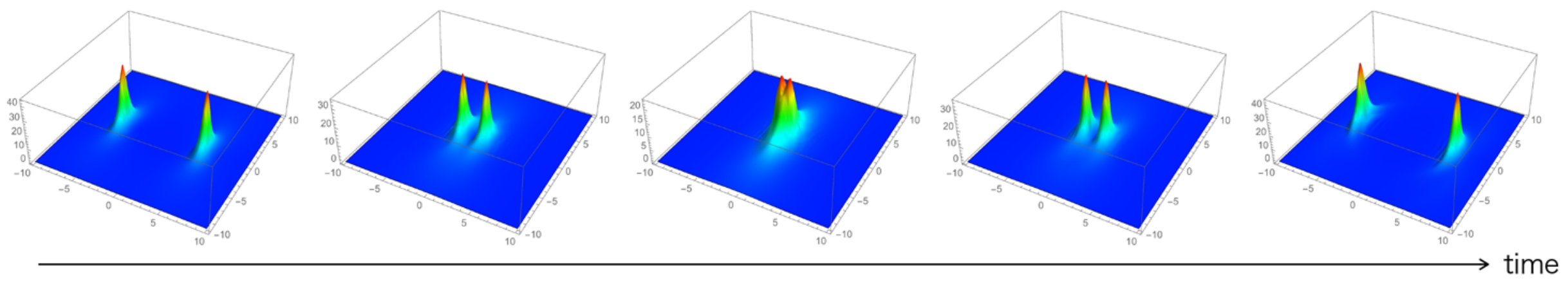}
\caption{Evolution of energy density for the two anti-parallel vortices scattering with $u=0.85$.}
\label{fig:E01_u085}
\end{center}
\end{figure}

\paragraph{anti-parallel vortices $(\alpha_1,\alpha_2) = \left(\frac{\pi}{2},-\frac{\pi}{2}\right)$:}
Let us next study the scattering of two anti-parallel vortices.
In addition to $u=0.6, 0.65, 0.7, 0.75, 0.8, 0.85$, in this case we also simulated the scattering with $u=0.9$.
The results are quite different from those of the parallel vortices 
in the sense that the vortices always backscatter, regardless of the initial velocity $u$,
see Fig.~\ref{fig:E01}.
We observed that the internal orientation is a constant in time, similarly to the parallel case, namely, $(\alpha_1,\alpha_2) = \left(\frac{\pi}{2},-\frac{\pi}{2}\right)$ holds throughout the simulation.
However, the amplitudes $\bm{\chi}(X_i(t),Y_i(t))$ vary in time by exciting massive modes.
The excitation is milder for relatively small $u$, while more important for relatively large $u$,
see Fig.~\ref{fig:E01_alpha}.
Again, this can be understood from how much the vortices are deformed from their
asymptotic individual configurations. When the vortices collide with a greater speed,
they get closer. Then the deformation becomes larger, which leads to greater excitations,
see Fig.~\ref{fig:E01_u085}.

As we have seen, the anti-parallel vortices bounce off even in the case of high initial
velocities $u=0.9$. This implies that the inter-vortex force between anti-parallel global vortices 
is a very strong repulsion.

\paragraph{Orthogonal vortices $(\alpha_1,\alpha_2) = \left(\frac{\pi}{2},0\right)$:}
As in the previous case, we simulated the scattering of orthogonal vortices for 
$u=0.6, 0.65, 0.7, 0.75, 0.8, 0.85, 0.9$. The resulting dynamics are qualitatively and partially
equal to those of the parallel vortices. Namely, there is a threshold velocity below (above) which
the vortices backscatter (scatter at right angle).
The threshold $u_c$
is in this case (for $a=10$) between $u=0.75$ and $u = 0.8$, which is higher than that of parallel vortices, marking stronger repulsive inter-vortex forces.
\begin{figure}[h]
\begin{center}
\includegraphics[width=13cm]{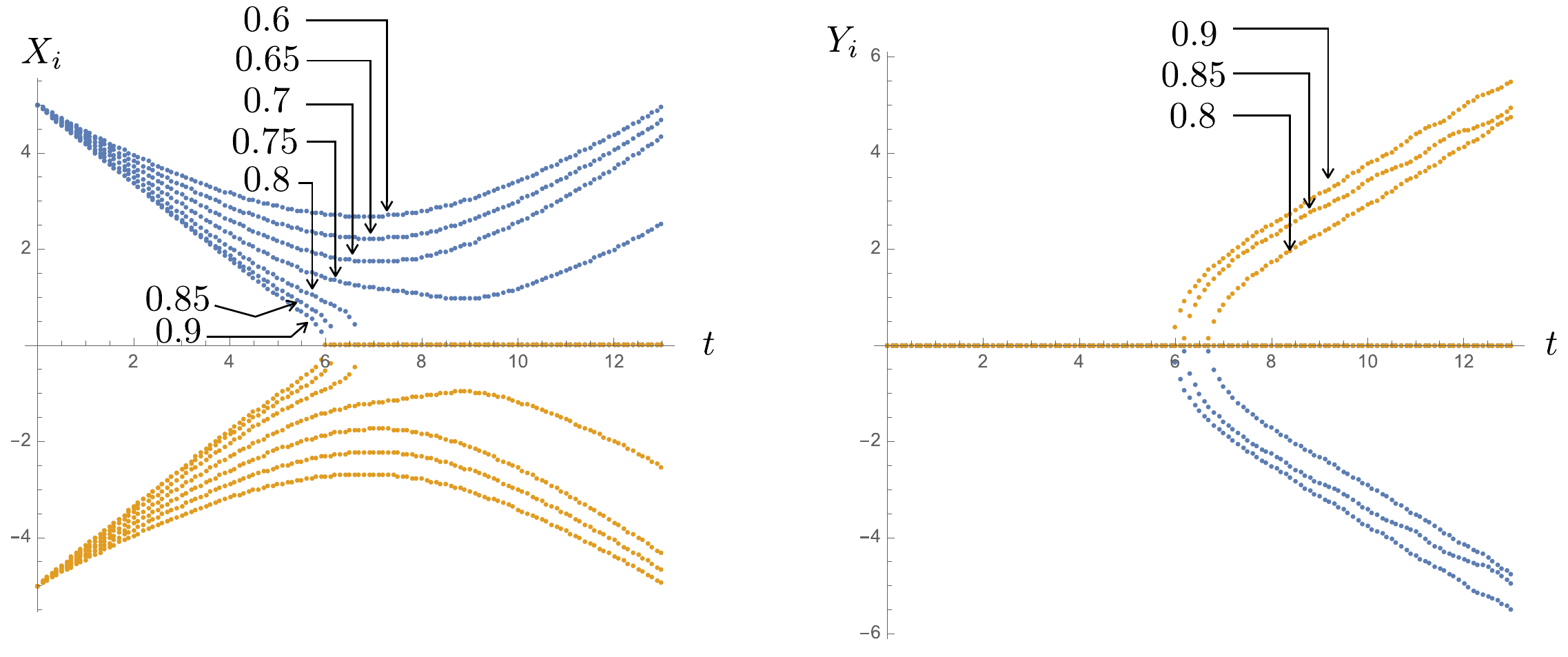}
\caption{The trajectories $(X_i(t),Y_i(t))$ of the two orthogonal vortices for $u=0.6, 0.65, 0.7,
0.75, 0.8, 0.85, 0.9$.}
\label{fig:A01}
\end{center}
\end{figure}
Fig.~\ref{fig:A01} shows the results of our numerical simulations, and is quite similar to
Fig.~\ref{fig:D01} for the parallel vortices. However, there is an important difference:
the internal orientations are no longer constants of motion.
\begin{figure}[h]
\begin{center}
\includegraphics[width=15cm]{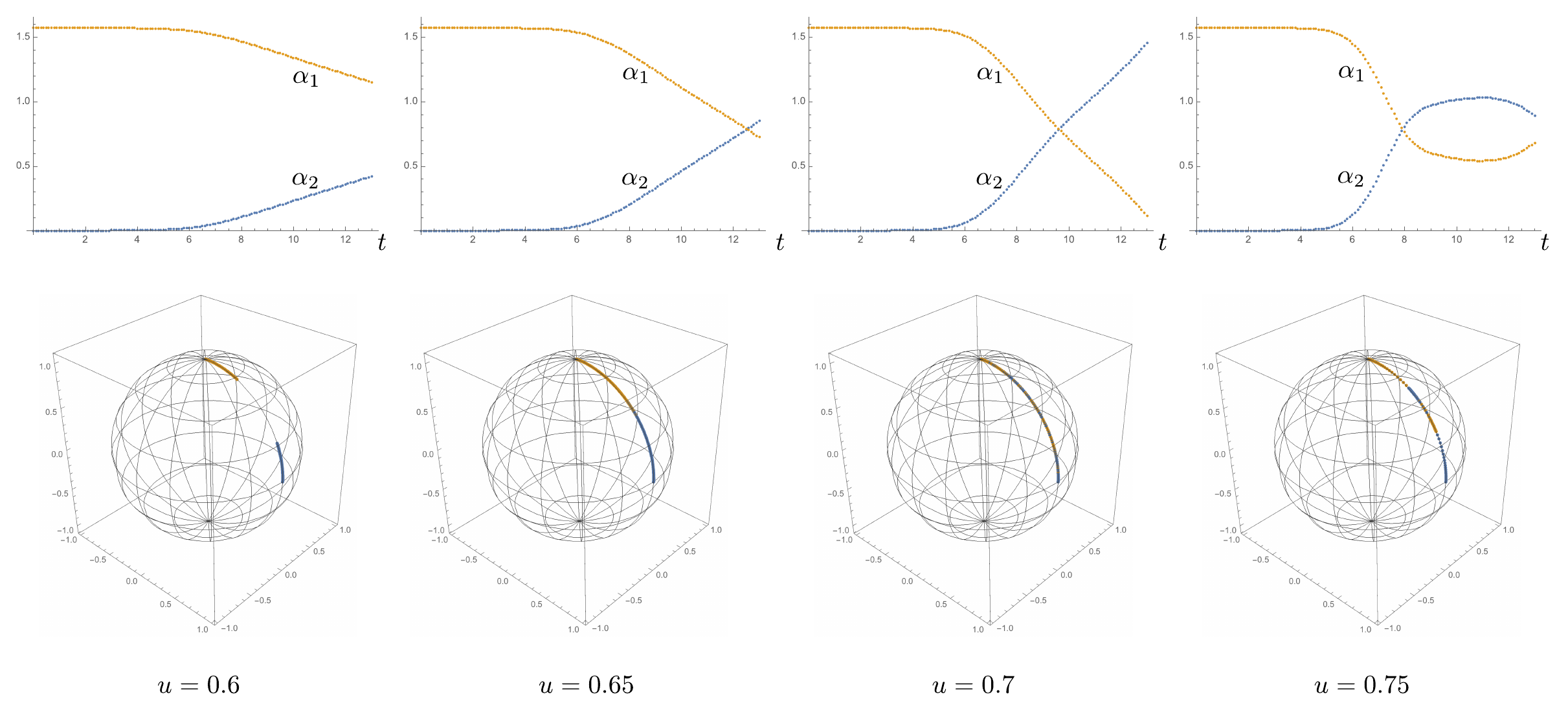}
\caption{Evolutions of internal orientations $\alpha_{1,2}$ of the orthogonal vortices
with the velocities $u=0.6, 0.65, 0.7, 0.75$ below the threshold.}
\label{fig:A01_alpha_1}
\end{center}
\end{figure}
Fig.~\ref{fig:A01_alpha_1} shows the evolution of the orientations $\alpha_{1,2}$ for the
cases that the vortices backscatter ($u=0.6, 0.65, 0.7, 0.75$). For all of these simulations,
the orientations initially do not change. However, they start to vary approximately when the vortices reach
their closest separation. $\alpha_1$ decreases toward $\frac{\pi}{4}$, and at the same time
$\alpha_2$ increases toward $\frac{\pi}{4}$. The decreasing of $\alpha_1$ and increasing of $\alpha_2$
linearly continues for relatively small $u = 0.6, 0.65$. Therefore, the dynamics of the internal orientation
is almost circular motion at constant velocity. However, this is not the case for $u$ near the
threshold. We found that the internal orientations oscillate around $\alpha_{1,2} = \frac{\pi}{4}$
for $u=0.75$, see the right-most panel of Fig.~\ref{fig:A01_alpha_1}.

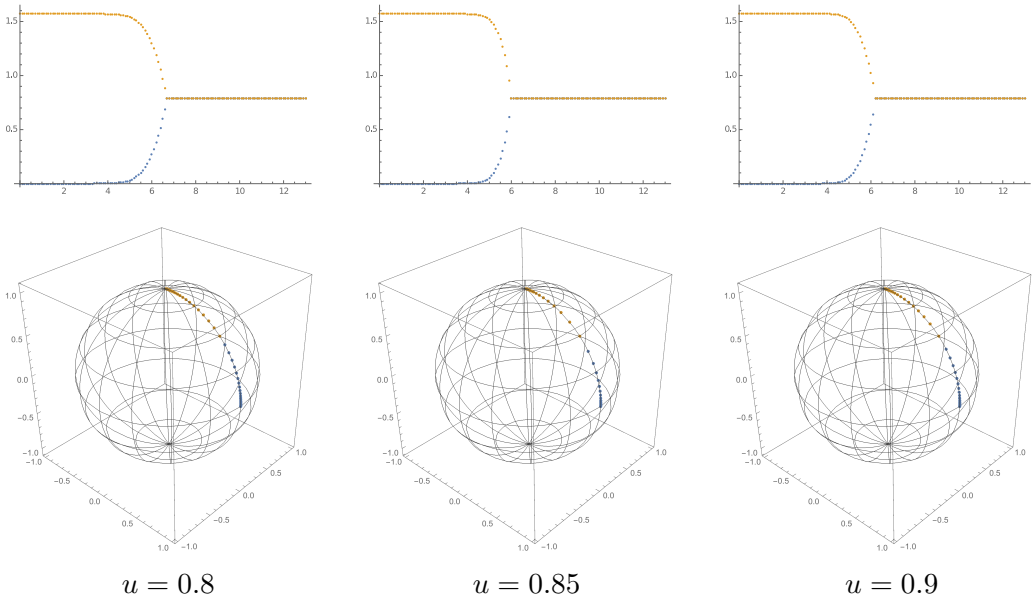
\begin{figure}[h]
\begin{center}
\def\svgwidth{14cm}
\begingroup%
  \makeatletter%
  \providecommand\color[2][]{%
    \errmessage{(Inkscape) Color is used for the text in Inkscape, but the package 'color.sty' is not loaded}%
    \renewcommand\color[2][]{}%
  }%
  \providecommand\transparent[1]{%
    \errmessage{(Inkscape) Transparency is used (non-zero) for the text in Inkscape, but the package 'transparent.sty' is not loaded}%
    \renewcommand\transparent[1]{}%
  }%
  \providecommand\rotatebox[2]{#2}%
  \newcommand*\fsize{\dimexpr\f@size pt\relax}%
  \newcommand*\lineheight[1]{\fontsize{\fsize}{#1\fsize}\selectfont}%
  \ifx\svgwidth\undefined%
    \setlength{\unitlength}{653.69668579bp}%
    \ifx\svgscale\undefined%
      \relax%
    \else%
      \setlength{\unitlength}{\unitlength * \real{\svgscale}}%
    \fi%
  \else%
    \setlength{\unitlength}{\svgwidth}%
  \fi%
  \global\let\svgwidth\undefined%
  \global\let\svgscale\undefined%
  \makeatother%
  \begin{picture}(1,0.58600356)%
    \lineheight{1}%
    \setlength\tabcolsep{0pt}%
    \put(0,0){\includegraphics[width=\unitlength,page=1]{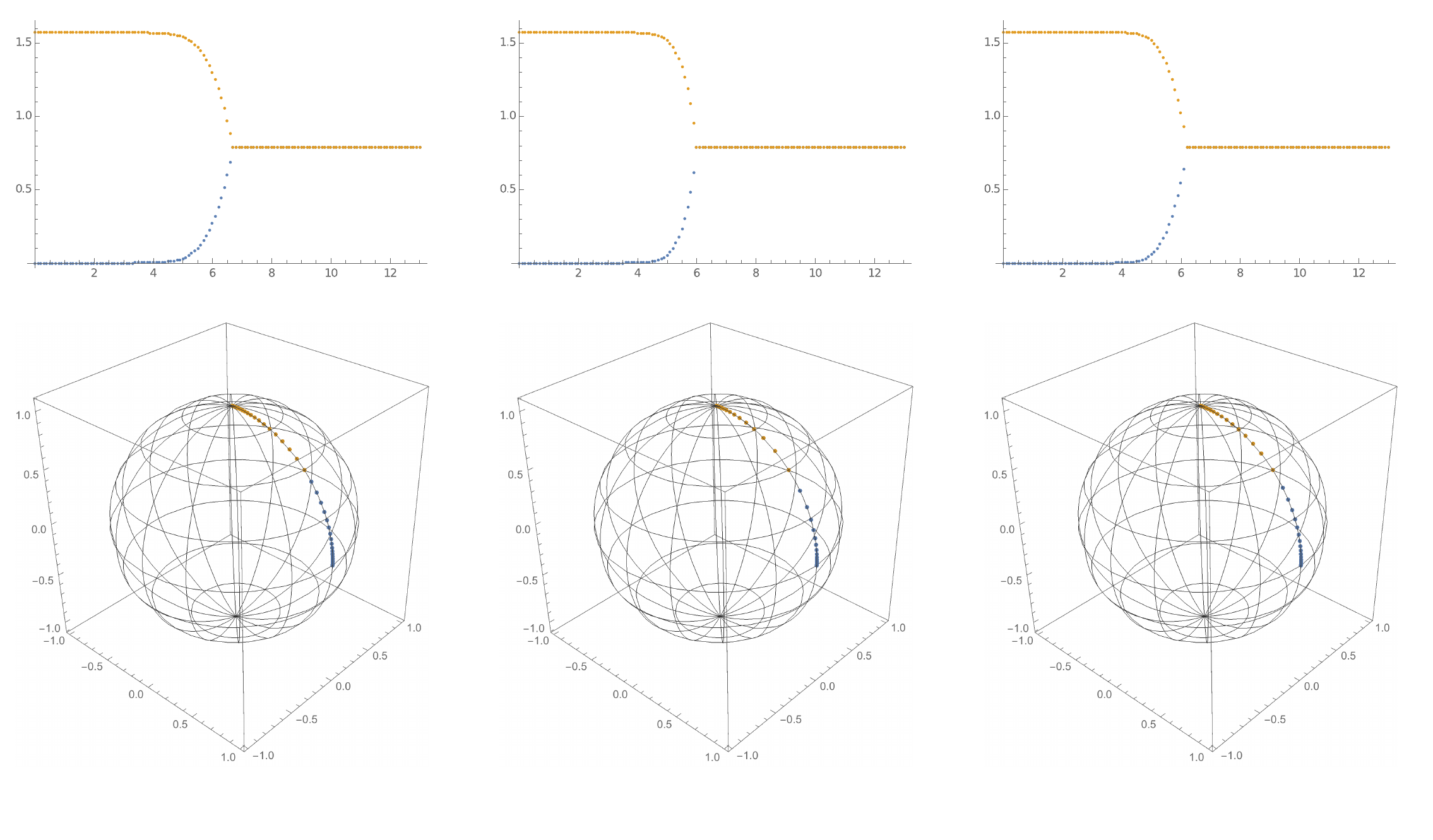}}%
    \put(0.12,0.015){\color[rgb]{0,0,0}\makebox(0,0)[lt]{\lineheight{1.25}\smash{\begin{tabular}[t]{l}$u=0.8$\end{tabular}}}}%
    \put(0.45,0.015){\color[rgb]{0,0,0}\makebox(0,0)[lt]{\lineheight{1.25}\smash{\begin{tabular}[t]{l}$u=0.85$\end{tabular}}}}%
    \put(0.80,0.015){\color[rgb]{0,0,0}\makebox(0,0)[lt]{\lineheight{1.25}\smash{\begin{tabular}[t]{l}$u=0.9$\end{tabular}}}}%
  \end{picture}%
\endgroup%

\caption{Evolutions of internal orientations $\alpha_{1,2}$ of the orthogonal vortices
with the velocities $u=0.8, 0.85, 0.9$ above the threshold.}
\label{fig:A01_alpha_2}
\end{center}
\end{figure}

Evolution of the internal orientations for $u$ beyond the threshold 
is significantly different from that below the threshold.
As shown in Fig.~\ref{fig:A01_alpha_2}, when the vortices collide on the real plane,
the orientations also collide on the internal sphere. While the vortices scatter off
at right angles in the real plane, the orientations become stuck 
at $\alpha_1 = \alpha_2 = \frac{\pi}{4}$. In other words, the orthogonal vortices
become parallel vortices after scattering at right angles.
This is an unexpected phenomenon compared to the previous study of head-on collisions of BPS non-Abelian vortices with $\mathbb{C}P^{N-1}$ non-Abelian moduli, where
the orientations also scatter at right angles in the internal space \cite{Eto:2006db}. 
Hence, the orientations becoming stuck in parallel orientations 
is specific to the non-Abelian vortices with $S^{N-1}$ orientational moduli.

\paragraph{Necessary condition for right angle scattering:} 
We observed that the two orientations tend to be parallel when
the vortices approach on the real plane. 
Then, we observed that right angle scattering can occur only when the orientations are 
parallel at least at the moment of the collision. This is also true for the scattering
of parallel vortices as we have seen in Fig.~\ref{fig:D01}.
Hence, {\it the necessary condition for right-angles scattering to occur is that
the two orientations should be parallel at the moment of the collision.}
This is reasonable in the following sense. For right angle scattering, we cannot
say which vortex scatters towards which direction since physical observables like
the energy density are symmetric under replacement of the two vortices.
Namely, at the moment of scattering, they completely merge into each other and scatter
off at right angles. This can happen only when the two vortices are indistinguishable.
Therefore, the orientations should be parallel if the vortices scatter at right angles.

\subsection{Scattering of orthogonal vortices with non-zero impact parameter}

\begin{figure}[h]
\begin{center}
\includegraphics[width=7cm]{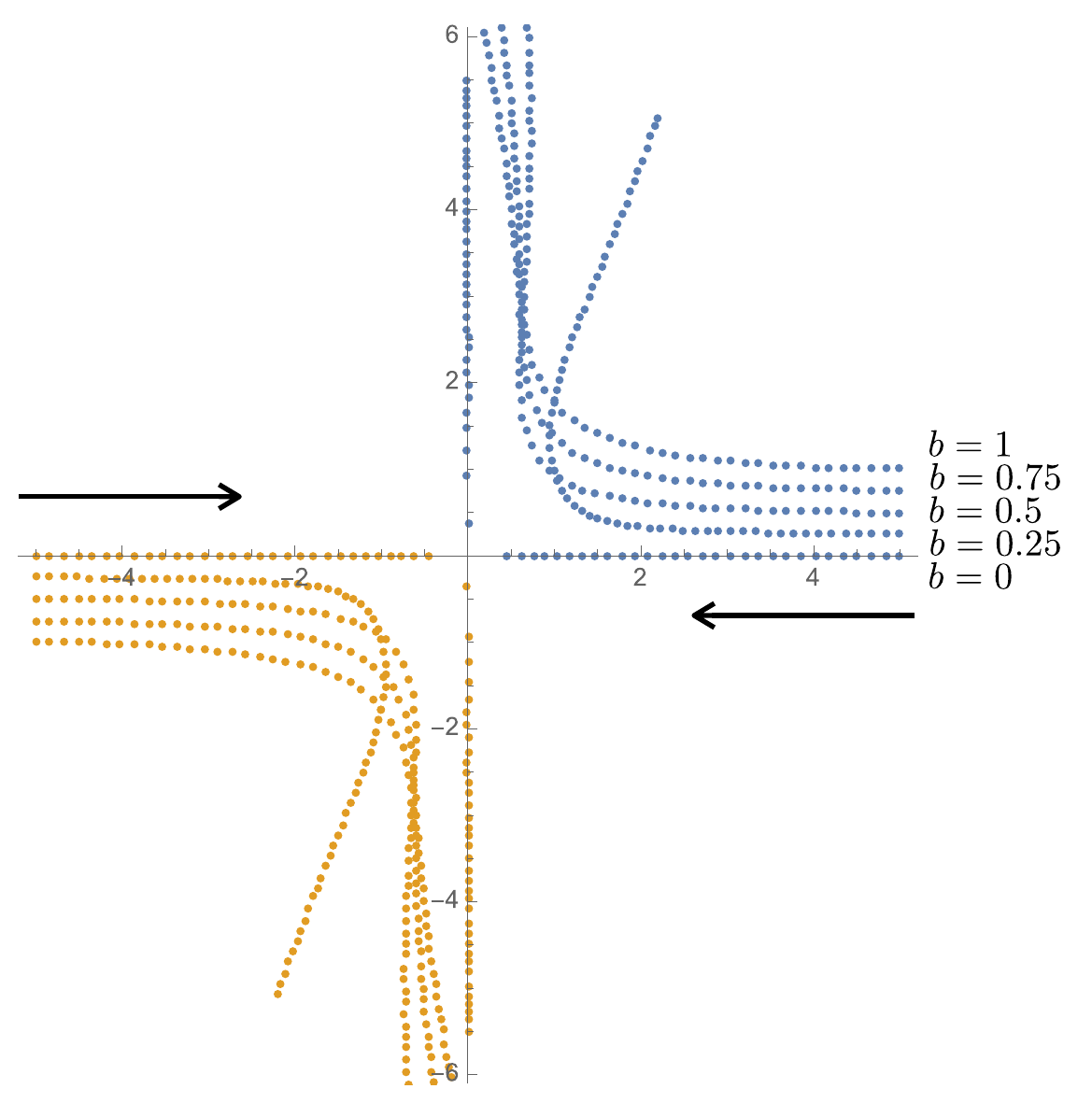}
\caption{Trajectories of the two initially orthogonal 
vortices with impact parameter $b=0, 0.25, 0.5, 0.75, 1$.}
\label{fig:A_b}
\end{center}
\end{figure}

We now briefly show the scattering of orthogonal vortices when turning on the impact
parameter $b$. Fig.~\ref{fig:A_b} summarizes our results. We observe that the impact parameter does not significantly affect the physics of the orientational moduli. Since we are primarily interested in this aspect, we shall not investigate this matter further until Section~\ref{exotic}.


\section{Static interaction of local non-Abelian vortices}
\label{sec:static_int_lnA}

We are now ready to proceed with the scattering of fully local non-Abelian vortices by 
turning on the $U(1)$ gauge coupling $e \neq 0$.

\subsection{Abelian v.s. non-Abelian vortices}

When the gauge coupling is turned on, the gauge fields become dynamical fields.
The full equations of motion  resulting from Lagrangian 
(\ref{eq:lag}) with potential (\ref{eq:pot}) are
\be
\p^\mu\p_\mu A^\nu - \p^\nu \p_\mu A^\mu + ie\left(\phi D^\nu\phi^* - \phi^* D^\nu \phi\right) = 0,\label{eq:f_eom_1}\\
D_\mu D^\mu \phi + \frac{\p V}{\p \phi^*} = 0,\label{eq:f_eom_2}\\
\p_\mu \p^\mu \chi^i  + \frac{1}{2}\frac{\p V}{\p \chi^i} = 0.\label{eq:f_eom_3}
\ee

To begin, we analyze the single non-Abelian local vortex. 
Let us make a common ansatz for the single vortex placed at the origin
\be
\phi = v f(r) e^{i\theta},\quad \bm{\chi} = v g(r) \bm{n},\quad A^i = \frac{1}{e}\epsilon^{ij}\frac{x^j}{r^2}(1-a(r)),
\label{eq:vortex_ansatz}
\ee
where $r$ and $\theta$ are the two-dimensional polar coordinates, and $\bm{n}$ is an arbitrary constant unit vector.
The equations of motion become
\be
f'' + \frac{1}{r}f' - \frac{a^2}{r^2}f - \frac{m_\phi^2}{2}(f^2+g^2-1)f  = 0,\\
g'' + \frac{1}{r}g' - \frac{m_\phi^2}{2}(f^2+g^2-1)g - m_\chi^2 g = 0,\\
a'' - \frac{1}{r}a' - m_\gamma^2 f^2 a = 0.
\ee
Here, primes stand for derivatives in terms of $r$.
We can further simplify these equations by introducing a dimensionless coordinate $\rho \equiv m_\phi r$, resulting in
\be
\ddot f + \frac{1}{\rho}\dot f - \frac{a^2}{\rho^2}f - \frac{1}{2}(f^2+g^2-1)f  = 0,\label{eq:ax_eom_1}\\
\ddot g + \frac{1}{\rho}\dot g - \frac{1}{2}(f^2+g^2-1)g - \left(\frac{m_\chi}{m_\phi}\right)^2 g = 0,\label{eq:ax_eom_2}\\
\ddot a - \frac{1}{\rho}\dot a - \left(\frac{m_\gamma}{m_\phi}\right)^2 f^2 a = 0,\label{eq:ax_eom_3}
\ee
where dots are now derivatives with respect to $\rho$. We are to solve these equations with the following boundary conditions
\be
(f,g,a) \to (1,0,0)\qquad \text{for}\quad \rho \to \infty,\\
(f, \dot g, a) \to (0,0,1)\qquad \text{for}\quad \rho \to 0.
\ee
Typical solutions are shown in Fig.~\ref{fig:vortex_profile}.

The vortex solutions essentially depend on the two mass ratios $m_\gamma/m_\phi$ and $m_\chi/m_\phi$.
Our main concern here is to clarify when the vortex becomes of the non-Abelian type.
Namely, we want to understand for which parameter region $\chi$-condensation at the vortex core is non-zero.
As mentioned below Eq.~(\ref{eq:condition_condensation}), we can roughly say that 
the $\chi$-condensation is non-zero when $\frac{m_\chi}{m_\phi} \ll \frac{1}{\sqrt2}$ holds ($\Omega^2 \ll 2\lambda v^2$) .
However, as we will discuss in the next subsection, an important property of the non-Abelian vortex is deeply related 
with the fact that the mass ratios $\frac{m_\chi}{m_\phi}$ and $\frac{m_\gamma}{m_\phi}$ are smaller or lager than $\frac{1}{2}$. 
It seems quite hard to analytically obtain a phase boundary dividing the parameter space into two regions, Abelian vortex
and non-Abelian vortex. Instead, here we perform a numerical survey.
\begin{figure}[t]
\begin{center}
\includegraphics[width=8cm]{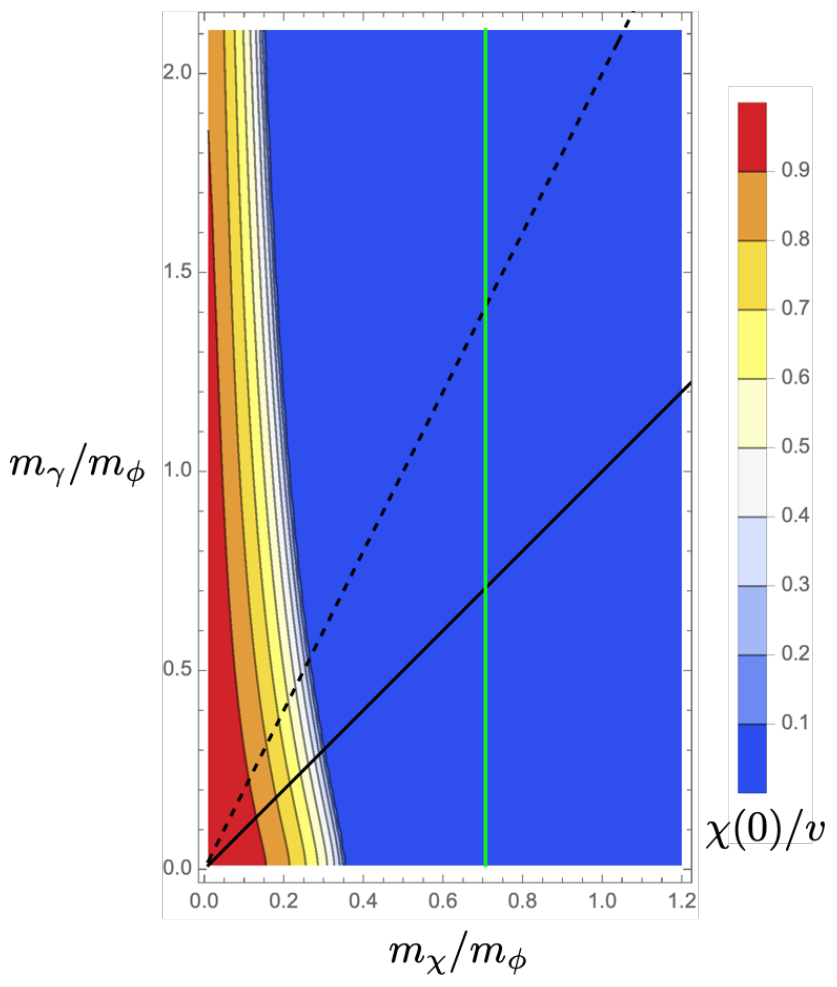}
\caption{The phase diagram of the single vortex. The vortex is of the Abelian type in the blue region, otherwise
it is of the non-Abelian type. The solid line $m_\gamma = m_\chi$ is the analytic estimation of the phase boundary between type I*A (above) and type II (below) non-Abelian vortices, whereas the dashed line $m_\gamma = 2m_\chi$ corresponds to the analytic estimation of the boundary between type I*A (below) and type I*B (above), see discussion around Table~\ref{tab:summary}.}
\label{fig:phase_boundary}
\end{center}
\end{figure}
We numerically solve the above vortex equations for points in the $\left(\frac{m_\chi}{m_\phi},\frac{m_\gamma}{m_\phi}\right)$ plane.
We choose $\frac{m_\chi}{m_\phi} = \frac{n_\chi}{100}$ and $\frac{m_\gamma}{m_\phi} = \frac{n_\gamma}{100}$ with
integers $1 \le n_\chi \le 121$ and $1 \le n_\gamma \le 211$. We call a vortex non-Abelian only if the $\chi$ condensation 
at the vortex center satisfies $g(0) > 0.01$, otherwise we regard it as Abelian.
The result is shown in Fig.~\ref{fig:phase_boundary}. The phase boundary between the Abelian and non-Abelian phases stands to the left side of the green vertical line $\frac{m_\chi}{m_\phi} = \frac{1}{\sqrt2}$, as expected.
Near the phase boundary, the $\chi$ condensation is very small. 
Hence, we expect the non-Abelian vortex can easily transform into the Abelian one under a dynamical process, and vice versa.
When we use non-Abelian vortices throughout the paper, we make sure to take parameters from deep within the non-Abelian phase.

\subsection{Static inter-vortex forces from a point vortex approximation}

Let us next derive the static vortex potential
which will be important to understand the various types of collisions of non-Abelian vortices.
It was previously studied in \cite{Tallarita:2017opp}, but here we analyze it in further detail,
finding the previously obtained formula of the inter-vortex interaction includes important higher order corrections.
Below we will use the techniques developed in Refs.~\cite{Speight:1996px} and \cite{Auzzi:2007wj}.

\subsubsection{The Abelian phase}

Before investigating the static vortex interaction of non-Abelian vortices, let us review that for the Abelian ones.
Therefore, we assume $g = 0$ for the time being.
Let us focus on the asymptotic behaviors of the profile functions by perturbing them as $f = 1- \delta f$, and
$a = \delta a$ at $r\to\infty$. 
Plugging these into the above equations of motion, and picking up only terms which are
of leading order at $r\to\infty$,
we have
\be
\delta f'' + \frac{\delta f'}{r} - m_\phi^2 \delta f &=& - \frac{\delta a^2}{r^2},
\label{eq:asy_Af}\\
\delta a'' - \frac{\delta a'}{r} - m_\gamma^2 \delta a &=& 0.
\label{eq:asy_Aa}
\ee
Note that we keep the quadratic term at the right hand side of Eq.~(\ref{eq:asy_Af}) since they can be of the same 
order as $\delta f$ \cite{Perivolaropoulos:1993uj,Eto:2009wq}, as we will see shortly below. 
Let us first solve Eq.~(\ref{eq:asy_Aa}).
Normalizable solutions are given by
\be
\delta a &=& \frac{1}{\sqrt2}\frac{\mu}{2\pi} m_\gamma r K_1(m_\gamma r) \simeq 
\frac{\mu}{4\sqrt\pi} \sqrt{m_\gamma r}\, e^{-m_\gamma r},
\label{eq:delta_Aa}
\ee
where $\mu$ is an undetermined constant which can be fixed numerically if necessary.
One can check that
$\mu$ is positive from the equations of motion. 
As expected, the decay constant is $m_\gamma^{-1}$.
With this asymptotic function in hand, we can now determine $\delta f$ by solving Eq.~(\ref{eq:asy_Af}).
If we ignore the quadratic term on the right hand side, we have $\delta f \simeq \frac{q}{4\sqrt{\pi}} \frac{1}{\sqrt{m_\phi r}} e^{-m_\phi r}$. 
However, the term on the right hand side is of order $-\frac{\mu^2 m_\gamma}{16 \pi r} e^{-2m_\gamma r}$.
Therefore, it can be ignored only when $m_\phi < 2 m_\gamma$ holds \cite{Perivolaropoulos:1993uj}.
When $m_\phi > 2 m_\gamma$ we have to take it into account, and the correct asymptotic behavior of $\delta f$ is then given by a piecewise expression
\be
\delta f 
\simeq 
\begin{cases}
\dfrac{q}{4 \sqrt\pi}
\dfrac{1}{\sqrt{m_\phi r}}\, e^{-m_\phi r} &\quad\text{for}\quad m_\phi < 2m_\gamma,\\
\dfrac{\mu^2}{16\pi} \dfrac{m_\gamma^2}{m_\phi^2-4m_\gamma^2}\dfrac{1}{m_\gamma r}e^{-2m_\gamma r} &\quad\text{for}\quad m_\phi > 2 m_\gamma,
\end{cases}
\ee
where $q$ is a positive constant. 
This can also be expressed in terms of the Bessel function $K_0$ as 
\be
\delta f 
\simeq \left\{
\begin{array}{lcl}
\dfrac{q}{2\sqrt2 \,\pi} K_0(m_\phi r)  & & \text{for}\quad m_\phi < 2m_\gamma,\\
\dfrac{m_\gamma^2}{m_\phi^2-4m_\gamma^2} \left(\dfrac{\mu}{2\sqrt2 \,\pi} K_0(m_\gamma r)\right)^2 & &
\text{for}\quad m_\phi > 2m_\gamma,
\end{array}
\right.\,.
\label{eq:asym_Acase2}
\ee

Now we are ready to figure out the asymptotic interaction between two static vortices.
To this end, let us move to a different gauge 
where $\phi$ becomes real by $\phi \to e^{-i\theta} \phi$ and $A_\mu \to A_\mu + \frac{1}{e}\p_\mu \theta$. 
Then, we introduce a real scalar field $\varphi$ through $\phi= 1 - \varphi/\sqrt2$.

\paragraph{The case of $m_\phi <2m_\gamma$}

For a single vortex, the asymptotic behaviors in the new gauge are given by
\be
\varphi &\to& v \frac{q}{2\pi} K_0(m_\phi r),
\label{eq:asym_Aphi}\\
A^i &\to&  
- \frac{1}{e}\epsilon^{ij}\frac{x^j}{r^2} \left(\frac{1}{\sqrt2}\frac{\mu}{2\pi} m_\gamma r K_1(m_\gamma r)\right)
= - \frac{1}{\sqrt2}\frac{\mu}{2\pi e}\left(\vec k \times \vec \nabla\right)^i K_0(m_\gamma r),
\label{eq:asym_Aa}
\ee
where we used $\frac{d}{dr} K_0(r) = - K_1(r)$ and introduced a constant vector $\vec k = (0,0,1)$.
Our next task is to replicate these asymptotic behaviors in a quadratic Lagrangian,
\be
{\cal L}_{\rm free} = - \frac{1}{4}F_{\mu\nu}F^{\mu\nu} + \frac{m_\gamma^2}{2}A_\mu A^\mu 
+ \frac{1}{2} \p_\mu \varphi \p^\mu \varphi - \frac{m_\phi^2}{2} \varphi^2.
\ee
In addition, we introduce source terms \cite{Speight:1996px}
\be
{\cal L}_{\rm s} = \varphi \rho_\phi  - A_\mu j^\mu.
\ee
Then the equations of motion are given by
\be
(\p_\mu \p^\mu + m_\gamma^2) a^\nu - \p^\nu \p_\mu a^\mu = j^\nu,\\
(\p_\mu \p^\mu + m_\phi^2) \varphi  = \rho_\phi.\label{eq:EOM_Avarphi}
\ee
In order to reproduce the asymptotic behaviors from the above linearized equations, we can make use of the 2 dimensional Green function
\be
\left(-\vec \nabla^2 + m^2\right) K_0(m r) = 2\pi \delta^{(2)}(\vec x).
\label{eq:green}
\ee
We end up with the following point sources to replicate a single Abelian vortex as a point object
\be
\rho_\phi = vq \, \delta^{(2)}(\vec x),\quad 
\big(j^0,\vec j\big) = \left(0,- \frac{1}{\sqrt2}\frac{\mu}{e} \vec k \times \vec\nabla \delta^{(2)}(\vec x)\right).
\ee
Now, we can place one Abelian vortex at $\vec x = \vec x_1$ and
the other at $\vec x = \vec x_2$. Namely, we have
\be
\rho_\phi^i &=& vq \, \delta^{(2)}(\vec x - \vec x_i),
\label{eq:scalar_Asource}\\
\big(j^0_i,\vec j_i\big) &=& \left(0,- \frac{1}{\sqrt2}\frac{\mu}{e} \vec k \times \vec\nabla \delta^{(2)}(\vec x - \vec x_i)\right),
\ee
with $i=1,2$, the total source being just the sum of two sources,
\be
(\rho_{\phi},j^\mu) = \sum_{i=1}^2 (\rho_{\phi}^i,j^\mu_i).
\ee
Since we work in the linearized theory, the created fields are simply superpositions of each field
\be
\varphi_i &=& v\frac{q}{2\pi}K_0(m_\phi|\vec x - \vec x_i|),
\label{eq:asymptotic_Ascalar_field}\\
\vec A_i &=& - \frac{1}{\sqrt2}\frac{\mu}{2\pi e} \vec k \times \vec \nabla K_0(m_\gamma |\vec x - \vec x_i|).
\ee
Hence, the interaction between two point sources mediated by the fields $\varphi$, and $\vec A$ 
reads
\be
L_{\rm int} = \int d^2x\, \left(\rho_\phi^1 \varphi_2 - j_\mu^1 A^\mu_2\right).
\ee
This can be understood as the interaction between the first source at $\vec x = \vec x_1$ and the fields created by the second source
(of course, one can interchange the indices $1 \leftrightarrow 2$, which leads to the same result).
One can easily obtain
\be
L_{\rm int}^\phi &=& v^2 \frac{q^2}{2\pi} K_0(m_\phi|\vec x_1 - \vec x_2|).
\label{eq:Lint_Aphi}
\ee
The interaction mediated by the massive photon can also be computed as
\be
L_{\rm int}^\gamma &=& \frac{\mu^2}{4\pi e^2} \int d^2x\, 
\left(\vec k \times \vec\nabla \delta^{(2)}(\vec x - \vec x_1)\right)
\cdot
\left(\vec k \times \vec \nabla K_0(m_\gamma |\vec x - \vec x_2|)\right) \nonumber\\
&=& - \frac{\mu^2}{4\pi e^2} (\vec\nabla_1)^2 K_0(m_\gamma|\vec x_1 - \vec x_2|) \nonumber\\
&=& - v^2 \frac{\mu^2}{2\pi} K_0(m_\gamma|\vec x_1 - \vec x_2|),
\label{eq:L_int_Agamma}
\ee
where we used Eq.~(\ref{eq:green}) and $\vec\nabla_1$ stands for the gradient operator in terms of $\vec x_1$.
Having these formulae at hand, we can call $q$ an Abelian scalar charge, 
and $\mu$ a magnetic dipole moment, respectively.
Hence, we reproduce the well-known formula for the Abelian vortices 
\be
v^{-2} V_{\rm int}(a) = 
- \frac{q^2}{2\pi} K_0(m_\phi a)
+ \frac{\mu^2}{2\pi} K_0(m_\gamma a),
\label{eq:V_Aint}
\ee
with $a = |\vec x_1 - \vec x_2|$.
Since the modified Bessel function is exponentially small as $K_0(r) \sim \sqrt{\frac{\pi}{2r}}e^{-r}$,
the interaction mediated only by the lightest field dominates whereas the rest are suppressed.
Therefore, when $m_\phi$ is the smallest ($m_\phi < m_\gamma < 2 m_\gamma$), the inter-vortex force is attractive
with $v^{-2} V_{\rm int}(a) = - \frac{q^2}{2\pi} K_0(m_\phi a)$. 
On the other hand, when $m_\gamma < m_\phi < 2 m_\gamma$, the interaction is dominated by the massive photon
with the repulsive potential $v^{-2} V_{\rm int}(a) = \frac{\mu^2}{2\pi} K_0(m_\gamma a)$. 

\begin{figure}[t]
\begin{center}
\includegraphics[width=10cm]{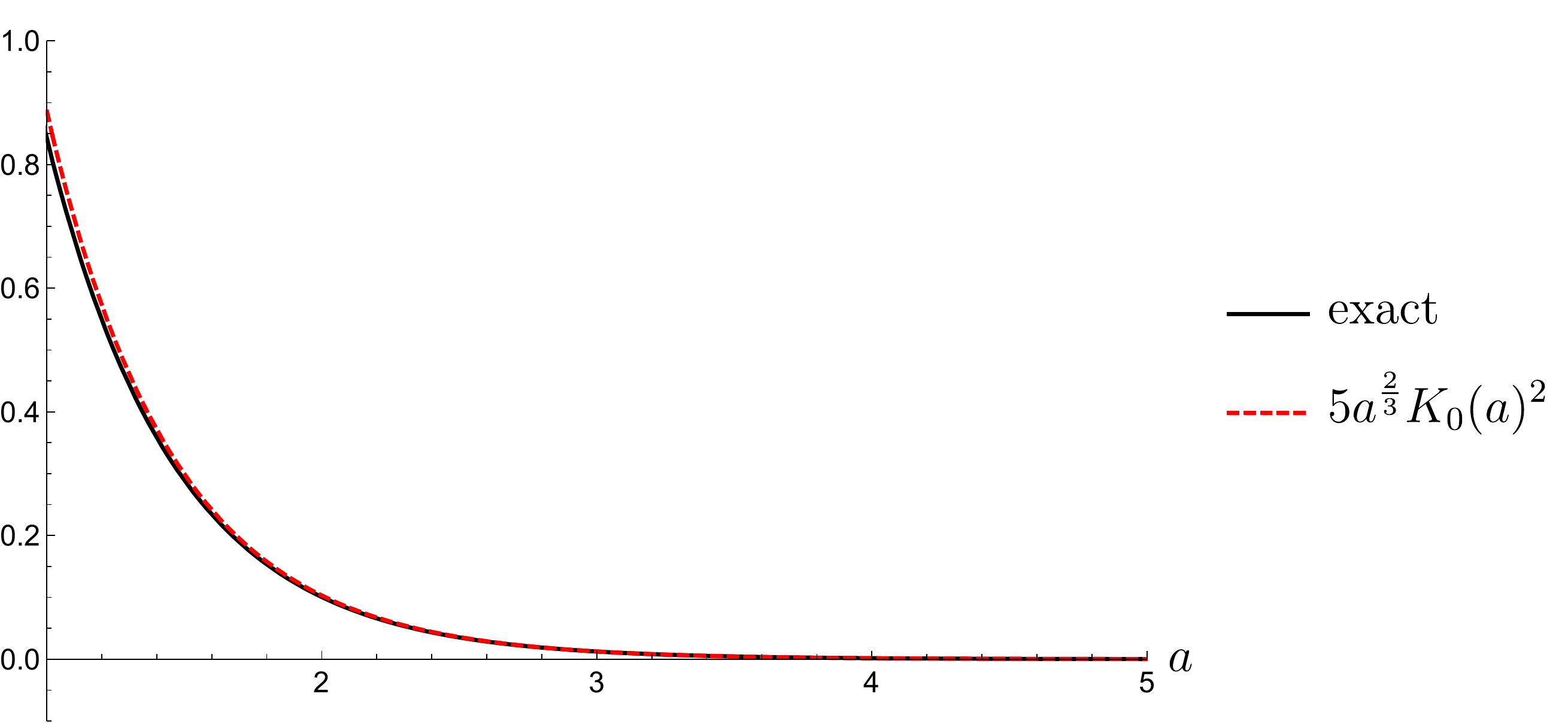}
\caption{Numerical evaluation of Eq.~(\ref{eq:K^2}).}
\label{fig:approx}
\end{center}
\end{figure}

However, we have to be careful in the region $m_\phi > 2 m_\gamma$ where Eq.~(\ref{eq:Lint_Aphi}) is no longer valid.
In order to get a correct formula, we have to take the lower asymptotic function in Eq.~(\ref{eq:asym_Acase2}).\footnote{
To the best of our knowledge, the following derivation of the static inter-vortex potential with $m_\phi > 2 m_\gamma$ does not exist in the literature.}
The asymptotic behavior of $\delta f$ in this case is not $\delta f \sim K_0(m_\phi r)$ but
$\delta f \sim K_0(m_\gamma r)^2$. 
To derive the correct interaction potential associated with $\varphi$, we have to go back to Eq.~(\ref{eq:EOM_Avarphi}).
Comparing this to Eq.~(\ref{eq:asy_Af}), we can read a scalar source
\be
\rho_\phi = 
\sqrt 2\,v\dfrac{\delta a^2}{r^2} \simeq \sqrt 2\,v m_\gamma^2 \left(\frac{\mu}{2\sqrt2\,\pi} K_0(m_\gamma r)\right)^2 \qquad \text{for}\quad m_\phi > 2m_\gamma
\ee
and the asymptotic field $\delta f$ generated by this source is nothing but the lower asymptotic function in Eq.~(\ref{eq:asym_Acase2}):
\be
\varphi = 
\sqrt 2\,v \dfrac{m_\gamma^2}{m_\phi^2-4m_\gamma^2} \left(\dfrac{\mu}{2\sqrt2 \,\pi} K_0(m_\gamma r)\right)^2\qquad
\text{for}\quad m_\phi > 2 m_\gamma.
\ee
Therefore, the $\phi$-interaction for $m_\phi > 2m_\gamma$ is given by
\be
L_{\rm int}^\phi = \frac{v^2 \mu^4}{64\pi^4} \frac{m_\gamma^4}{m_\phi^2 - 4 m_\gamma^2} \int d^2x\, 
K_0(m_\gamma|\vec x - \vec x_1|)^2 K_0(m_\gamma|\vec x - \vec x_2|)^2.
\ee
Though we do not have an analytic formula for the integral in the above equation, we can find a simple approximation
for it numerically with a good accuracy
\be
\int d^2x\, 
K_0(|\vec x - \vec x_1|)^2 K_0(|\vec x - \vec x_2|)^2 \simeq 5 a^\frac{2}{3} K_0(a)^2,
\label{eq:K^2}
\ee
for $a \gg 1$. Fig.~\ref{fig:approx} shows the numerical evaluation of the integral and the above analytic approximation.
By using this formula, we finally find the asymptotic $\phi$-interaction
\be
v^{-2} V_{\rm int}^\phi \simeq 
-\dfrac{5 \mu^4}{64\pi^4} \dfrac{m_\gamma^2}{m_\phi^2 - 4 m_\gamma^2} (m_\gamma a)^\frac{2}{3} K_0(m_\gamma a)^2 
\qquad \text{for}\quad m_\phi> 2m_\gamma.
\label{eq:V_int_phi_b}
\ee
Thus, the asymptotic potential is in this case given by $V_{\rm int}^\phi \propto (m_\gamma d)^{-\frac{1}{3}} e^{-2m_\gamma d}$.
This is an attractive force, but much smaller than $V_{\rm int}^\gamma$. 
Therefore, the repulsive force due to the massive photon dominates the scalar one for all the region $m_\gamma < m_\phi$.

In summary, we reproduced the well-known fact that the Abelian vortices attract (repel) each other when $m_\phi < m_\gamma$ ($m_\phi> m_\gamma$).
The case $m_\phi < m_\gamma$ is commonly called type I, while the opposite is type II.

\subsubsection{The non-Abelian phase}\label{sec:nonAbelianphase}

Let us next investigate the static vortex interactions of non-Abelian vortices. We assume $(m_\chi/m_\phi,m_\gamma/m_\phi)$
is chosen as a point well within the non-Abelian phase.
Again, we focus on the asymptotic behaviors of the profile functions by perturbing them as $f = 1- \delta f$, $g=\delta g$, and
$a = \delta a$ at $r\to\infty$. Note that now we have $\delta g \neq 0$. 
Plugging these into the equations of motion, and picking up only terms which are
of leading order at $r\to\infty$,
we have
\be
\delta f'' + \frac{\delta f'}{r} - m_\phi^2 \delta f &=& - \frac{\delta a^2}{r^2} - \frac{m_\phi^2}{2}\delta g^2,
\label{eq:asy_f}\\
\delta g'' + \frac{\delta g'}{r} - m_\chi^2 \delta g &=& 0.
\label{eq:asy_g}
\ee
We have an additional term proportional to $\delta g^2$ in the first equation (\ref{eq:asy_f}) but
the linearized equation of $\delta a$ is unchanged from Eq.~(\ref{eq:asy_Aa}). Therefore, $\delta a$ is given by Eq.~(\ref{eq:delta_Aa}).
On the other hand, Eq.~(\ref{eq:asy_g}) is solved by
\be
\delta g &=& \frac{1}{\sqrt2}\frac{p}{2\pi} K_0(m_\chi r) 
\simeq 
\frac{p}{4\sqrt{\pi}} \frac{1}{\sqrt{m_\chi r}}\, e^{-m_\chi r},
\label{eq:delta_g}
\ee
where $p$ is an undetermined constant, which without loss of generality we take to be positive since its sign can be absorbed in $\bm{n}$. 
As expected, the decay constant of $\delta g$ is $m_\chi^{-1}$.

Now we are ready to solve the linearized equation for $\delta f$.
Compared to the Abelian case, we have the additional contribution from $\delta g^2$. Therefore, we have to
compare three masses $m_\phi$, $2m_\gamma$ and $2m_\chi$.
Since we are interested in non-Abelian vortices, we impose $2 m_\chi < m_\phi$ hereafter, see Fig.~\ref{fig:phase_boundary}.
We can divide the parameter space into two regions.
One has $2m_\gamma = \min\{2m_\gamma,2m_\chi,m_\phi\}$, so that $2m_\gamma < 2m_\chi < m_\phi$, and the other has
$2m_\chi = \min\{2m_\gamma,2m_\chi,m_\phi\}$, which can happen if $2m_\chi < 2 m_\gamma < m_\phi$ or $2m_\chi < m_\phi < 2 m_\gamma$.
We already solved the former case, the result is shown in the lower line of Eq.~(\ref{eq:asym_Acase2}).
The latter case can be treated along similar lines, and we have
\be
\delta f 
\simeq \left\{
\begin{array}{ccl}
\dfrac{m_\gamma^2}{m_\phi^2-4m_\gamma^2} \left(\dfrac{\mu}{2\sqrt2 \,\pi} K_0(m_\gamma r)\right)^2 & &
\text{for}\quad 2m_\gamma = \min\{m_\phi, 2m_\chi,2m_\gamma\},\\
\dfrac{1}{2} \dfrac{m_\phi^2}{m_\phi^2 - 4 m_\chi^2} \left(\dfrac{p}{2\sqrt2 \,\pi} K_0(m_\chi r)\right)^2 
& & \text{for}\quad 2m_\chi = \min\{m_\phi, 2m_\chi,2m_\gamma\}.
\end{array}
\right.
\label{eq:asym_case2}
\ee

Now we are ready to discuss the asymptotic interactions between two static non-Abelian vortices.
To this end, as was done in the Abelian case, let us move to a different gauge 
where $\phi$ becomes real by $\phi \to e^{-i\theta} \phi$ and $A_\mu \to A_\mu + \frac{1}{e}\p_\mu \theta$. 
We then introduce a real scalar field $\varphi$ through $\phi= 1 - \varphi/\sqrt2$ and use the renormalized field
$\bm{\zeta} = \sqrt2\, \bm{\chi}$.
Introducing a quadratic Lagrangian 
\be
{\cal L}_{\rm free} = - \frac{1}{4}F_{\mu\nu}F^{\mu\nu} + \frac{m_\gamma^2}{2}A_\mu A^\mu 
+ \frac{1}{2} \p_\mu \varphi \p^\mu \varphi - \frac{m_\phi^2}{2} \varphi^2 + \frac{1}{2} \p_\mu \bm{\zeta} \cdot \p^\mu \bm{\zeta} - 
\frac{m_\chi^2}{2} \bm{\zeta}\cdot\bm{\zeta},
\ee
together with sources
\be
{\cal L}_{\rm s} = \varphi \rho_\phi + \bm{\zeta} \cdot \bm{\rho}_\chi - A_\mu j^\mu\,,
\ee
the equations of motion are given by
\be
(\p_\mu \p^\mu + m_\gamma^2) a^\nu - \p^\nu \p_\mu a^\mu = j^\nu,\\
(\p_\mu \p^\mu + m_\phi^2) \varphi  = \rho_\phi,\label{eq:EOM_varphi}\\
(\p_\mu \p^\mu + m_\chi^2) \bm{\zeta}  = \bm{\rho}_\chi.
\ee
The massive photon sector is exactly the same as the Abelian case. 
Similarly, the $\varphi$ sector with $2 m_\gamma = \min\{m_\phi,2m_\gamma,2m_\chi\}$ is again identical to the Abelian case when $m_\phi> 2m_\gamma$. Therefore, we do not repeat their analysis here, the results are given in Eqs.~(\ref{eq:L_int_Agamma}) and (\ref{eq:V_int_phi_b}).
On the other hand, the $\varphi$ sector with $2 m_\chi = \min\{m_\phi,2m_\gamma,2m_\chi\}$ is new, the difference with 
the case $2 m_\gamma = \min\{m_\phi,2m_\gamma,2m_\chi\}$ being in the coefficients in front of $K_0(m_\gamma r)^2$ and
$K_0(m_\chi r)^2$. Thus, we immediately find
\be
V_{\rm int}^\gamma 
= v^2 \frac{\mu^2}{2\pi} K_0(m_\gamma d),
\ee
and
\be
\frac{V_{\rm int}^\phi}{v^2} \simeq \left\{
\begin{array}{ccl}
-\dfrac{5 \mu^4}{64\pi^4} \dfrac{m_\gamma^2}{m_\phi^2 - 4 m_\gamma^2} (m_\gamma d)^\frac{2}{3} K_0(m_\gamma d)^2 & & \text{for}\quad 2m_\gamma = \min\{m_\phi, 2m_\chi,2m_\gamma\}\\
-\dfrac{5 p^4}{128\pi^4} \dfrac{m_\phi^4}{(m_\phi^2 - 4 m_\chi^2)m_\chi^2}  (m_\chi d)^\frac{2}{3}K_0(m_\chi d)^2 & & \text{for}\quad 2m_\chi = \min\{m_\phi, 2m_\chi,2m_\gamma\}
\end{array}
\right.
\ee
Note that these expressions are insensitive to the relative orientation of the non-Abelian vortices.

The final task is finding the source for ${\bm \chi}$, but this is mostly straightforward and the result is
\be
\bm{\rho}_\chi = vp\bm{n} \, \delta^{(2)}(\vec x)\,.
\ee
Now let us prepare one non-Abelian vortex with $\bm{n}_1$ at $\vec x = \vec x_1$ and
the other with $\bm{n}_2$ at $\vec x = \vec x_2$. Namely, we have
$\bm{\rho}_\chi^i = vp\bm{n}_i \, \delta^{(2)}(\vec x-\vec x_i)$
with $i=1,2$, and 
the total source is just a sum of two sources.
Since we work in the linearized theory, the resulting fields are just superpositions of each field
\be
\bm{\zeta}_i(\vec x) = v\frac{p}{2\pi} \bm{n}_i K_0(m_\phi|\vec x - \vec x_i|).
\label{eq:asymptotic_scalar_field}
\ee
Hence, the interaction between two point sources mediated by the field $\bm{\zeta}$
reads
\be
v^{-2} V_{\rm int}^\chi = - v^{-2}\int d^2x\,  \bm{\rho}_{\chi}^1 \cdot \bm{\zeta}_2 = 
-\frac{p^2}{2\pi} \bm{n}_1 \cdot \bm{n}_2 K_0(m_\chi d).
\label{eq:L_int_chi}
\ee
In contrast to the interaction associated with $\phi$ and $\gamma$, this expression depends on the relative
orientation ${\bm n}_1 \cdot {\bm n}_2 = \cos (\alpha_1 - \alpha_2)$.

In summary, we have two regimes for the full interaction potential among the two non-Abelian vortices.
One corresponds to $2m_\gamma = \min\{m_\phi,2m_\gamma,2m_\chi\}$, and we will refer to it as the type~II non-Abelian phase,
\be
\frac{V_{\rm int}(d)}{v^2} = 
-\dfrac{5 \mu^4}{64\pi^4} \dfrac{m_\gamma^2}{m_\phi^2 - 4 m_\gamma^2} (m_\gamma d)^\frac{2}{3} K_0(m_\gamma d)^2
- \bm{n}_1 \cdot \bm{n}_2 \frac{p^2}{2\pi}  K_0(m_\chi d)
+ \frac{\mu^2}{2\pi} K_0(m_\gamma d).\nonumber\\
\label{eq:V_int1}
\ee
The other corresponds to $2m_\chi = \min\{m_\phi,2m_\gamma,2m_\chi\}$, and we will refer to it as the type~I* non-Abelian phase,
\be
\frac{V_{\rm int}(d)}{v^2} = 
-\dfrac{5 p^4}{128\pi^4} \dfrac{m_\phi^4}{(m_\phi^2 - 4 m_\chi^2)m_\chi^2}  (m_\chi d)^\frac{2}{3}K_0(m_\chi d)^2
- \bm{n}_1 \cdot \bm{n}_2 \frac{p^2}{2\pi}  K_0(m_\chi d)
+ \frac{\mu^2}{2\pi} K_0(m_\gamma d).\nonumber\\
\label{eq:V_int2}
\ee
We would like to emphasize at this point that the first terms in Eqs.~(\ref{eq:V_int1}) and (\ref{eq:V_int2}) are not usual
since the scalar field $\phi$'s mass is not $m_{\gamma,\chi}$ but rather $m_\phi$.

Now we are ready to correctly add up the results and present the static inter-vortex forces.

\paragraph{The type II non-Abelian vortex}
When $m_\gamma$ is the smallest among $\{m_\gamma, 2m_\gamma, m_\chi, 2m_\chi, m_\phi\}$ we say vortices are type II non-Abelian.
The corresponding inter-vortex potential is given by (\ref{eq:V_int1}).
More concretely, there are two cases: $m_\gamma < 2 m_\gamma < m_\chi < 2m_\chi < m_\phi$ and
$m_\gamma  < m_\chi < 2 m_\gamma < 2m_\chi < m_\phi$.
For both of them the lightest mass is $m_\gamma$ and the inter-vortex force is dominated by the repulsive potential, \textit{i.e.} the third term of Eq.~(\ref{eq:V_int1}),
$V_{\rm int}^{\gamma} \sim  d^{-\frac{1}{2}} e^{-m_\gamma d}$.
Strictly speaking we should distinguish both cases since the subdominant interaction is
$V_{\rm int}^\phi \sim - d^{-\frac{1}{3}}e^{-2m_\gamma d}$ for the former, while it is
$V_{\rm int}^\chi \sim - \bm{n}_1 \cdot \bm{n}_2 d^{-\frac{1}{2}}e^{-m_\chi d}$ for the latter. However, these are much smaller than
$V_{\rm int}^\gamma$, hence we can ignore the difference. 
{\it Namely, irrespective of the relative internal orientations
the vortices repel in the type II case.}

\paragraph{The type I* non-Abelian vortex}
The alternative parameter choice for the non-Abelian vortex is 
$m_\chi = \min\{m_\gamma, 2m_\gamma, m_\chi, 2m_\chi, m_\phi\}$, corresponding to the inter-vortex potential
in Eq.~(\ref{eq:V_int2}).
The dominant interaction is mediated by the lightest field ${\bm \chi}$, and the corresponding potential is
$V_{\rm int}^\chi \sim - \bm{n}_1 \cdot \bm{n}_2 d^{-\frac{1}{2}}e^{-m_\chi d}$, which comes from the second term
of Eq.~(\ref{eq:V_int2}).
{\it Interestingly, this is an attractive force when the non-Abelian vortices are parallel, ${\bm n}_1\cdot{\bm n}_2 > 0$, but becomes repulsive for anti-parallel vortices with ${\bm n}_1\cdot{\bm n}_2 < 0$.}
This is a distinctive property of the local non-Abelian vortices in comparison with local Abelian vortices or global non-Abelian vortices.
Moreover, when the orientations are orthogonal $\bm{n}_1 \cdot \bm{n}_2 = 0$ 
the dominant term vanishes and we have to look at
the subdominant part. There are then two cases according to the ordering of the masses.
The first case, which we call type I*A, corresponds to 
$m_\chi < m_\gamma < 2 m_\chi < \min\{m_\phi , 2 m_\gamma\}$.  
The subdominant interaction for the orthogonal vortices $\bm{n}_1 \cdot \bm{n}_2 = 0$ 
is then repulsive, being mediated by the massive photon with the corresponding potential given by $V_{\rm int}^{\gamma} \sim  d^{-\frac{1}{2}} e^{-m_\gamma d}$.
The second case, which we call type I*B, arises for
$m_\chi < 2 m_\chi < \min\{m_\gamma , 2 m_\gamma , m_\phi\}$.
Then the interaction between orthogonal vortices is attractive, mediated by the $\phi$ field with
a potential $V^\phi \sim d^{-\frac{1}{3}} e^{-2m\chi d}$.

~

The analytic estimation of the phase boundaries discussed above is shown in Fig.~\ref{fig:phase_boundary}, where
the solid line ($m_\gamma = m_\chi$) is the phase boundary between types I*A and II, whereas the dashed line 
($m_\gamma = 2m_\chi$) corresponds to the boundary between types I*A and I*B.
The intervortex-forces found above for the three types of vortices, type II, type I*A and type I*B, are summarized in Table~\ref{tab:summary}.
\begin{table}[h]
\begin{center}
\begin{tabular}{c||c|c|c|c}
\hline
v-v & lightest mass & parallel & orthogonal & anti-parallel\\
\hline
\hline
type II & $m_\gamma$ & repulsion & repulsion & repulsion \\
\hline
type I*A & $m_\chi$ & attraction & repulsion & repulsion\\
\hline
type I*B & $m_\chi$ & attraction & attraction & repulsion\\
\hline
\end{tabular}
\caption{Static inter-vortex force between non-Abelian vortices.}
\label{tab:summary}
\end{center}
\end{table}

So far, we have concentrated on understanding the asymptotic inter-vortex force between two non-Abelian vortices with
winding numbers $+1$ and $+1$.
However, the above derivation also holds with minimal changes for the vortex-antivortex pair with winding numbers $+1$ and $-1$.
When we replace a vortex by an anti-vortex, the ansatz given in Eq.~(\ref{eq:vortex_ansatz}) is changed to
\be
\phi = v f(r) e^{-i\theta},\quad \bm{\chi} = v g(r) \bm{n},\quad A^i = \frac{1}{e}\epsilon^{ij}\frac{x^j}{r^2}(1+a(r)).
\ee
While the scalar profile functions $f$ and $g$ are unchanged, signs are changed in $e^{i\theta} \to e^{-i\theta}$ and $a \to -a$.
This implies that the asymptotic inter-vortex forces mediated by the scalar fields $\varphi$ and $\bm{\chi}$ are the same as those
for the vortex-vortex pairs. On the other hand, the magnetic interaction changes its sign, namely it is always attractive.
\begin{table}[h]
\begin{center}
\begin{tabular}{c||c|c|c|c}
\hline
v-$\bar{\text{v}}$ & lightest mass & parallel & orthogonal & anti-parallel\\
\hline
\hline
type II & $m_\gamma$ & attraction & attraction & attraction \\
\hline
type I*A & $m_\chi$ & attraction & repulsion & repulsion\\
\hline
type I*B & $m_\chi$ & attraction & attraction & repulsion\\
\hline
\end{tabular}
\caption{Static inter-vortex force between a non-Abelian vortex and antivortex.}
\label{tab:summary2}
\end{center}
\end{table}
The static inter-vortex force between a non-Abelian vortex and antivortex is summarized in Table~\ref{tab:summary2}.
A vortex and an antivortex of the Abelian type always attract each other irrespective of being either type I or type II. 
In contrast, the anti-parallel non-Abelian vortex and antivortex repel if they are of type I*.

A comment is now in order. In the previous section we studied global vortices and saw that the interaction in that case is also dependent on the relative orientational moduli.
However, it was a short-range interaction and therefore subdominant for well-separated vortices.
Indeed, there exists a massless Nambu-Goldstone (NG) mode associated with the broken global $U(1)$ symmetry which gives rise to a long-range repulsion between two global vortices. 
This washes out all the short-range forces associated with the massive fields, as can be seen in Figs.~\ref{fig:D01}, \ref{fig:E01}, and \ref{fig:A01}, where the vortices with relatively slow
initial scattering velocity always, irrespective of the relative orientations, backscatter due to the long-range repulsion. 
In contrast, when we turn on the gauge coupling $e$ as done in this section, the NG mode is absent since it is eaten by the gauge field.
Hence, there are no massless fields leading to a long-range interaction. This is the reason why the inter-vortex forces
critically depend on the internal orientations for the type I* non-Abelian vortices.

\subsection{Numerical simulations}
\label{sec:numerics}

The classification of the non-Abelian vortex types obtained in the previous subsection essentially relies on the point-vortex
approximation which is valid only for very well-separated non-Abelian vortices.
In this subsection we numerically confirm it.
This is not only necessary as a consistency check of our numerical results, comparing them with the previously found analytical formulae, but
also to shed new light towards understanding the vortex interactions beyond the asymptotic and static approximation.
To this end, we need to numerically solve the full equations of motion (\ref{eq:f_eom_1})--(\ref{eq:f_eom_3})
for an initial configuration with two static vortices at relative distance $a$.

First let us prepare all the formulae which we will use throughout this and the next sections.
In what follows we will impose the Lorentz gauge condition,
\be
\p_\mu A^\mu = 0,
\label{eq:GF}
\ee
so that the equation for the gauge field (\ref{eq:f_eom_1}) reduces to
\be
\p^\mu\p_\mu A^\nu + ie\left(\phi D^\nu\phi^* - \phi^* D^\nu \phi\right) = 0.
\ee
Divergence of this equation reads
\be
\p^2 \left(\p_\mu A^\mu\right) = 0,
\label{eq:LT}
\ee
where we used $\p_\nu \left(\phi D^\nu\phi^* - \phi^* D^\nu \phi\right) = \left(\phi D_\nu D^\nu\phi^* - \phi^* D^\nu D^\nu \phi\right)
= \phi \frac{\p V}{\p \phi} - \phi^* \frac{\p V}{\phi^*} = 0$.
We will numerically solve the equations of motion under the Lorentz gauge condition Eq.~(\ref{eq:GF}).
To perform this consistently, we will choose initial configurations which satisfy Eq.~(\ref{eq:GF})
at an initial time, say $t=0$. Then the longitudinal component $\p_\mu A^\mu$ propagates from $\p_\mu A^\mu = 0$ by 
Eq.~(\ref{eq:LT}), so that it remains zero and the Lorentz gauge condition is automatically satisfied at any time after $t=0$.

For numerical purposes let us write down the full set of equations of motion in terms of real fields, expressing
$\phi = \phi_{\rm r} + i \phi_{\rm i}$. We have
\be\label{eoms}
\nabla^2\phi_{\rm r}-2\lambda(|\phi|^2+\chi_i\chi^i-v^2)\phi_{\rm r} - 2e \vec A \cdot \vec \nabla \phi_{\rm i} \,-\, e^2 \vec A\,{}^2\, \phi_{\rm r} =0,\\
\nabla^2\phi_{\rm i}-2\lambda(|\phi|^2+\chi_i\chi^i-v^2)\phi_{\rm i}+2e \vec A \cdot \vec\nabla\phi_{\rm r} \,-\, e^2 \vec A\,{}^2\, \phi_{\rm i} =0,\\
\nabla^2\chi_i-2\lambda(|\phi|^2+\chi_i\chi^i-v^2)\chi_i-\Omega^2\chi_i =0,\\
\nabla^2A_x-2e^2|\phi|^2 A_x - 2e \left(\phi_{\rm r}\partial_x\phi_{\rm i}-\phi_{\rm i}\partial_x\phi_{\rm r}\right)=0,\\
\nabla^2A_y-2e^2|\phi|^2 A_y - 2e \left(\phi_{\rm r} \partial_y\phi_{\rm i}-\phi_{\rm i}\partial_y\phi_{\rm r}\right)=0,
\ee
where we have set $\p_0 = 0$ and used the Lorentz gauge condition $\vec\nabla\cdot \vec A= 0$ with $A_0 = 0$.\footnote{
Our notation is $x^\mu = (t,\vec x),\ \p^\mu = (\p_t,-\vec\nabla),\ A^\mu = (A_t,-\vec A),\ x_\mu = (t,-\vec x),\ \p_\mu = (\p_t,\vec\nabla),\ 
A_\mu = (A_t,\vec A)$.}
These are essentially the same as Eqs.~(\ref{eq:ax_eom_1})--(\ref{eq:ax_eom_3}) under the axial symmetry condition.

Once a static solution to these equations is found, we next make an ansatz for an initial state of our numerical simulation.
The superposition of scalar fields $\phi$ and $\chi_i$ proceeds as explained in the previous Section \ref{sec:model}.
For the gauge field,
\begin{enumerate}[1)]
\item 
Let $A_x^{(0)}$ and $A_y^{(0)}$ be the solution for the static single non-Abelian vortex at the origin
\be
A_0 = 0,\quad A_x = A_x^{(0)}(x,y),\quad A_y = A_y^{(0)}(x,y).
\ee
\item We shift the static solution by $\delta x =\pm a/2$ and $\delta y = \pm b/2$, 
and boost it with velocity $u$ along the $x$ axis as
\be
&&\left\{
\begin{array}{l}
A_0^{(1)}(t,x,y) = -\gamma u A_x^{(0)}\left(\gamma \left(x-\frac{a}{2} + ut\right), y-\frac{b}{2}\right)\\
A_x^{(1)}(t,x,y) = \gamma  A_x^{(0)}\left(\gamma \left(x-\frac{a}{2} + ut\right), y-\frac{b}{2}\right)\\
A_y^{(1)}(t,x,y) = A_y^{(0)}\left(\gamma \left(x-\frac{a}{2} + ut\right), y-\frac{b}{2}\right)
\end{array}
\right.,
\label{eq:ini_A_1}\\
&&\left\{
\begin{array}{l}
A_0^{(2)}(t,x,y) = \gamma u A_x^{(0)}\left(\gamma \left(x+\frac{a}{2} - ut\right), y+\frac{b}{2}\right)\\
A_x^{(2)}(t,x,y) = \gamma  A_x^{(0)}\left(\gamma \left(x+\frac{a}{2} - ut\right), y+\frac{b}{2}\right)\\
A_y^{(2)}(t,x,y) = A_y^{(0)}\left(\gamma \left(x+\frac{a}{2} - ut\right), y+\frac{b}{2}\right)
\end{array}
\right..
\label{eq:ini_A_2}
\ee
Note that we generate a time component of the gauge field which comes from boosting the vector field from the static solution. 
\item We prepare the initial configuration for integrating the equations of motion
in real time \textit{\`a la} Abrikosov as
\be
A_\mu (0,x,y) = A_\mu^{(1)}(0,x,y) + A_\mu^{(2)}(0,x,y)
\label{gaugeini}
\ee
and
\be
\dot A_\mu (0,x,y) = \dot A_\mu^{(1)}(0,x,y) + \dot A_\mu^{(2)}(0,x,y).
\ee
\end{enumerate}

This initial configuration satisfies the Lorentz gauge condition $\partial_\mu A^\mu\big|_{t=0} =0$ since the static configuration before boosting is prepared 
under the Lorentz gauge condition $\vec\nabla \cdot \vec A^{(0)} = 0$. As mentioned above, 
as long as the initial fields satisfy this condition, we are guaranteed that it
will be satisfied throughout the time evolution of the system. We made sure to check this numerically for the solutions presented in this paper. In practice, with typical simulation parameters (see below) we find the mean absolute error in the gauge condition remains $\mathcal{O}(10^{-2})$ during our simulations, in full accordance with the numerical accuracy of our discrete methods.

After these preliminaries, we must now solve the full dynamical equations of motion of the system, which are expressed with
respect to the real fields as
\be\label{full_eoms_first}
- \p_\mu\p^\mu \phi_{\rm r}-2\lambda(|\phi|^2+\chi_i\chi^i-v^2)\phi_{\rm r} + 2e A_\mu\p^\mu \phi_{\rm i} + e^2 A_\mu A^\mu\, \phi_{\rm r} =0,\\
- \p_\mu\p^\mu \phi_{\rm i}-2\lambda(|\phi|^2+\chi_i\chi^i-v^2)\phi_{\rm i} - 2e A_\mu \p^\mu \phi_{\rm r} + e^2 A_\mu A^\mu \phi_{\rm i} =0,\\
- \p_\mu\p^\mu \chi_i-2\lambda(|\phi|^2+\chi_i\chi^i-v^2)\chi_i-\Omega^2\chi_i =0,\\
\p_\mu\p^\mu A_0 + 2 e^2 |\phi|^2 A_0 + 2e\left(\phi_{\rm r} \p_t \phi_{\rm i} - \phi_{\rm i}\p_t\phi_{\rm r} \right)  = 0,\\
-\p_\mu\p^\mu A_x - 2 e^2 |\phi|^2 A_x - 2e\left(\phi_{\rm r} \p_x \phi_{\rm i} - \phi_{\rm i}\p_x\phi_{\rm r} \right)  = 0,\\
-\p_\mu\p^\mu A_y - 2 e^2 |\phi|^2 A_y - 2e\left(\phi_{\rm r} \p_y \phi_{\rm i} - \phi_{\rm i}\p_y\phi_{\rm r} \right)  = 0.
\label{full_eoms_last}\ee
These are the full equations of motion which we will work with hereafter.\newline

We solve Eqs.~\eqref{full_eoms_first}--\eqref{full_eoms_last} by first discretizing the $(x,y)$ plane with a regular grid, computing
differential operators using a second order finite element scheme with central differences in the bulk and backwards ones on the boundaries. Time evolution is then achieved using a fourth order Runge-Kutta method. To ensure a smooth time-evolution of all the physical fields, we take Neumann boundary conditions everywhere. Typical parameters for all of our simulations were grid-sizes of $401 \times 401$ points and finite time-steps of size $dt = 10^{-3}$, corresponding to grid-spacings of order $\mathcal{O}(10^{-2})$ and $\mathcal{O}(10^4)$ time-iterations to simulate each scattering process. Accuracy is generally of order $\mathcal{O}(10^{-2})$, and when necessary we adjusted the discretization parameters to ensure it remains at this level.

For the sake of speed, the full algorithm is implemented in Mathematica so that it can be run on graphical processing units. For this purpose, we transform Eqs.~\eqref{full_eoms_first}--\eqref{full_eoms_last} into a fixed computation graph acting on raw numerical arrays, making use of Mathematica's convenient neural network framework. Note that for maximal efficiency the Runge-Kutta time iteration is also implemented in a similar way.

~

\begin{figure}[t]
\begin{center}
\includegraphics[width=12cm]{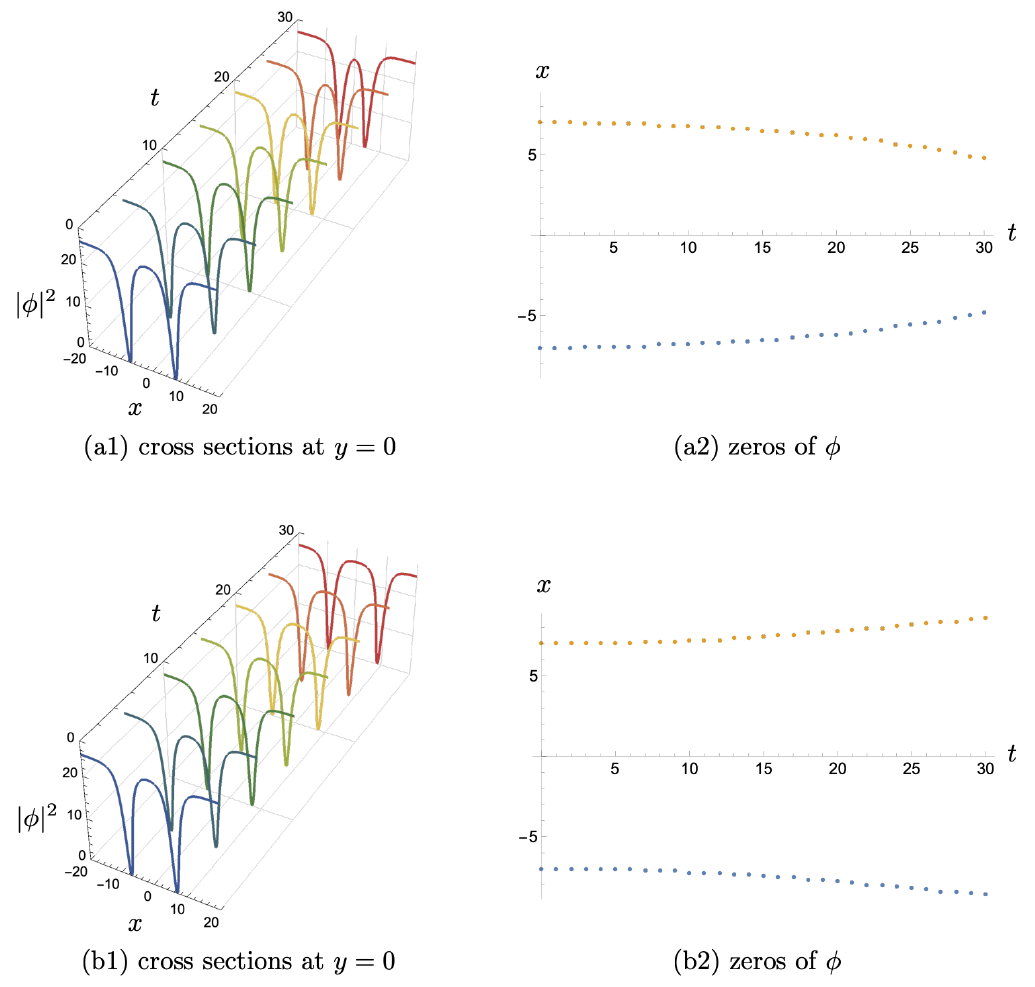}
\caption{The inter-vortex forces for the type I* non-Abelian vortices with
$(m_\chi/m_\phi, m_\gamma/m_\phi) = (0.05, 0.3)$, or more explicitly ($v = 5$, $\lambda = 1/4$ , $e = 3/10\sqrt{2}$, $\Omega = 1/4$).
Panels (a1) and (a2) correspond to parallel vortices, whereas panels (b1) and (b2) correspond to anti-parallel vortices.
Panels on the left show the time evolution of $|\phi|^2$ (cross sections at $y=0$), while panels on the right present the $x$-position of the vortex centers from $t=0$ to $t=30$.}
\label{fig:force_1s}
\end{center}
\end{figure}

\begin{figure}[t]
\begin{center}
\includegraphics[width=12cm]{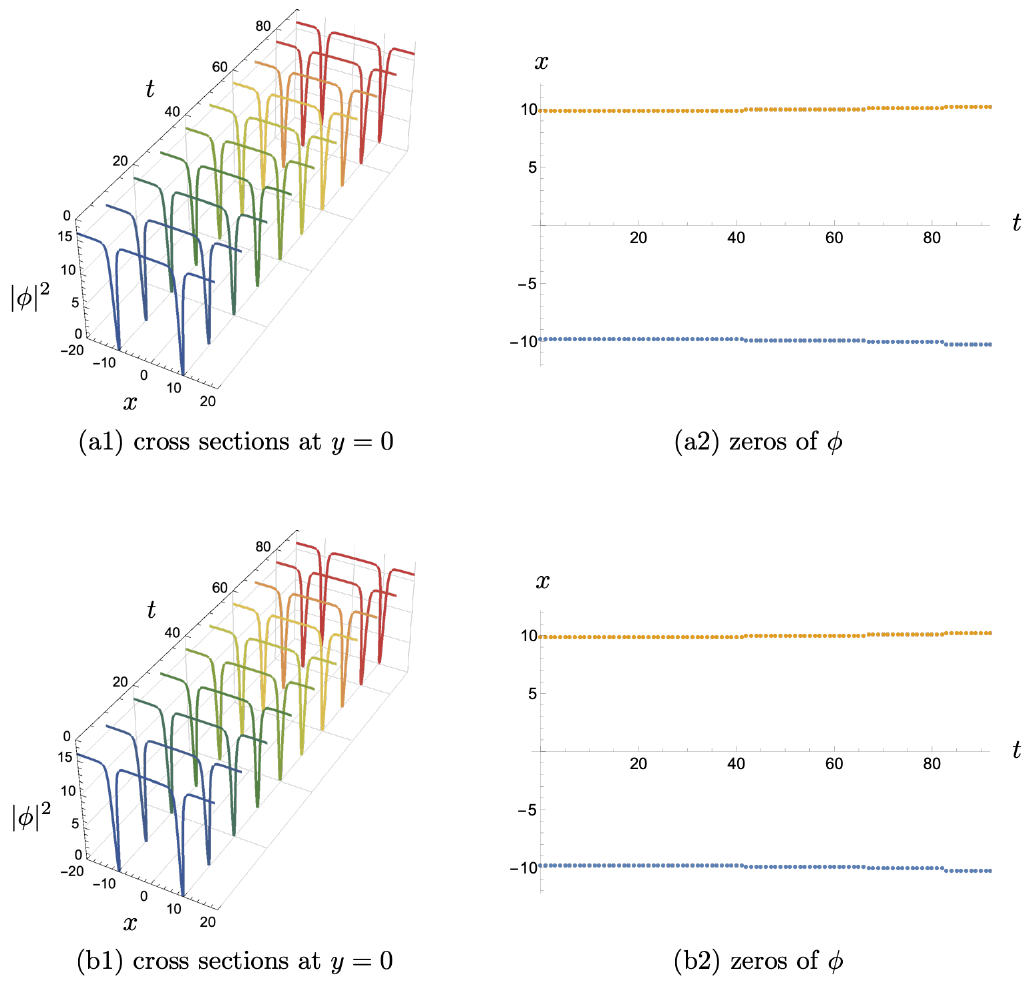}
\caption{The inter-vortex forces for type II non-Abelian vortices with
$(m_\chi/m_\phi, m_\gamma/m_\phi) = (0.25, 0.1)$, or more explicitly 
($v = 4$, $\lambda = 1/4$ , $e = 1/10\sqrt{2}$, $\Omega = 1$).
Panels (a1) and (a2) are for parallel vortices, whereas panels (b1) and (b2) are for anti-parallel vortices.
Panels on the left show time evolution of $|\phi|^2$ (cross sections at $y=0$), while panels on the right present the $x$-position of the vortex centers from $t=0$ to $t=90$.}
\label{fig:force_2}
\end{center}
\end{figure}

For our initial purpose of studying the static inter-vortex forces we simply set the initial boost velocity $u=0$, as well as the impact parameter $b=0$. Namely, we superpose two static non-Abelian vortices at a distance $a$ at initial time $t=0$,
and numerically integrate the equations of motion. We then observe whether the vortices approach or separate each other.
In order to choose the initial distance $a$ we follow the following procedure: we first measure a characteristic vortex size $L$,
defined as the maximal half-life among the fields $\phi$, $\chi$ and $A_\theta$. Then we set $a = 6L$, so that vortex individuality is well preserved.
Fig.~\ref{fig:force_1s} gives examples of our simulations of the type I* vortices with
$(m_\chi/m_\phi, m_\gamma/m_\phi) = (0.05, 0.3)$.
As expected, the parallel vortices attract whereas
antiparallel ones repel, see the panels (a1) and (a2) of Fig.~\ref{fig:force_1s} for the former, and (b1) and (b2) for the latter.
Similarly, Fig.~\ref{fig:force_2} shows an example of our numerical simulations for type II non-Abelian vortices with
$(m_\chi/m_\phi, m_\gamma/m_\phi) = (0.25, 0.1)$.
The numerical results are consistent with the analytic result that the type II vortices repel each other irrespective of whether 
the relative orientation is parallel or antiparallel.

\begin{figure}[t]
\begin{center}
\includegraphics[width=15.5cm]{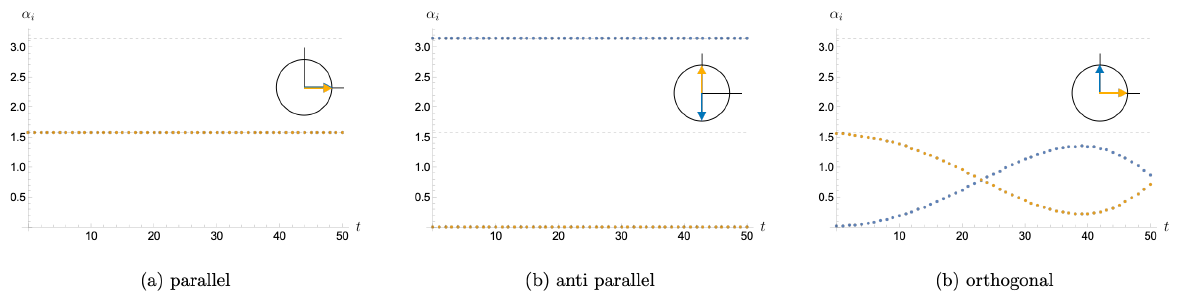}
\caption{Time evolution of the internal orientations for the parallel, antiparallel and orthogonal cases, 
for parameters $(m_\chi/m_\phi, m_\gamma/m_\phi) = (0.05, 0.3)$ or more explicitly ($v = 5$, $\lambda = 1/4$ , $e = 3/10\sqrt{2}$, $\Omega = 1/4$).}
\label{fig:internal_force}
\end{center}
\end{figure}

Another check of the analytic formulae for the static inter-vortex forces is given by looking at the evolution of the internal orientations. 
We can define a force which acts between orientational moduli $\bm{n}_1$ and $\bm{n}_2$. According to the analytic formulae (\ref{eq:V_int1})
and (\ref{eq:V_int2}), it is natural to introduce an internal force
\be
F(\alpha) = - \frac{d}{d\alpha} V_{\rm int} = -\frac{p^2}{2\pi}K_0(m_\chi a) \sin \alpha ,
\ee
where $\alpha$ stands for the relative angle $\alpha = \alpha_1 - \alpha_2$, such that $\bm{n}_1\cdot\bm{n}_2 = \cos \alpha$.
This force vanishes at $\alpha = 0$ and $\pi$, meaning the relative orientation should not vary under time evolution when the initial configuration
is a superposition of a pair of static (unboosted) vortices whose orientations are parallel or anti-parallel. 
Indeed, we observed that the relative orientations are kept
parallel/anti-parallel during the whole simulation periods in all our numerical simulations.
In contrast, the internal force is not zero for $\alpha \neq 0,\pi$, pointing towards the parallel direction $\alpha = 0$
in the internal space.
Fig.~\ref{fig:internal_force} shows examples of time evolution of the internal orientations for the parallel and anti-parallel cases 
(corresponding to Fig.~\ref{fig:force_1s}), and also the orthogonal case for the parameter choice $(m_\chi/m_\phi, m_\gamma/m_\phi) = (0.05, 0.3)$.

\begin{figure}[th]
\begin{center}
\includegraphics[width=6cm]{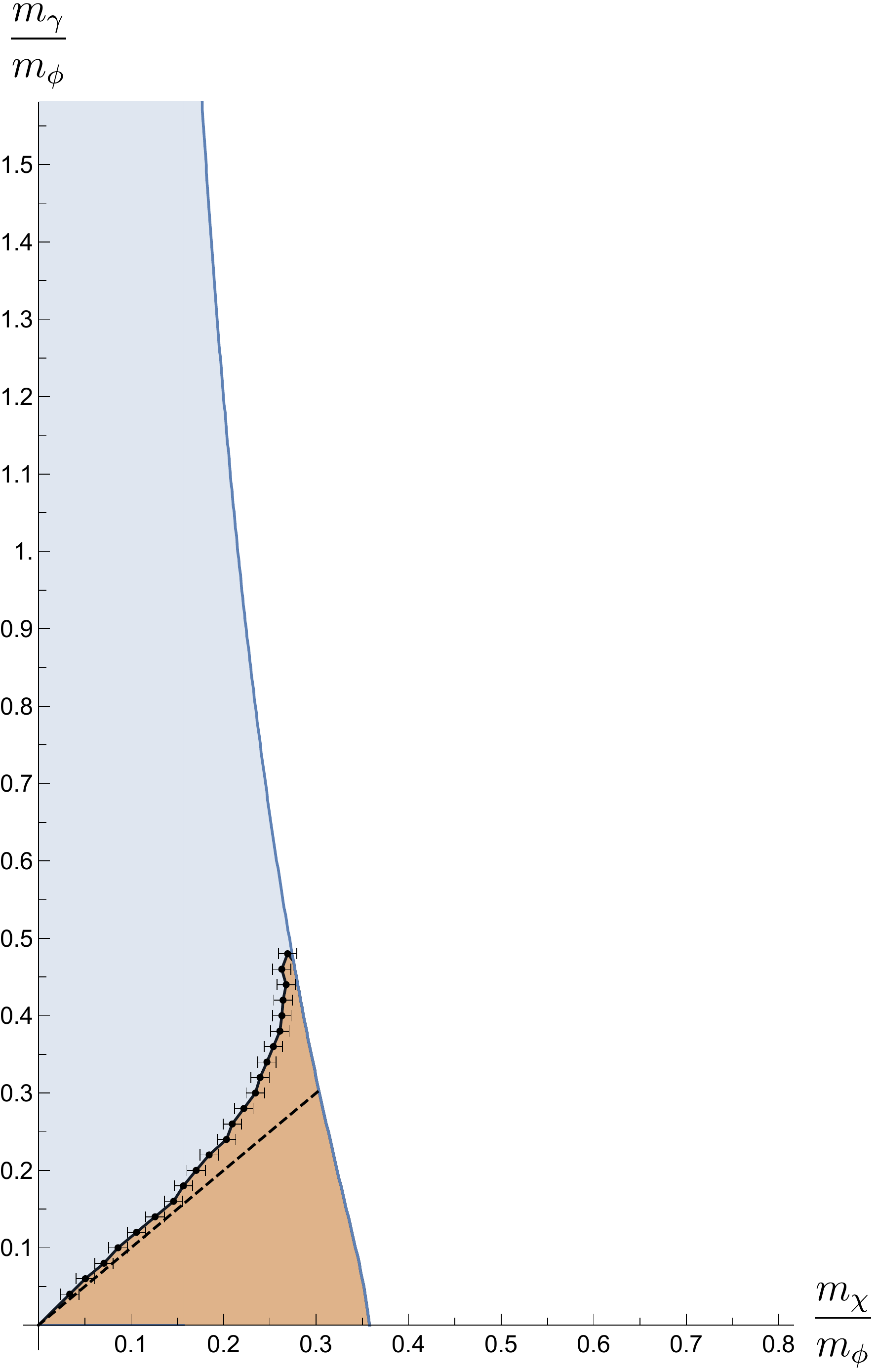}
\caption{Numerical survey of vortex types in parameter space $(m_\chi/m_\phi,m_\gamma/m_\phi)$. 
The white region on the right hand side corresponds to Abelian vortices (of type II since $m_\gamma/m_\phi < 1$), whereas shaded regions to its left correspond to non-Abelian ones. Type I* non-Abelian vortices are found in the upper-left region (light-blue shading), while type II non-Abelian vortices are found in the lower-left one (orange shading). The dashed black line marks the analytic approximation to the type I* -- type II non-Abelian boundary found in Section~\ref{sec:nonAbelianphase}, with the actual boundary being shifted to higher $m_\gamma/m_\phi$ values.}
\label{fig:phase_boundary_num}
\end{center}
\end{figure}

Finally, we performed a numerical survey in the parameter space $(m_\chi/m_\phi,m_\gamma/m_\phi)$
to check the validity of our analytic classification of non-Abelian vortices.
We repeatedly simulated the initial motion of two static vortices with parallel orientations for various points of the parameter space.
We classify the vortices as type I* (II) if the parallel vortices attract (repel).\footnote{It is hard to numerically distinguish
type I*A and I*B vortices because the orthogonal orientations change very quickly, certainly before the vortices start to move appreciably.} 
Fig.~\ref{fig:phase_boundary_num} shows the result of this survey, and agrees nicely with the (approximate) analytic result stating that the phase boundary between type I* and type II vortices falls in the line $m_\gamma = m_\chi$.
Therefore, we are reassured that our numerical code works correctly.

\section{The scattering of non-Abelian local vortices}
\label{sec:scattering_lnA}

After having determined the phase diagram of non-Abelian vortex types, we now proceed to investigate their scattering dynamics.

\subsection{Head-on collisions: type II non-Abelian vortices}

Let us begin with head-on collisions of type II non-Abelian vortices.
A common property of this type of vortices for any set of parameters is
the orientation-independent repulsive force between static vortices.
The static repulsive force should also have consequences in the dynamics of vortices, such as the
head-on collisions we will study below. In other words, the repulsive force can qualitatively 
explain most of the simulation results.
Due to this fact, dynamics of the type II non-Abelian local vortices will be mostly similar to that of the non-Abelian global vortices. However, the repulsion between global vortices is a long-range force mediated by
the massless NG field, whereas forces between local vortices correspond to a short-range force which is in competition with
the other forces mediated by the massive scalar fields $\phi$ and $\chi$. This makes the dynamics of
the local vortices different from that of the global vortices in some cases.

For our first example, let us take the point $(m_\chi/m_\phi,m_\gamma/m_\phi) = (0.25,0.1)$ which is deep enough within the phase boundary between type I* and type II, see Fig.~\ref{fig:phase_boundary_num}.
Therefore, the vortices we consider here are solidly of type II.
More explicitly, we take $v = 4$, $\lambda = 1/4$, $e = 1/10\sqrt{2}$, $\Omega = 1$. 
For this choice of parameters, the vortex size defined as the half-life length of the gauge field is found to be $L \sim 3.25$.
We prepare our initial configuration with $a = 4L$ so that initial separation is $\sim 13$, and set the impact parameter $b=0$
for Eqs.~(\ref{eq:ini_phi_chi_1}), (\ref{eq:ini_phi_chi_2}), (\ref{eq:ini_A_1}) and (\ref{eq:ini_A_2}).
We performed the simulation for various initial velocities from $u=0.3$ to $u=0.9$, with steps of $\delta u = 0.1$.
As before, for each initial velocity we consider three different initial orientations, namely parallel, anti-parallel, and orthogonal.

\begin{figure}[ht]
\begin{center}
\def\svgwidth{15cm}
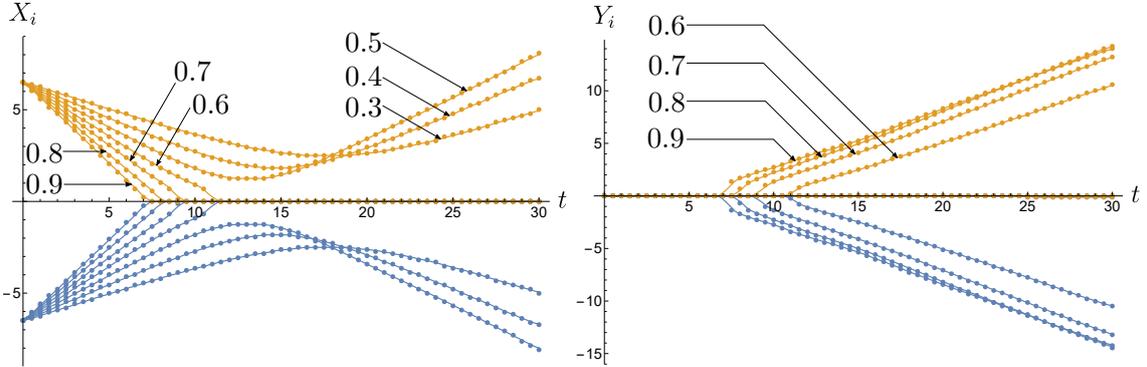
\caption{Trajectories $X_i(t)$ and $Y_i(t)$ of two parallel non-Abelian vortices of type II colliding head-on with initial velocities $u=0.3, 0.4, \cdots, 0.9$ (marked on the plots for clarity).}
\label{fig:XYvsT_type2_para_01}
\end{center}
\end{figure}

\begin{figure}[ht]
\begin{center}
\includegraphics[width=15cm]{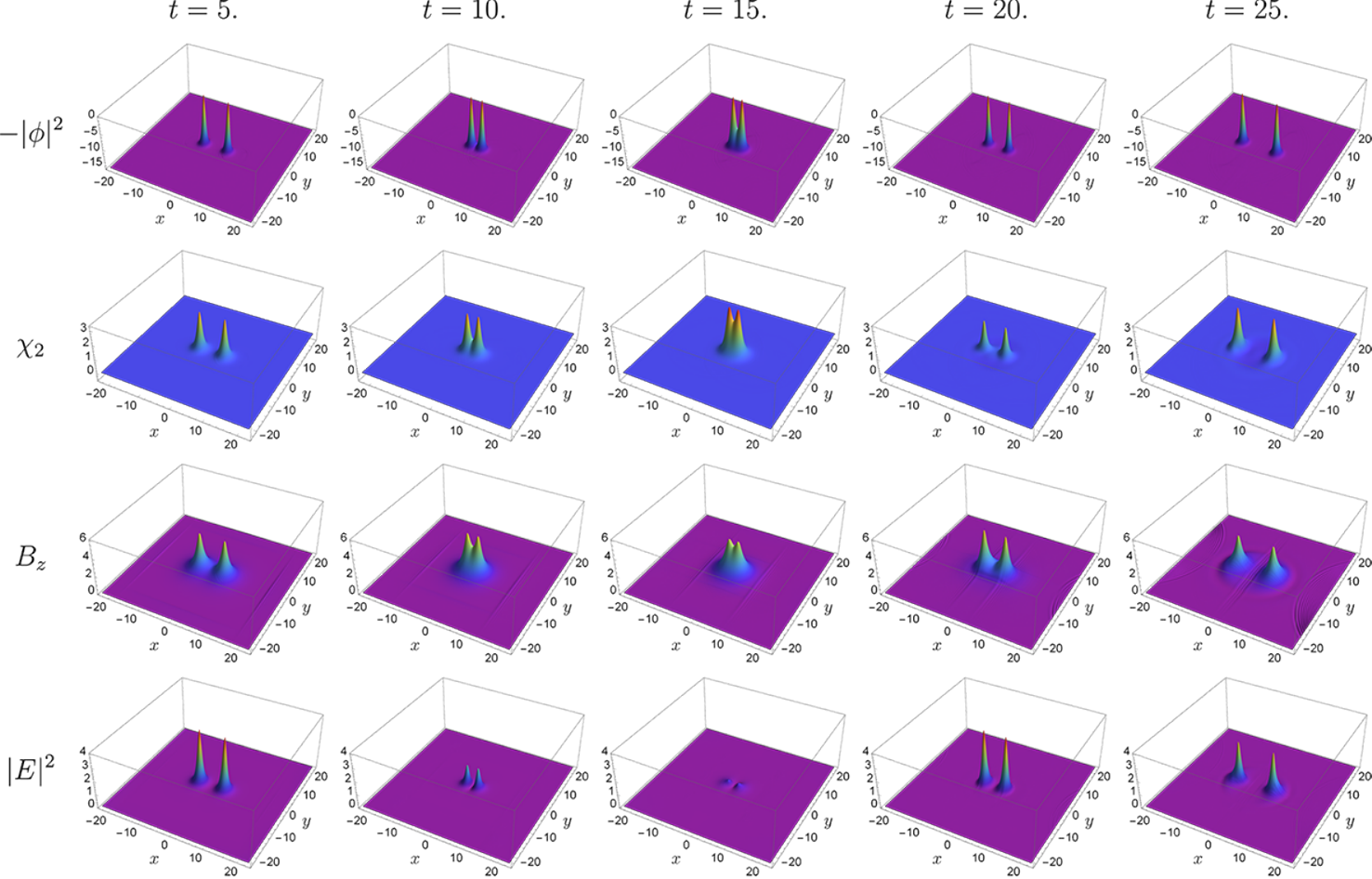}
\caption{Snapshots of $-|\phi|^2$, $\chi_2$, $B_z$ and $|E|^2$ for head-on scattering of type II parallel non-Abelian vorticet with initial velocity $u=0.5$.}
\label{fig:3D_type2_para_u5_01}
\end{center}
\end{figure}

\begin{figure}[ht]
\begin{center}
\includegraphics[width=15cm]{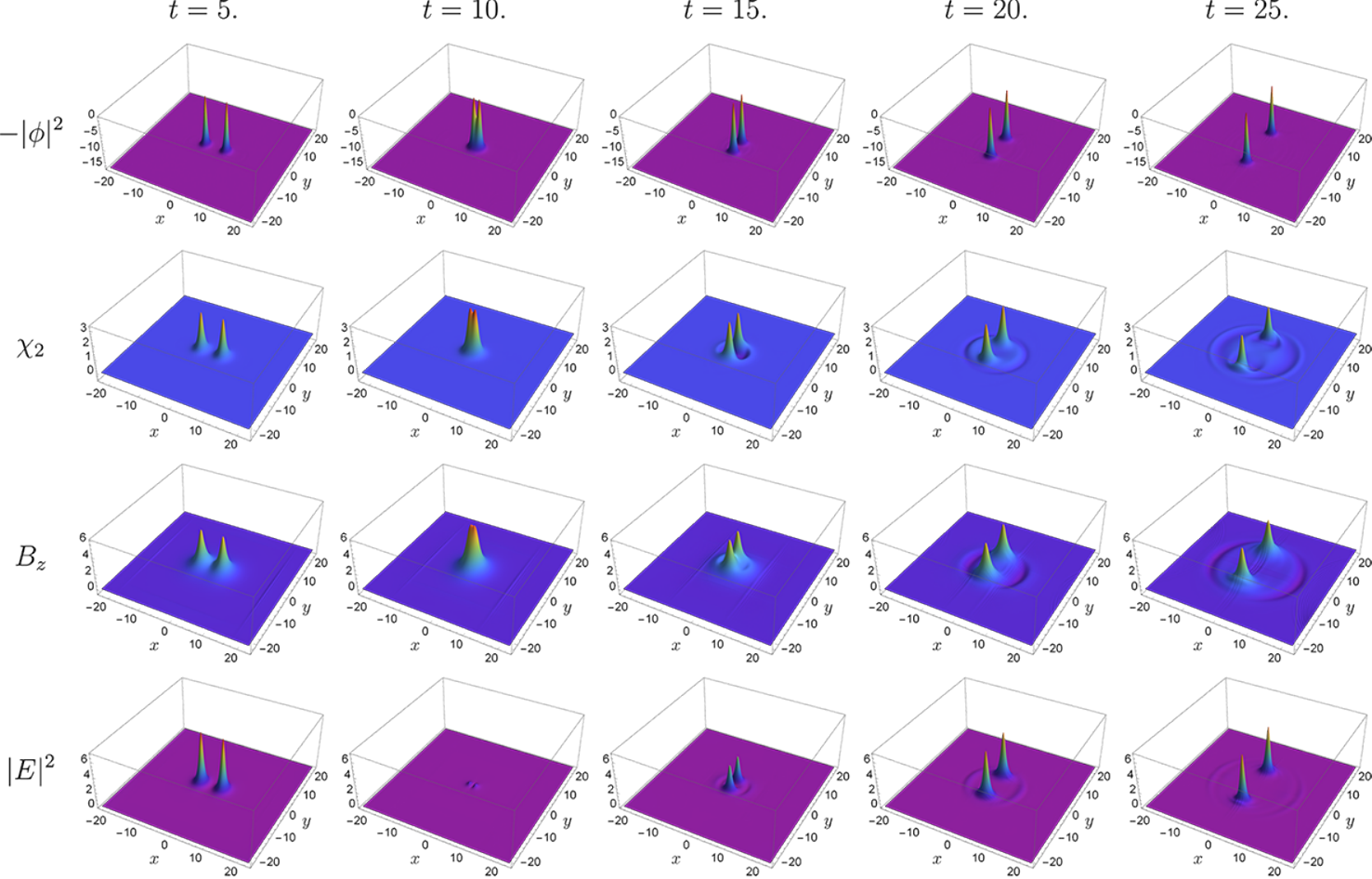}
\caption{Snapshots of $-|\phi|^2$, $\chi_2$, $B_z$ and $|E|^2$ for head-on collision of type II parallel non-Abelian vortices with initial velocity $u=0.6$.}
\label{fig:3D_type2_para_u6_01}
\end{center}
\end{figure}

\paragraph{Parallel vortices:}
First we describe the simulations scattering two vortices with parallel orientations.
Fig.~\ref{fig:XYvsT_type2_para_01} shows the positions $(X_i(t),Y_i(t))$ of the vortices 
determined by $\phi(X_i(t),Y_i(t)) = 0$. 
The simulation results are qualitatively very similar to those of 
the head-on collisions of the parallel non-Abelian global vortices given in Fig.~\ref{fig:D01}.
The vortices bounce off on the $x$ axis for relatively small $u$, whereas they collide head-on and 
scatter at right angles for large initial velocities.
We also observed that internal orientations are preserved at $\alpha_1= \alpha_2 = \frac{\pi}{2}$, namely $\chi_1 = \chi_3 = 0$, 
during all the simulations. The critical velocity separating bounce-back and right-angles scattering behaviours lies between $u= 0.5$ and $u = 0.6$. We show snapshots of $-|\phi|^2$, $\chi_2$ and $B_z$ for the bounce-back case in Fig.~\ref{fig:3D_type2_para_u5_01}, and
for the right-angles scattering case in Fig.~\ref{fig:3D_type2_para_u6_01}. Note that since we are dealing with moving vortices an electric field is also generated, and we also display it in the figures.
The peaks of $-|\phi|^2$, $\chi_2$, and $B_z$ interlock during the simulations.
In the figures, one can also see that the magnetic field $B_z$ has peaks with the largest width compared to $\phi$ and $\chi_2$.
This is a feature of type II vortices, reflecting the order of masses $m_\gamma < m_{\phi,\chi}$, and explains that the dominant inter-vortex force is repulsive, namely the repulsion occurring between magnetic fluxes.

\begin{figure}[ht]
\begin{center}
\def\svgwidth{15cm}
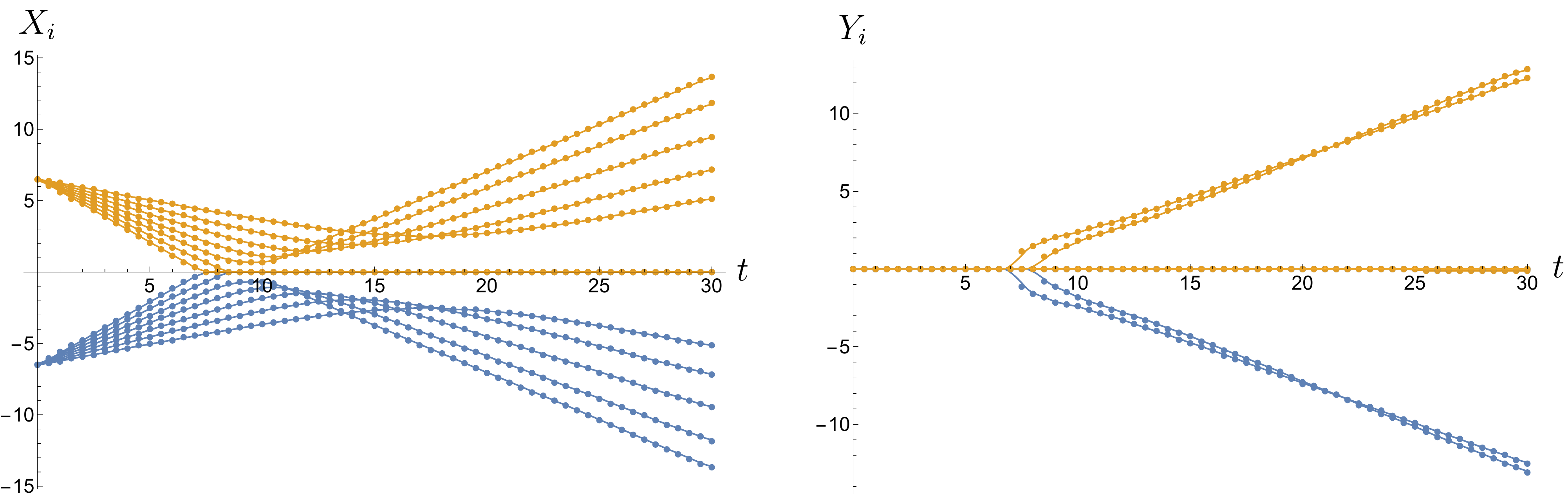
\caption{Trajectories $X_i(t)$ and $Y_i(t)$ of two anti-parallel non-Abelian vortices of type II colliding head-on with initial velocities $u=0.3, 0.4, \cdots, 0.9$ (marked on the plots for clarity).}
\label{fig:XYvsT_type2_apara_01}
\end{center}
\end{figure}

\begin{figure}[ht]
\begin{center}
\includegraphics[width=15cm]{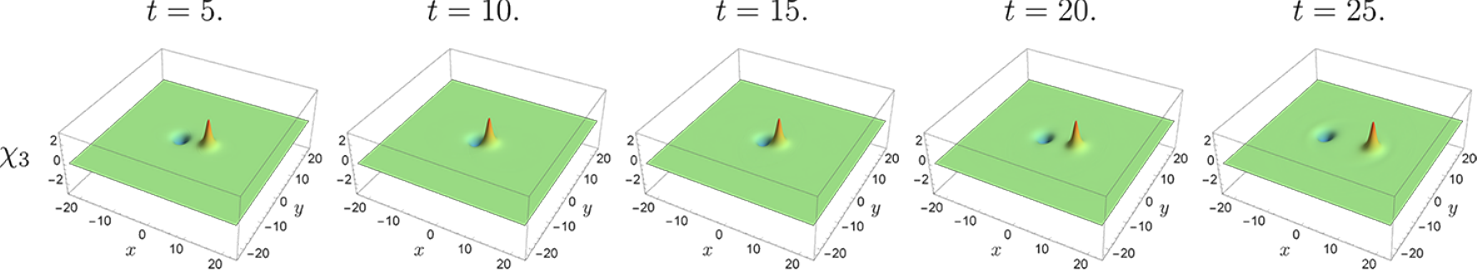}
\caption{Snapshots of the $\chi_3$ field for the head-on collision of type II anti-parallel non-Abelian vortices with initial velocity $u=0.5$. The vortices bounce-back due to the repulsive force acting between type II vortices, .}
\label{fig:3D_type2_apara_u5_01}
\end{center}
\end{figure}

\paragraph{Anti-Parallel vortices:}
Let us next examine head-on collisions of type II anti-parallel vortices.
We expect to see evidence for a stronger repulsive force in comparison with parallel vortices, because the $\bm{\chi}$ field also contributes with a
repulsive channel in addition to the dominant magnetic repulsion.
Our simulation results support this nicely, as  shown in Fig.~\ref{fig:XYvsT_type2_apara_01}.
The critical velocity separating bounce-back and right-angles scattering is now found to be between $u=0.7$ and $u = 0.8$, \textit{i.e.} it is larger compared to the parallel case, implying that indeed the repulsive force for the anti-parallel vortices is stronger than that between the parallel ones.
The bounce-back scatterings of the anti-parallel vortices for $u < 0.8$ are quite similar to those found in the parallel cases
shown in Fig.~\ref{fig:3D_type2_para_u5_01}.
A difference appears only for the $\bm{\chi}$ field. For the anti-parallel cases, the orientations of the bounce-back scatterings are preserved at
$(\alpha_1,\alpha_2) = (\pi/2,-\pi/2)$, namely $\chi_1 = \chi_2 = 0$ during the whole simulations.
In Fig.~\ref{fig:3D_type2_apara_u5_01} we show only the evolution of $\chi_3$ for initial velocity $u=0.5$, since evolution of $\phi$ and $B_z$ are essentially the same as
those shown in Fig.~\ref{fig:3D_type2_para_u5_01}.

\begin{figure}[ht]
\begin{center}
\includegraphics[width=15cm]{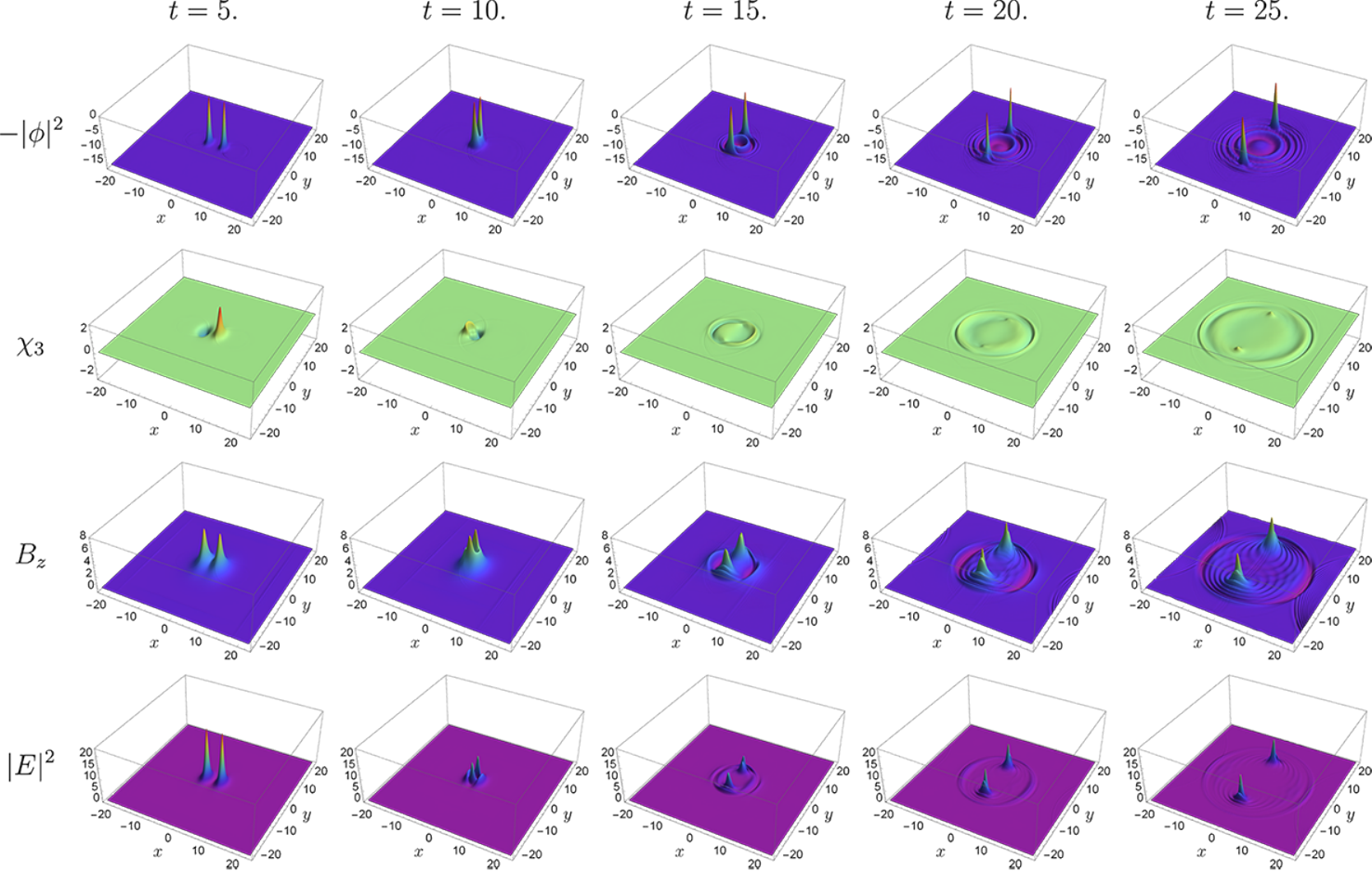}
\caption{Snapshots of $-|\phi|^2$, $\chi_3$, $B_z$ and $|E|^2$ for the head-on scattering of type II anti-parallel vortices with the initial velocity $u=0.8$.}
\label{fig:3D_type2_apara_u8_01}
\end{center}
\end{figure}
In contrast to the parallel case, the anti-parallel scatterings of local vortices have significantly different properties
from those of the global vortices. Indeed, let us compare Fig.~\ref{fig:E01} for the global case and Fig.~\ref{fig:XYvsT_type2_apara_01} for the local case.
In the former, the global vortices are protected by the strong repulsive interaction, so that
the vortices bounce-back even at high initial velocity $u=0.9$. 
On the other hand, since the long range repulsive force is absent due to gauging of the $U(1)$ symmetry, 
the local vortices at high velocity $u=0.8$ and $u = 0.9$ scatter at right angles
by overcoming a repulsive potential barrier. As we have seen before in the previous studies both on global and local vortices, 
right-angles scattering occurs only when the orientations are parallel
at the moment of collision. In other words, the non-Abelian vortices exhibit right-angles scattering only when they are identical (parallel) and therefore
indistinguishable at the instant of impact. Indeed, it is widely known that 
right-angles scattering is a peculiar feature which commonly appears in head-on scattering of identical solitons, 
such as  Abelian vortices, monopoles and skyrmions. 
Yet the non-Abelian vortices are not identical solitons if their orientations are different.
In some sense, the anti-parallel vortices we are considering are in fact the furthest away possible from indistinguishability, and as we have already mentioned their orientations remain anti-parallel during
the scatterings. This poses no problems for bounce-back scenarios, and indeed
we found the anti-parallel vortices bounce-back with their orientations preserved for $u \leq 0.7$ as shown in Fig.~\ref{fig:3D_type2_apara_u5_01}.
Nevertheless, we observe that anti-parallel vortices with $u=0.8$ and $u = 0.9$ scatter at right-angles when colliding head-on, immediately posing the question: how can we explain this, if the vortices are not indistinguishable? The answer is depicted in Fig.~\ref{fig:3D_type2_apara_u8_01}.
The evolution of $\phi$ and $B_z$ are qualitatively the same as ordinary scattering at right angles, shown for example in Fig.~\ref{fig:3D_type2_para_u6_01}. 
However, $\chi_3$ is seen to evolve very differently
from the usual scatterings. Around the moment of collision, the $\chi_3$ fields are no longer interlocked to the vortices.
The impact of the collision makes the $\bm{\chi}$-condensation scatter off the vortex core, and we cannot clearly say the vortices are either parallel or anti-parallel.
At that point, they have become something intermediate between Abelian and non-Abelian vortices. From the fact that our numerical simulations exhibit right-angles scattering,
we should interpret that the vortices are identical solitons at the instant of collision. Some time after impact the vortices again acquire well-defined orientations, but now
they are parallel and come accompanied by extensive ripples in the $\chi_3$ component.

\begin{figure}[t]
\begin{center}
\includegraphics[width=15cm]{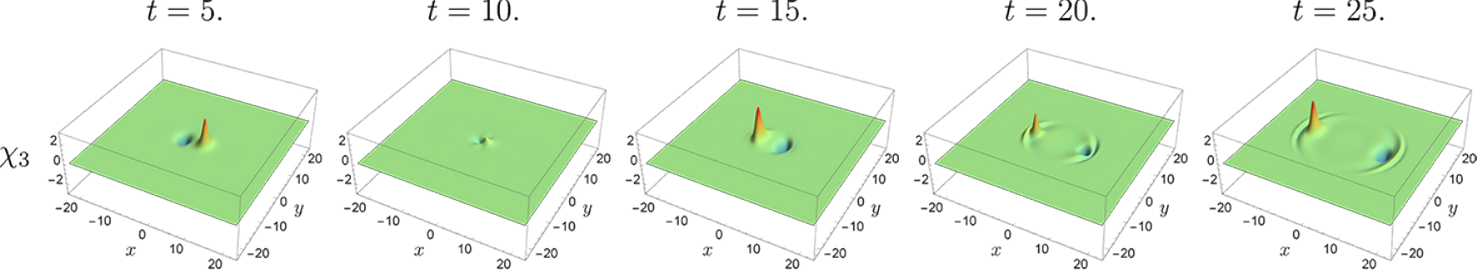}
\caption{Snapshots of $\chi_3$ for the head-on collision of type II anti-parallel non-Abelian vortices with initial velocity $u=0.7$.
The vortices bounce-back due to the repulsive force dominating in the type II case, however the orientations are exchanged at the moment of collision.}
\label{fig:3D_type2_apara_u7_01}
\end{center}
\end{figure}

Finally, we observe a new, strange behaviour for initial velocities $u$ close but not quite large enough to achieve right-angles scattering.
Let us concentrate on the case of $u=0.7$ for which the vortices bounce-back as shown in Fig.~\ref{fig:XYvsT_type2_apara_01}.
Their kinetic energy is only slightly below the threshold required to overcome the potential barrier, so they come very close to each other before bouncing back.
This leads to a new phenomenon where the vortices exchange their orientations as $(\alpha_1,\alpha_2) = (\pi/2,-\pi/2) \to (-\pi/2,\pi/2)$ during the bounce-back dynamics,
see Fig.~\ref{fig:3D_type2_apara_u7_01}.

\begin{figure}[ht]
\begin{center}
\def\svgwidth{15cm}
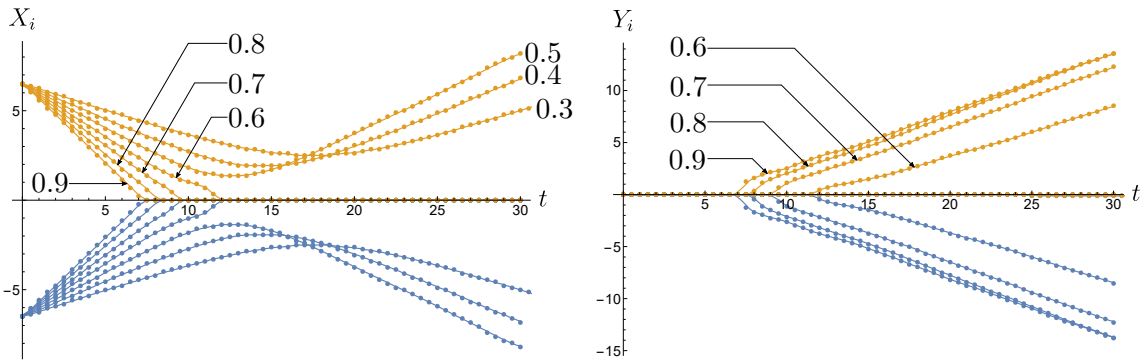
\caption{Trajectories $X_i(t)$ and $Y_i(t)$ of two initially orthogonal non-Abelian vortices of type II, scattering head-on with initial velocities $u=0.3, 0.4, \cdots, 0.9$ (marked on the plots for clarity).}
\label{fig:XYvsT_type2_ortho_01}
\end{center}
\end{figure}

\begin{figure}[t]
\begin{center}
\includegraphics[width=9cm]{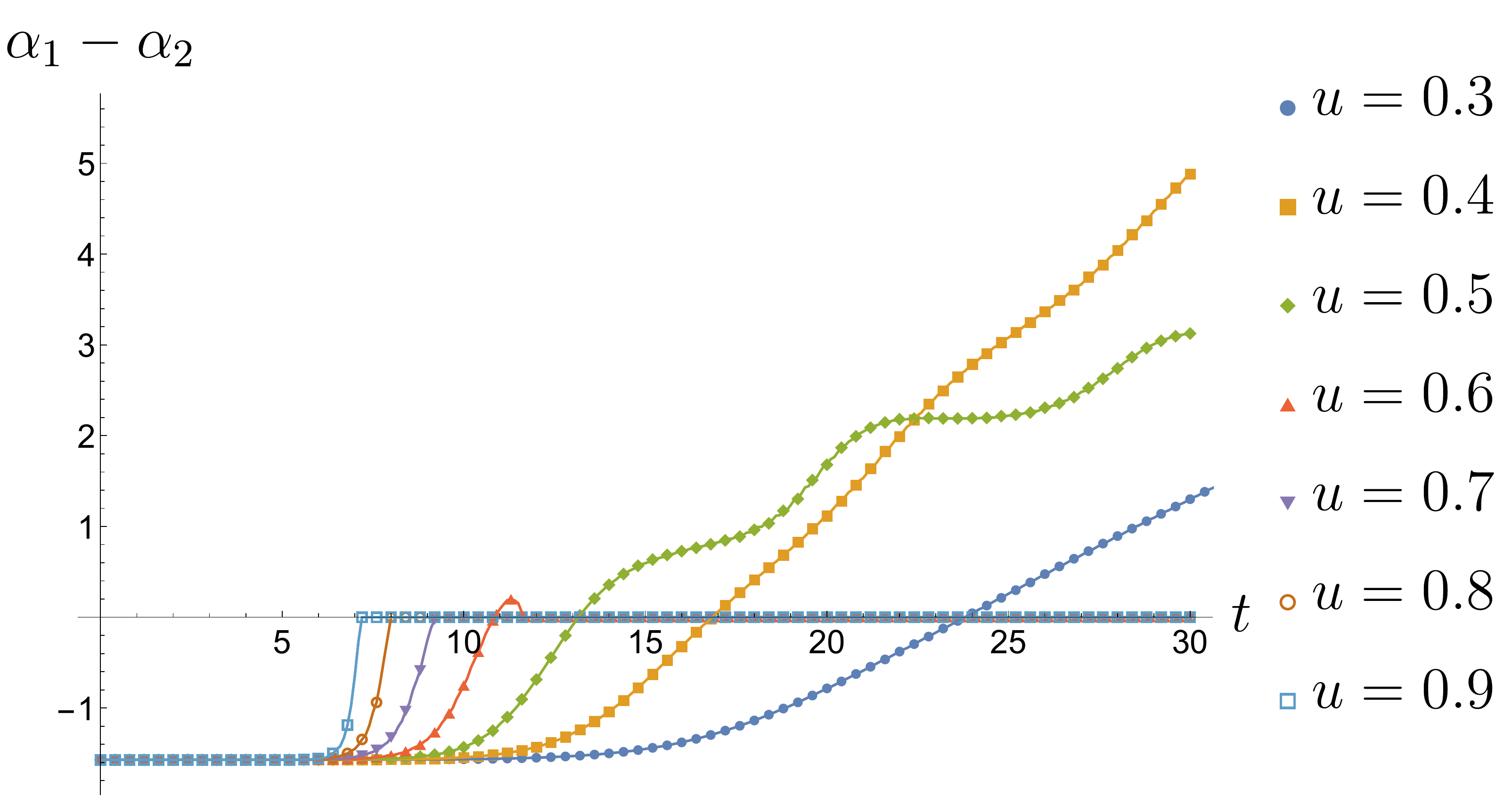}
\caption{Evolution of the relative orientations of type II non-Abelian vortices during head-on scatterings with initially orthogonal orientations and velocities $u=0.3,\cdots,0.9$.}
\label{fig:rel_ori_type2}
\end{center}
\end{figure}

\paragraph{Orthogonal vortices:}
The final category of type II non-Abelian vortex scatterings we analyze corresponds to head-on collisions of two initially orthogonal vortices, $(\alpha_1,\alpha_2) = (0,\pi/2)$.
Fig.~\ref{fig:XYvsT_type2_ortho_01} shows the vortex orbits for various initial velocities from $u=0.3$ to $u=0.9$.
The scattering orbits of the initially orthogonal vortices are found to be quite similar to those of the parallel vortices shown in Fig.~\ref{fig:XYvsT_type2_para_01}.
This observation implies that the internal orientations do not play a major role during the scatterings, which is expected for the type II case.
Yet evolution in the internal orientation space is quite different. While orientations for the parallel
cases are preserved throughout the scattering processes, those for the orthogonal scatterings evolve non-trivially and exhibit two distinct behaviours depending on whether 
the scattering is of the bounce-back or right-angles type. From Fig.~\ref{fig:XYvsT_type2_ortho_01}, the threshold between these two behaviours is found somewhere between
$u=0.5$ and $u=0.6$. 
\begin{figure}[t]
\begin{center}
\includegraphics[width=15cm]{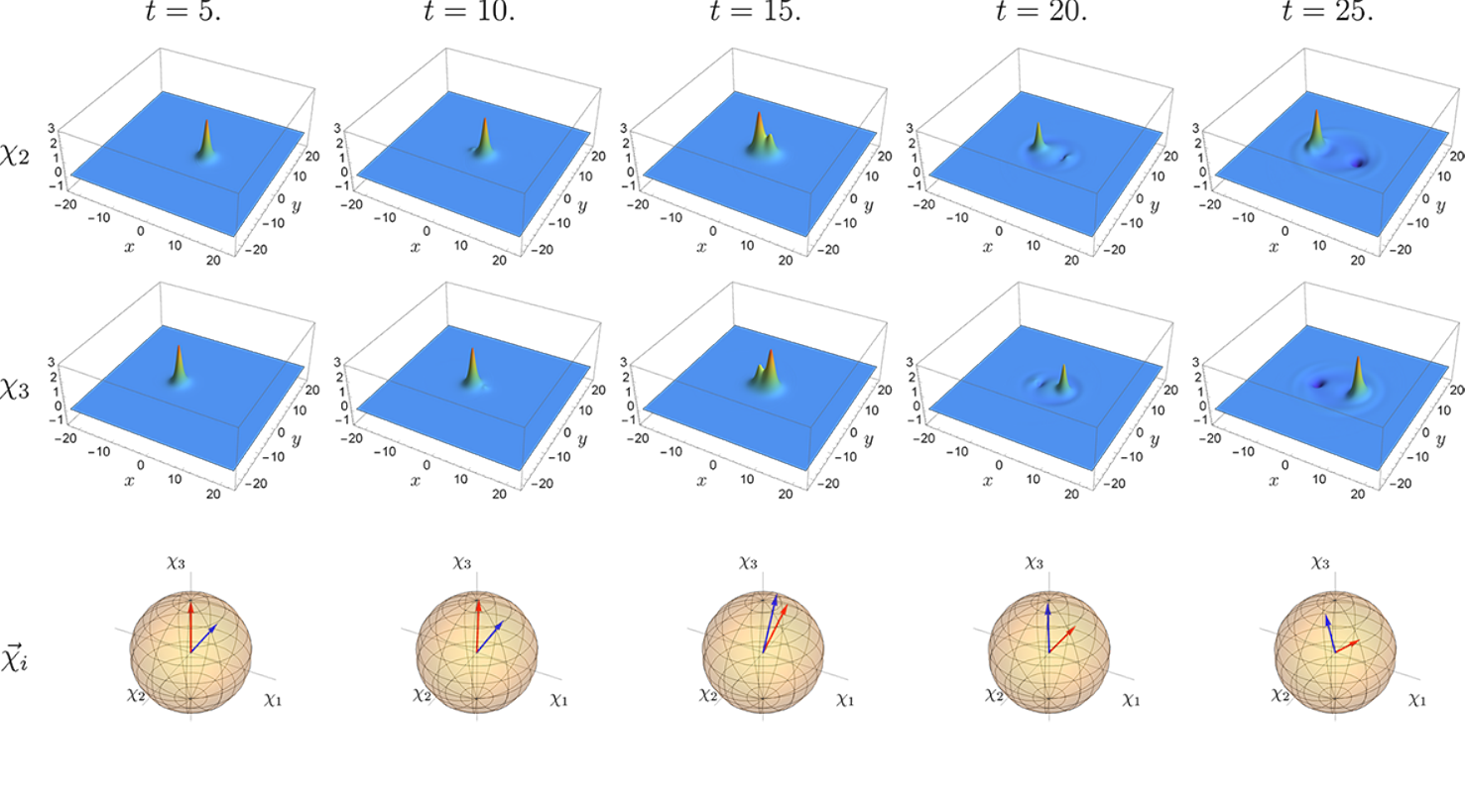}
\caption{Snapshots of $\chi_2$ and $\chi_3$ for the head-on collision of type II non-Abelian vortices, with initially orthogonal  orientations and initial velocity $u=0.5$.
The vortices bounce-back due to the type II repulsive force, the orientations dynamically change during and after the scattering. Internal orientations in $S^3$ are represented in the last row.}
\label{fig:3D_type2_ortho_u5_01}
\end{center}
\end{figure}
Evolution of $\chi_2$ and $\chi_3$ for $u=0.5$ (bounce-back) is shown in Fig.~\ref{fig:3D_type2_ortho_u5_01}.
Initially, the orientations are orthogonal since $(\chi_1,\chi_2,\chi_3) = (0,\chi_*,0)$ at one vortex center
and $(0,0,\chi_*)$ at the other vortex center, where $\chi_*$ stands for some positive constant.
The evolution of $\chi_2$ and $\chi_3$ is identical upon the reflections $x \leftrightarrow - x$ and $\chi_2 \leftrightarrow \chi_3$.
The positive peaks of $\chi_2$ and $\chi_3$ come close and eventually {\it pass through} each other, although the host vortices bounce-back. 
At the same time, in the aftermath of the scattering, negative condensations are generated. 
Fig.~\ref{fig:rel_ori_type2} shows the evolution of the relative orientation $\alpha_1 - \alpha_2$, and we find qualitatively similar
motion for all the scatterings below the threshold velocity. For low initial velocities, one vortex's internal orientation rotates clockwise in the $(\chi_2, \chi_3)$ plane, whereas the other vortex's orientation rotates anti-clockwise. The (de-compactified) angle between them increases with time, as shown in Fig.~\ref{fig:rel_ori_type2}.

Scatterings with greater initial velocities, $u=0.6,\cdots, 0.9$, which are all above the right-angles scattering threshold, are also qualitatively similiar to each other.
The vortices scatter at right angles, and the initially orthogonal internal orientations gradually come closer to each other and
become parallel at the instant of collision. After the collision, they remain parallel as the vortices move away from the collision point. The right-angles scattering is again 
a consequence of indistinguishability of the vortices at the moment of impact, see Fig.~\ref{fig:rel_ori_type2}.
\begin{figure}[t]
\begin{center}
\includegraphics[width=15cm]{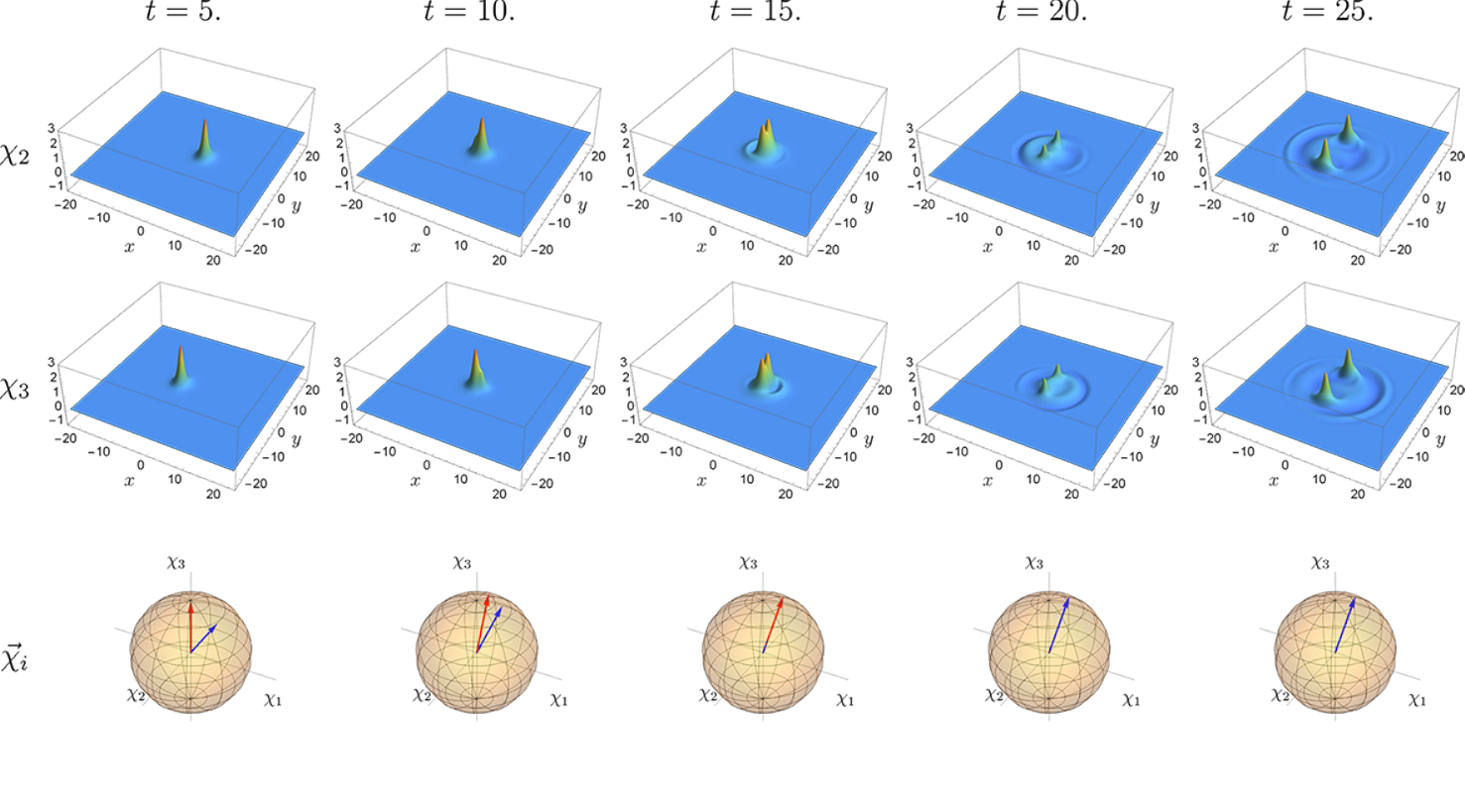}
\caption{Snapshots of $\chi_2$ and $\chi_3$ for the head-on collision of type II non-Abelian orthogonal vortices with initial velocity $u=0.6$.
The vortices scatter at right angles, their orientations having become parallel to each other at the time of collision, and continuing to be so in the asymptotic outgoing state. Internal orientations in $S^3$ are represented in the last row.}
\label{fig:3D_type2_ortho_u6_01}
\end{center}
\end{figure}
We show the evolution of $\chi_2$ and $\chi_3$ for the scattering of initially orthogonal vortices prepared with velocity $u=0.6$ in Fig.~\ref{fig:3D_type2_ortho_u6_01}.
Note that we have met qualitatively similar dynamics in the internal orientations for the non-Abelian global vortices, see for example Fig.~\ref{fig:A01_alpha_2}.
As mentioned before in the study of global vortices, the orientations becoming stuck in parallel directions after a right-angles scattering is a distinctive feature of the non-Abelian vortices with $S^{N-1}$ orientational moduli.
In contrast, BPS non-Abelian vortices with $\mathbb{C}P^{N-1}$ non-Abelian moduli exhibit
scattering at right angles both in real and internal spaces \cite{Eto:2006db}.

\subsection{Head-on collisions: type I* non-Abelian vortices}

Let us next study head-on scatterings of type I* non-Abelian vortices. We will focus on one representative point in parameter space,
corresponding to $(m_\chi/m_\phi, m_\gamma/m_\phi) = (0.05, 0.3)$ or more explicitly ($v = 5$, $\lambda = 1/4$ , $e = 3/10\sqrt{2}$, $\Omega = 1/4$).
For this choice, the vortex size is found to be $L \sim 2.33$ and the initial distance between two vortices is set to be $a = 6L$.
As before, in our simulations the initial incoming velocities are taken in the range $u=0.3$ to $u=0.9$.

\begin{figure}[ht]
\begin{center}
\def\svgwidth{15cm}
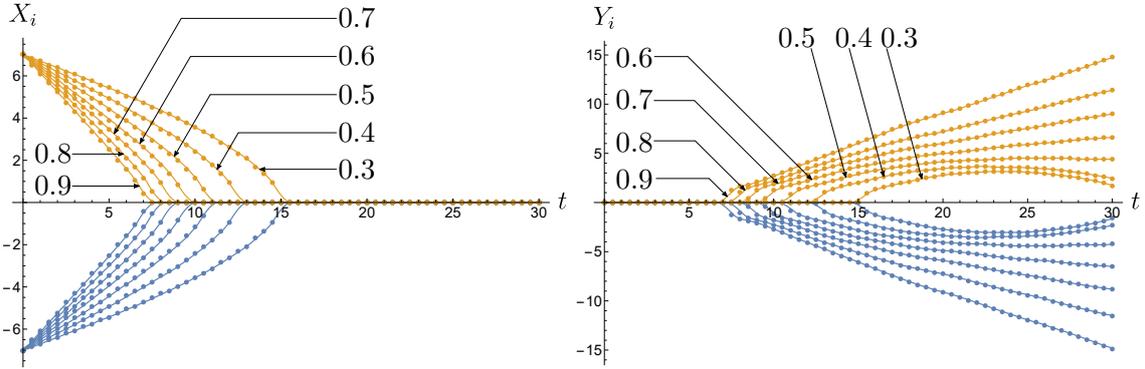
\caption{Trajectories $X_i(t)$ and $Y_i(t)$ of the parallel non-Abelian vortices of type I*, for initial velocities $u=0.3, 0.4, \cdots, 0.9$ (marked on the plots for clarity).}
\label{fig:XYvsT_type1_para_01}
\end{center}
\end{figure}

\paragraph{Parallel vortices:}
Results for head-on collisions of type I* vortices with parallel orientations and various initial velocities are summarized in Fig.~\ref{fig:XYvsT_type1_para_01}.

In all previous examples, vortices bounce backwards when the initial incoming velocity is small enough.
However, in this case we observe that the vortices always scatter at right angles, irrespective of the initial velocities.
This happens because the inter-vortex force between parallel type I* vortices is strongly attractive.
The effects of the attractive force are clearly seen in Fig.~\ref{fig:XYvsT_type1_para_01}. 
The vortices are accelerated towards the impact point before the collision, and they slow down after the collision.
The simulation shows that the vortices are bound to each other for low-lying $u$, and they will coalesce 
after several bounces by emitting  kinetic energy. On the other hand, the vortices seem not to form a bound state
and move away towards infinity for large $u$. We cannot make a prediction only from the numerical simulations, but together with considerations
based on the analytical formula Eq.~(\ref{eq:V_int2}), it is plausible that there exists an escaping velocity which
separates bound states and scattering states.

\begin{figure}[ht]
\begin{center}
\def\svgwidth{15cm}
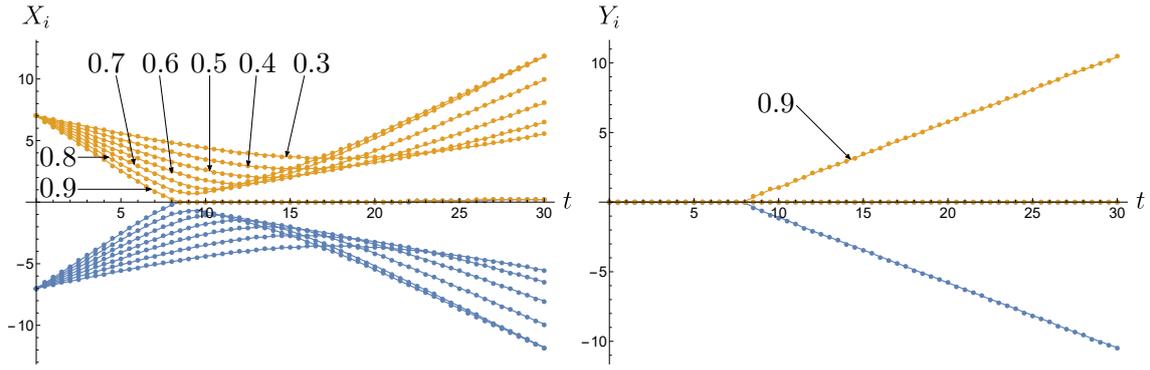
\caption{Trajectories $X_i(t)$ and $Y_i(t)$ of two anti-parallel non-Abelian vortices of type I*, for initial velocities $u=0.3, 0.4, \cdots, 0.9$ (marked on the plots for clarity).}
\label{fig:XYvsT_type1_apara_01}
\end{center}
\end{figure}

\paragraph{Anti-parallel vortices:}
Results for the head-on collisions of type I* non-Abelian vortices with anti-parallel orientations are summarized in Fig.~\ref{fig:XYvsT_type1_apara_01}.
The scatterings are quite similar to those of the type II anti-parallel vortices shown in Fig.~\ref{fig:XYvsT_type2_apara_01}.
Bounce-back scatterings for small velocities, bounce-back scattering while exchanging orientations for intermediate velocities, and
scattering at right angles for large velocities, are all observed here. Therefore, we do not elaborate further on this case to avoid repetitions.

\begin{figure}[ht]
\begin{center}
\def\svgwidth{15cm}
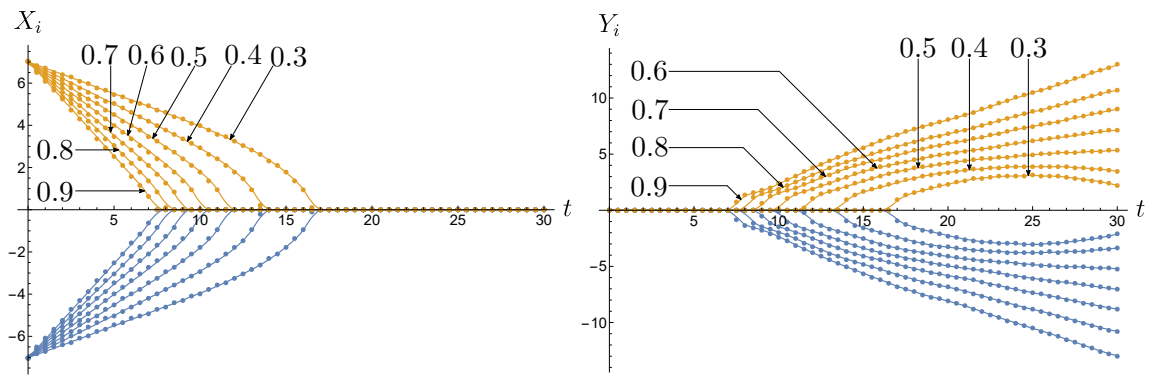
\caption{Trajectories $X_i(t)$ and $Y_i(t)$ of two orthogonal non-Abelian vortices of type I*, for initial velocities $u=0.3, 0.4, \cdots, 0.9$ (marked on the plots for clarity).}
\label{fig:XYvsT_type1_ortho_01}
\end{center}
\end{figure}

\begin{figure}[ht]
\begin{center}
\includegraphics[width=9cm]{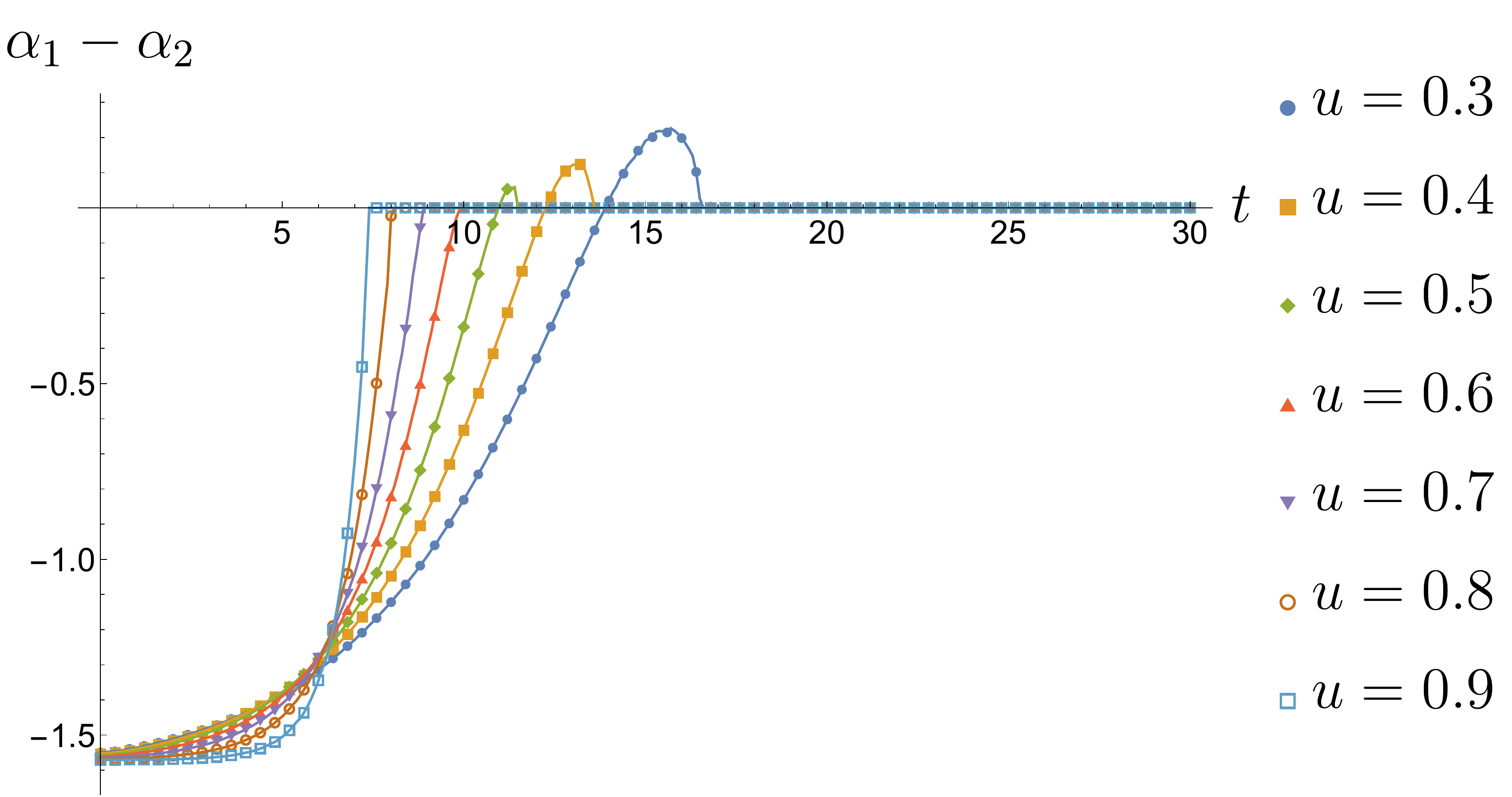}
\caption{Evolution of the relative orientations of type I* non-Abelian vortices during head-on scatterings with initially orthogonal orientations and velocities $u=0.3,\cdots,0.9$.}
\label{fig:ori_t1_para_ortho}
\end{center}
\end{figure}

\paragraph{Orthogonal vortices:}
Results for the head-on collisions of type I* non-Abelian vortices with orthogonal orientations are summarized in Fig.~\ref{fig:XYvsT_type1_ortho_01}.
Trajectories of orthogonal vortices in the $(x,y)$ plane are very similar to those of the parallel cases shown in Fig.~\ref{fig:XYvsT_type1_para_01}.
We encountered a similar situation for type II vortices, and as before the matching between the trajectories of parallel and orthogonal vortices implies that
the internal orientations do not significantly affect vortex motion. However, evolution of the internal orientations is quite different between the parallel and orthogonal cases as shown in Fig.~\ref{fig:ori_t1_para_ortho}.
The orientations are preserved to be parallel for the former while they approach each other and become parallel after the impact for the latter.
The internal parallelism again ensures right-angle scattering is possible.
Again, this is a unique property of non-Abelian vortices with $S^{N-1}$ orientational moduli.

\section{More general scatterings}
\label{exotic}

Our numerical methods are completely general, and as such prove to be well-suited to analyze a vast array of situations beyond the
two-vortex head-on scatterings considered in Section~\ref{sec:scattering_lnA}. In this section we present partial results for other 
cases that may be of interest. We have no pretensions of exhaustivity, and instead aim to convey the general flavour of some of the 
more prominent avenues one can immediately follow. More detailed studies of some of these situations could be warranted, but are left for future publications for the sake of brevity.

\paragraph{Scatterings with non-vanishing impact parameter:} All along Section~\ref{sec:scattering_lnA} the impact parameter $b$ introduced in \eqref{eq:ini_A_1}-\eqref{eq:ini_A_2} has remained turned off. We now consider cases where $b$ is non-zero, corresponding to scatterings in which vortices approach each other in the $x$ direction, but have a relative displacement in the $y$ direction.

\begin{figure}[h]
\begin{center}
\begin{tabular}{cc}
\includegraphics[width=7cm]{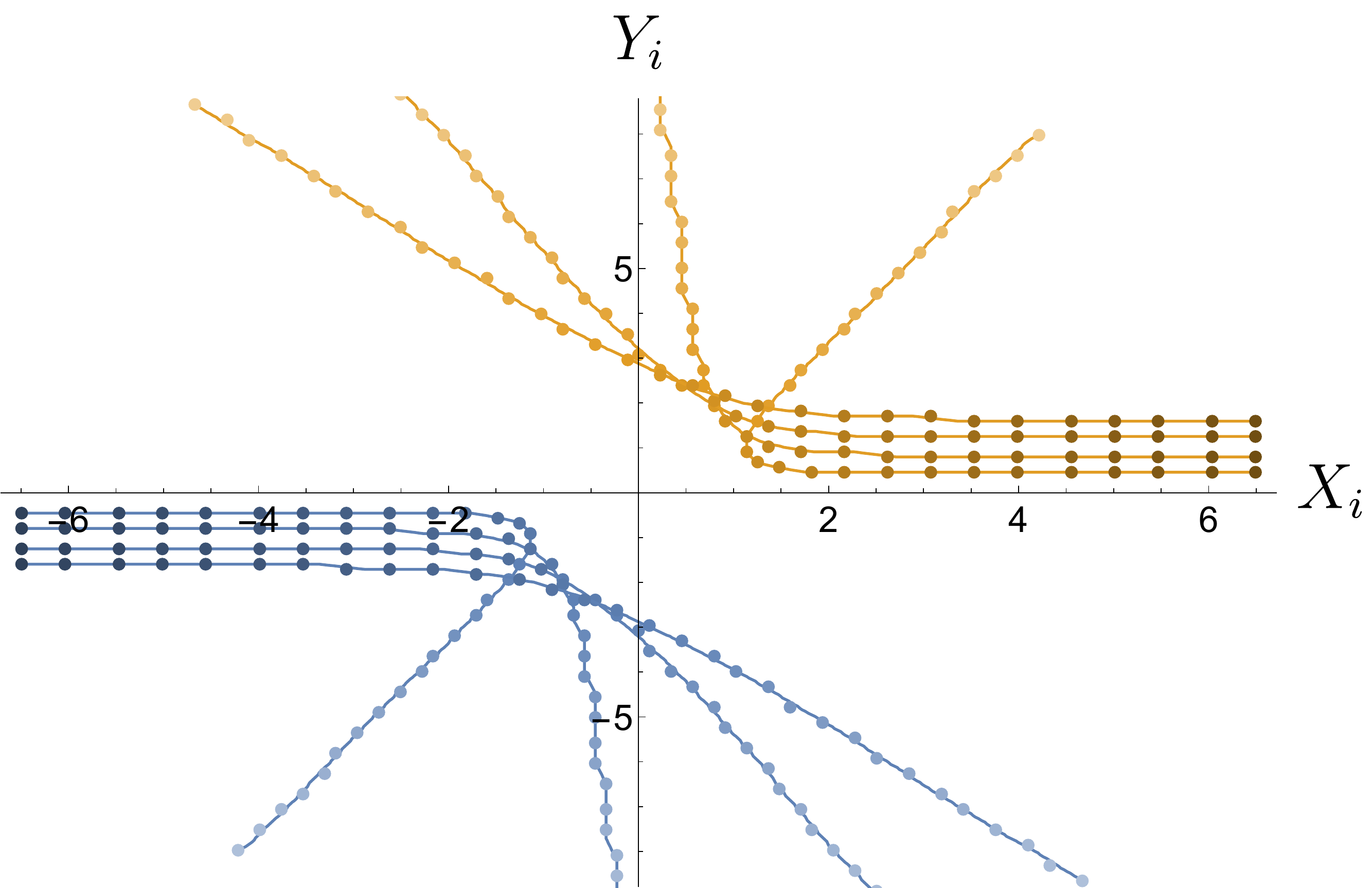} & \includegraphics[width=7cm]{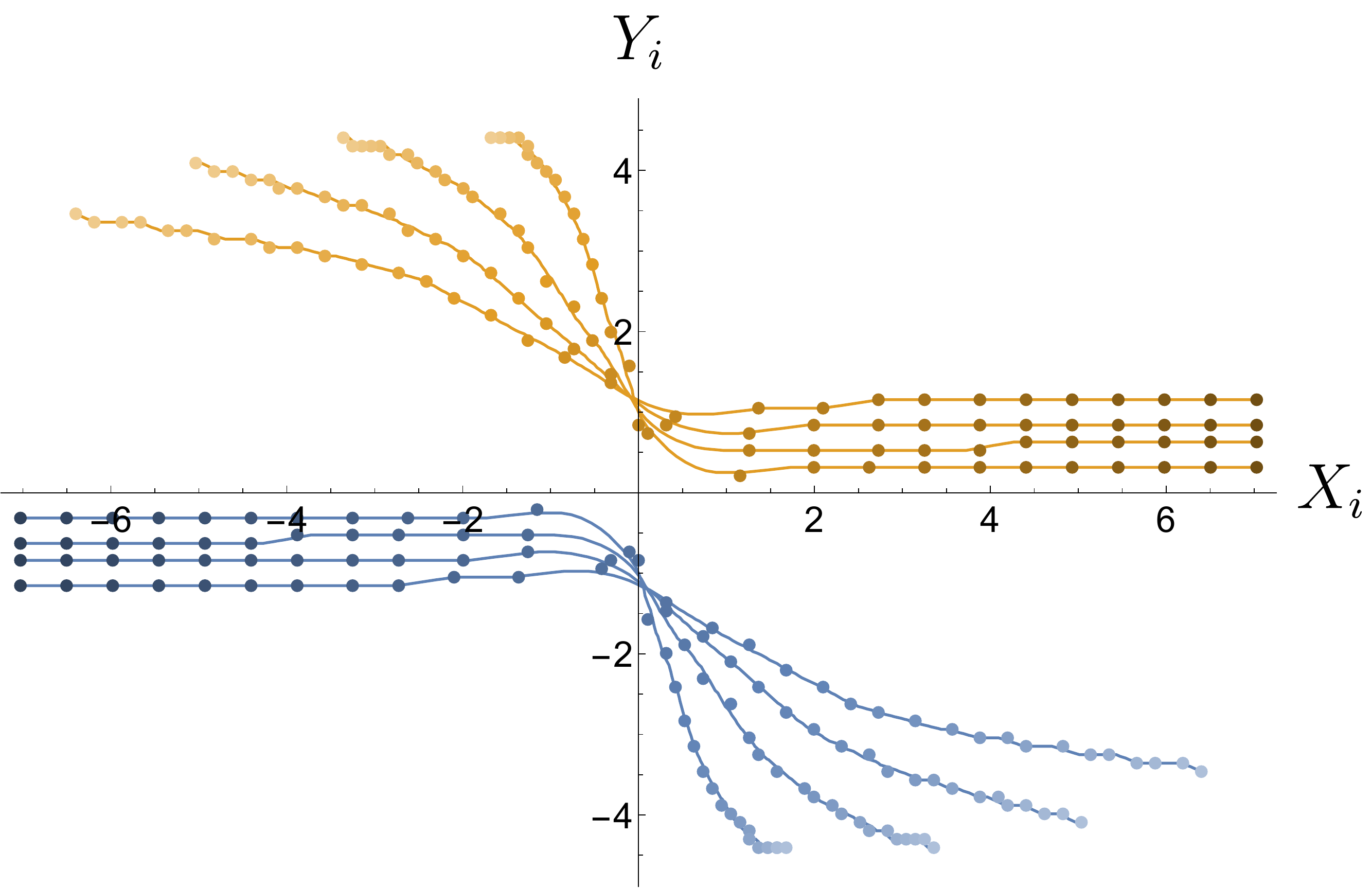}\\
(a) & (b)
\end{tabular}
\caption{Trajectories $(X_i(t), Y_i(t))$ of type II (a) and type I* (b) non-Abelian vortices with parallel internal orientations, scattering with impact parameters $b/L = 0.25, 0.50, 0.75, 1$. The vortices start at large values of $|X_i|$ (darker points) moving towards the origin with initial velocity $u = 0.50$, and scatter at increasingly obtuse angles as $b/L$ increases.}
\label{fig:impact_parameter}
\end{center}
\end{figure}

We illustrate the typical situation when the force between vortices is repulsive in Fig.~\ref{fig:impact_parameter}a, where we consider the scattering of two type II non-Abelian vortices with parallel orientations, initial velocities $u = 0.5$ and impact parameter $b/L = 0.25, 0.50, 0.75, 1$, where $L$ is again the vortex size. For comparison, typical results for scatterings with attractive inter-vortex forces and non-vanishing impact parameter are illustrated in Fig.~\ref{fig:impact_parameter}b, using in this case type I* non-Abelian vortices with otherwise identical parameters. We see in the latter case that the vortices attract each other when they are far away, but repel when they become very close. Thus, the interactions occurring near the origin push the vortices away, before the attractive forces regain their prominence once more, slowing the vortices and bending their trajectories as they move away.

\begin{figure}[h]
\begin{center}
\begin{tabular}{cc}
\includegraphics[width=7cm]{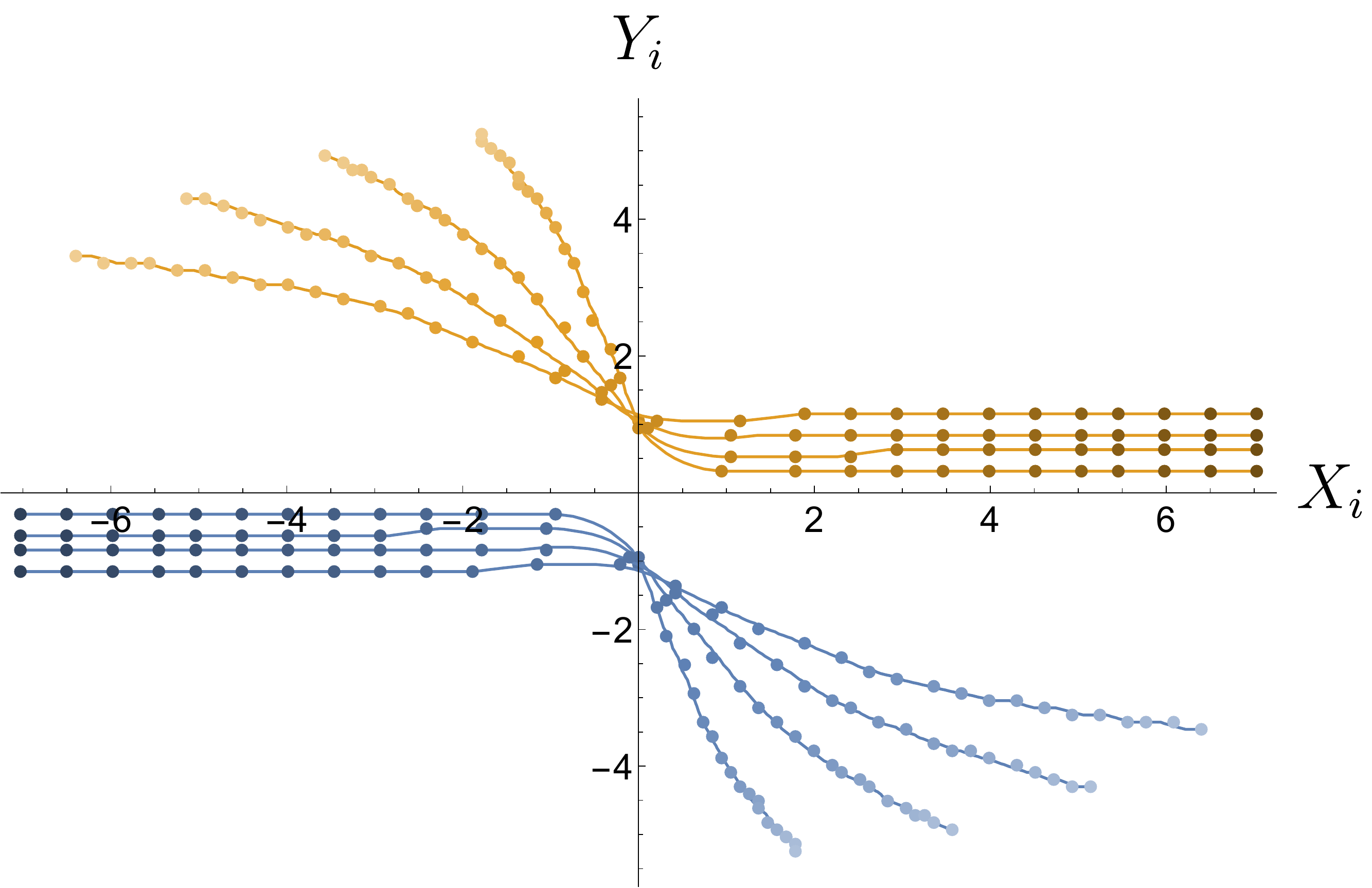} & \includegraphics[width=7cm]{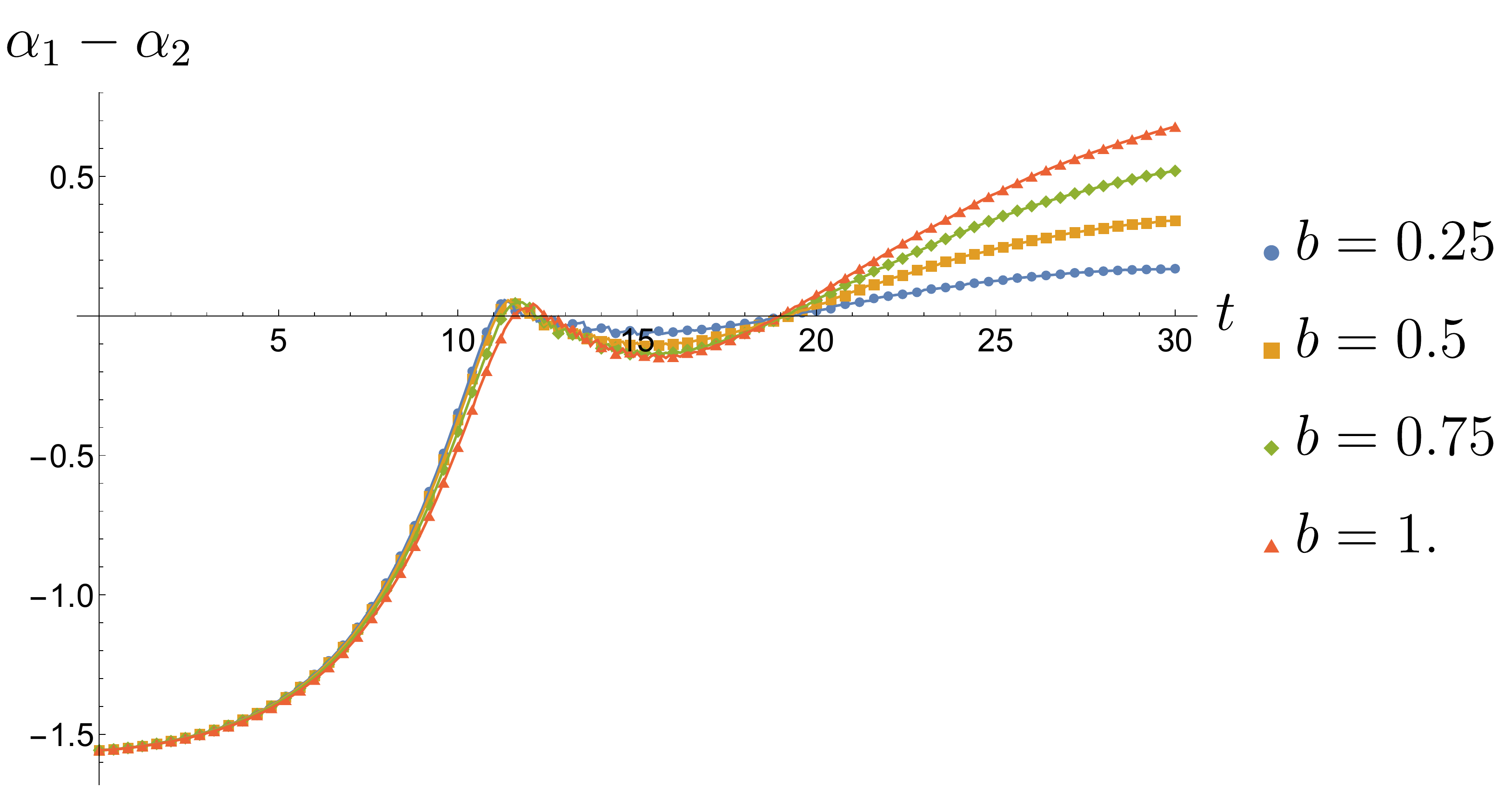}\\
(a) & (b)
\end{tabular}
\caption{Scattering of type I* non-Abelian vortices with initially orthogonal internal orientations and impact parameters $b/L = 0.25, 0.50, 0.75, 1$. (a) Trajectories $(X_i(t), Y_i(t))$ of the vortices: starting at large values of $|X_i|$ (darker points) and moving towards the origin with initial velocity $u = 0.50$, they scatter at increasingly obtuse angles as $b/L$ increases. (b) Angle formed between the internal orientations of the vortices during the scattering process.}
\label{fig:impact_parameter2}
\end{center}
\end{figure}

As in the previous Section, for parallel and antiparallel vortices the internal orientations are preserved during the whole simulations,  but in the case of initially orthogonal internal orientations these evolve with time in a non-trivial fashion. We illustrate such a situation in Fig.~\ref{fig:impact_parameter2}, where we plot the trajectories (a) and the angle between the internal orientations of the vortices (b),  for the non-Abelian type I* case with initial velocity $u = 0.5$ and impact parameters $b/L = 0.25, 0.50, 0.75, 1$.

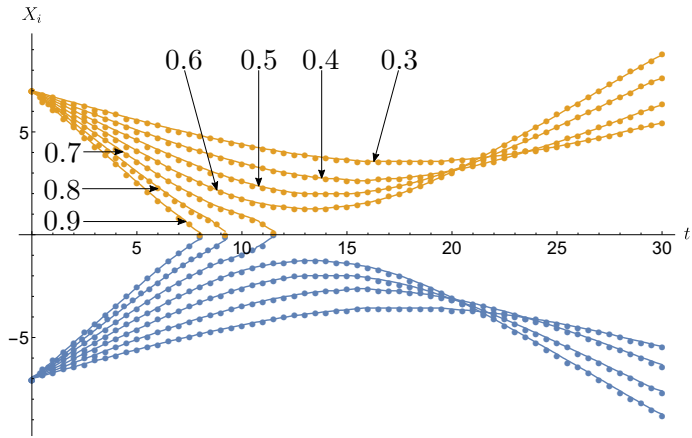
\begin{figure}[h]
\begin{center}
\def\svgwidth{9cm}
\begingroup%
  \makeatletter%
  \providecommand\color[2][]{%
    \errmessage{(Inkscape) Color is used for the text in Inkscape, but the package 'color.sty' is not loaded}%
    \renewcommand\color[2][]{}%
  }%
  \providecommand\transparent[1]{%
    \errmessage{(Inkscape) Transparency is used (non-zero) for the text in Inkscape, but the package 'transparent.sty' is not loaded}%
    \renewcommand\transparent[1]{}%
  }%
  \providecommand\rotatebox[2]{#2}%
  \newcommand*\fsize{\dimexpr\f@size pt\relax}%
  \newcommand*\lineheight[1]{\fontsize{\fsize}{#1\fsize}\selectfont}%
  \ifx\svgwidth\undefined%
    \setlength{\unitlength}{768bp}%
    \ifx\svgscale\undefined%
      \relax%
    \else%
      \setlength{\unitlength}{\unitlength * \real{\svgscale}}%
    \fi%
  \else%
    \setlength{\unitlength}{\svgwidth}%
  \fi%
  \global\let\svgwidth\undefined%
  \global\let\svgscale\undefined%
  \makeatother%
  \begin{picture}(1,0.63932294)%
    \lineheight{1}%
    \setlength\tabcolsep{0pt}%
    \put(0,0){\includegraphics[width=\unitlength,page=1]{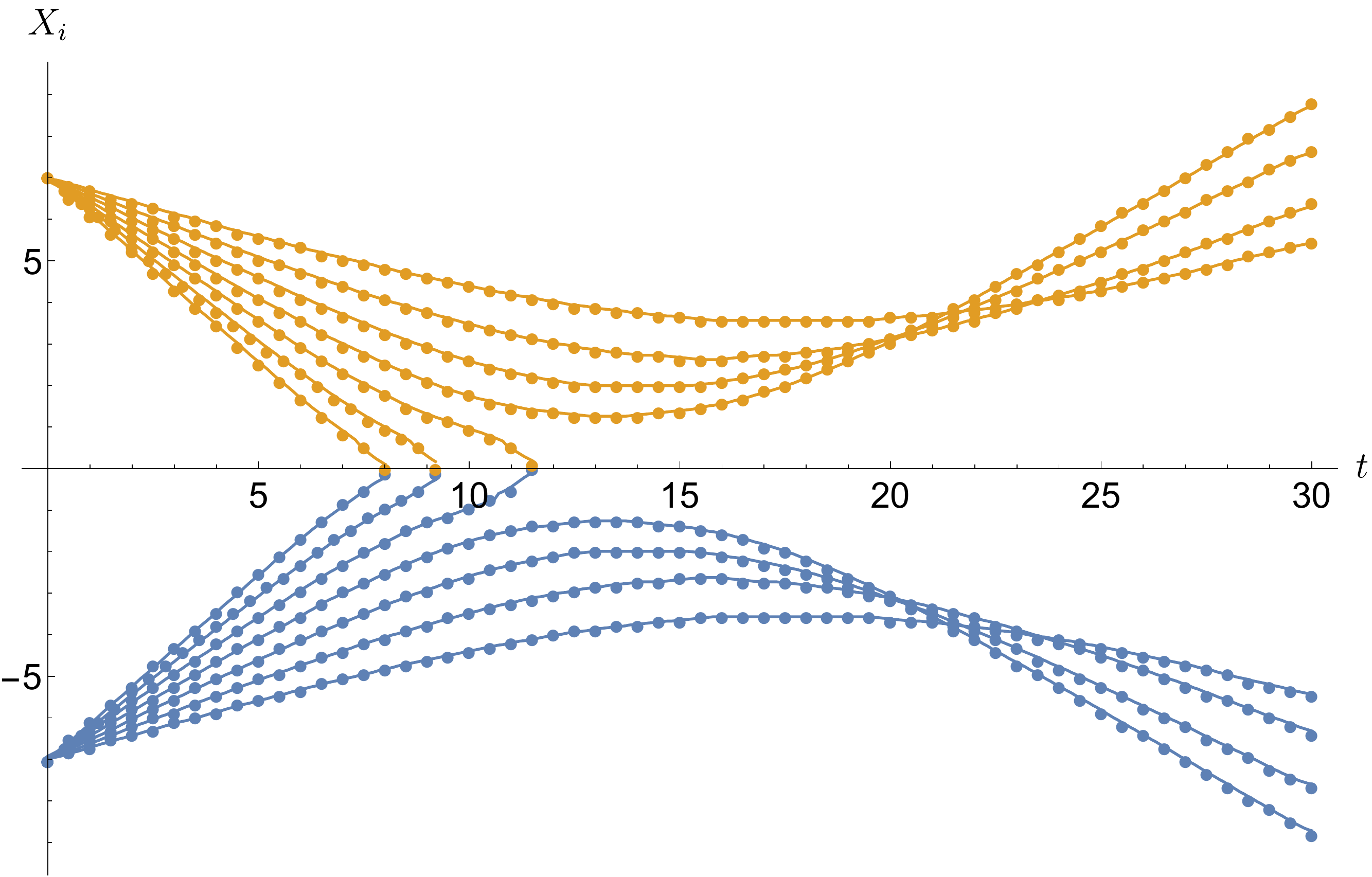}}%
    \put(0,0){\includegraphics[width=\unitlength,page=1]{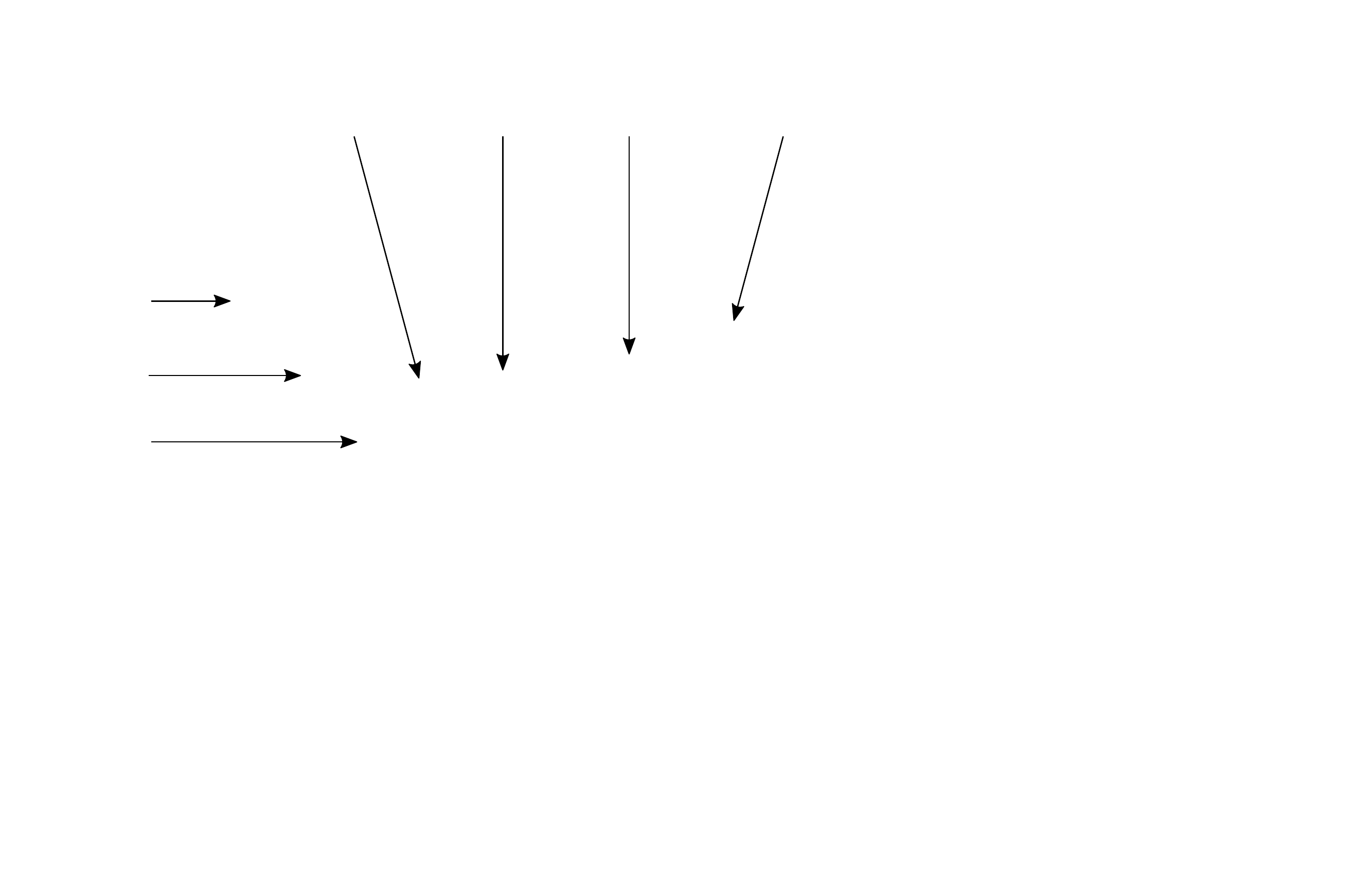}}%
    \put(0.052,0.305){\color[rgb]{0,0,0}\makebox(0,0)[lt]{\lineheight{1.25}\smash{\begin{tabular}[t]{l}$0.9$\end{tabular}}}}%
    \put(0.052,0.35){\color[rgb]{0,0,0}\makebox(0,0)[lt]{\lineheight{1.25}\smash{\begin{tabular}[t]{l}$0.8$\end{tabular}}}}%
    \put(0.053,0.405){\color[rgb]{0,0,0}\makebox(0,0)[lt]{\lineheight{1.25}\smash{\begin{tabular}[t]{l}$0.7$\end{tabular}}}}%
    \put(0.23,0.5425){\color[rgb]{0,0,0}\makebox(0,0)[lt]{\lineheight{1.25}\smash{\begin{tabular}[t]{l}$0.6$\end{tabular}}}}%
    \put(0.34,0.5425){\color[rgb]{0,0,0}\makebox(0,0)[lt]{\lineheight{1.25}\smash{\begin{tabular}[t]{l}$0.5$\end{tabular}}}}%
    \put(0.43,0.5425){\color[rgb]{0,0,0}\makebox(0,0)[lt]{\lineheight{1.25}\smash{\begin{tabular}[t]{l}$0.4$\end{tabular}}}}%
    \put(0.545,0.5425){\color[rgb]{0,0,0}\makebox(0,0)[lt]{\lineheight{1.25}\smash{\begin{tabular}[t]{l}$0.3$\end{tabular}}}}%
  \end{picture}%
\endgroup%

\caption{Trajectories for non-Abelian vortex-antivortex head-on scatterings of type I*, with initial velocities $u=0.3, 0.4, \cdots, 0.9$ (marked on the plots for clarity). In all cases, $Y_i(t) = 0$ at all times. The curves for $u = 0.7, 0.8, 0.9$ are truncated before the end of the simulation ($t = 30$) because the vortex-antivortex pair annihilates during the collision.}
\label{fig:XYvsT_type1_vvbar_apara_01}
\end{center}
\end{figure}

\begin{figure}[h]
\begin{center}
\includegraphics[width=15cm]{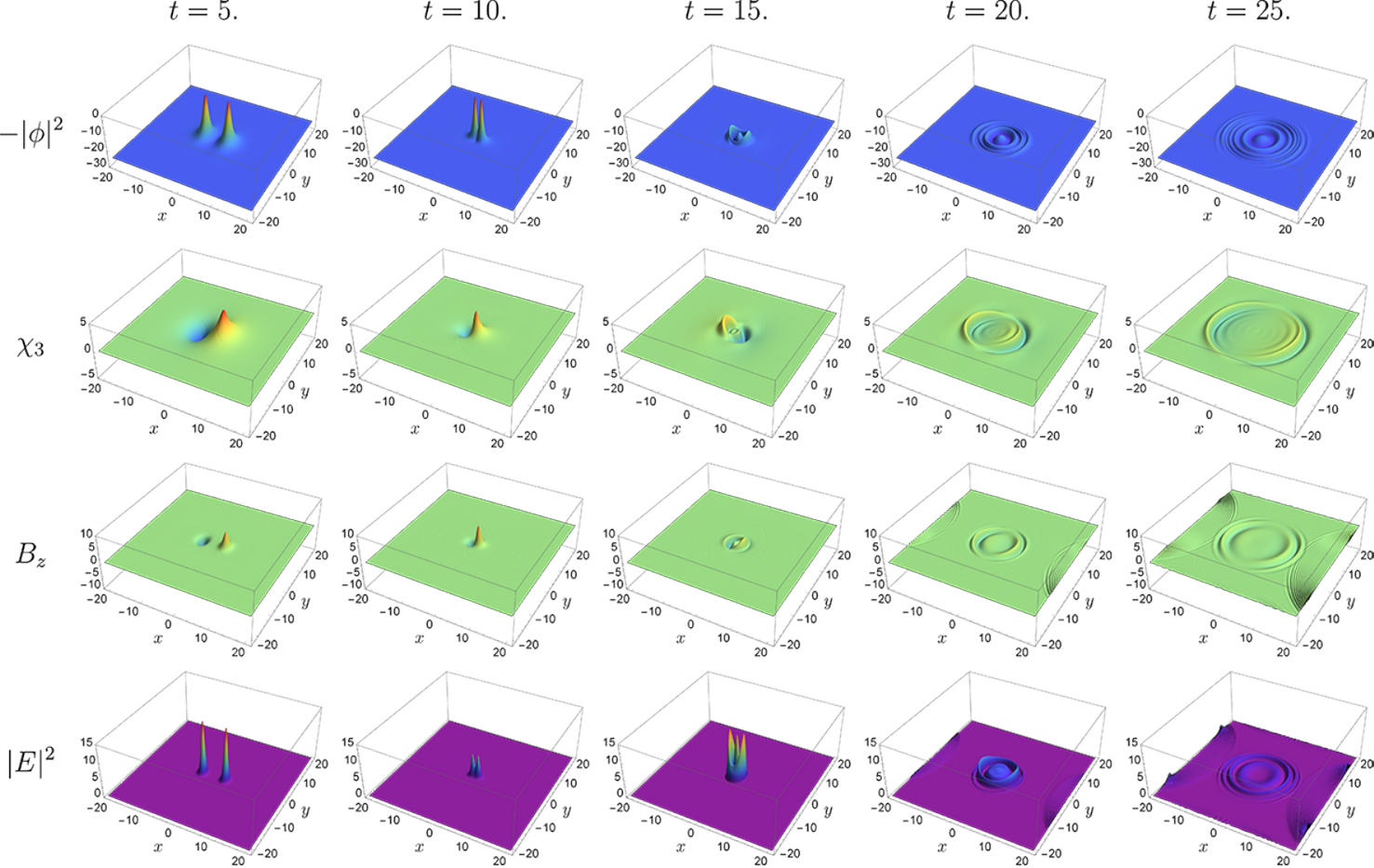}
\caption{Snapshots of $-|\phi|^2$, $\chi_3$, $B_z$ and $|E|^2$ for the head-on scattering of a type I* anti-parallel vortex-antivortex state with initial velocity $u=0.7$. Note the vortices annihilate during the collision occurring around $t=11.5$, and only waves are present thereafter.}
\label{fig:3D_type1_vvbar_apara_u7_01}
\end{center}
\end{figure}

\paragraph{Scatterings of vortices and anti-vortices:} We can also study scattering processes involving a vortex and an antivortex. A particularly interesting situation arises for type I* non-Abelian anti-parallel vortex-antivortex collisions, since in this case the corresponding forces are repulsive, see Table~\ref{tab:summary2}. We therefore expect to observe qualitatively similar behavior to that of the head-on collision of type II non-Abelian vortices with anti-parallel orientations, see \textit{e.g.} Figs.~\ref{fig:XYvsT_type2_apara_01}-\ref{fig:3D_type2_apara_u7_01}, at least for low initial velocities. Our results are shown in Fig.~\ref{fig:XYvsT_type1_vvbar_apara_01} where as expected we see that for low enough initial velocities a vortex and an antivortex colliding head-on bounce back due to the repulsion. The threshold velocity is in this case somewhere between $u = 0.6$ and $u = 0.7$, and for velocities higher than this threshold we observe on the other hand a markedly different situation. Indeed, upon colliding the vortex and antivortex do not scatter but annihilate each other, leaving behind a pulsating core that decays by emitting waves. Snapshots of the collision with initial velocity $u=0.7$ at various times, corresponding to such an annihilation process, are shown in Fig.~\ref{fig:3D_type1_vvbar_apara_u7_01}. Comparing with Fig.~\ref{fig:3D_type2_apara_u8_01} we see that although the evolution looks qualitatively similar before the collision, after it occurs there are no vortices but just waves.

\paragraph{Multi-vortex scatterings:} A simple generalization of the procedure described in Section~\ref{sec:numerics} allows us to prepare states with more than two vortices in the initial configuration. Time evolution then proceeds exactly as described there, so we are essentially limited only by our imagination and computational resources when it comes to analyzing more complicated vortex interactions.

For illustration purposes we consider here $n = 3$ and $n=4$ scattering vortices, and for concreteness we prepare the initial state with the $n$ vortices regularly distributed along an origin-centered circle of diameter $a = 8L$. Each vortex is boosted with the same initial velocity $u$ towards the origin, so that the whole configuration is invariant under rotations of $2\pi/n$ around the origin, up to differences in the internal orientations of each vortex.

\begin{figure}[h]
\begin{center}
\begin{tabular}{cc}
\includegraphics[width=5cm]{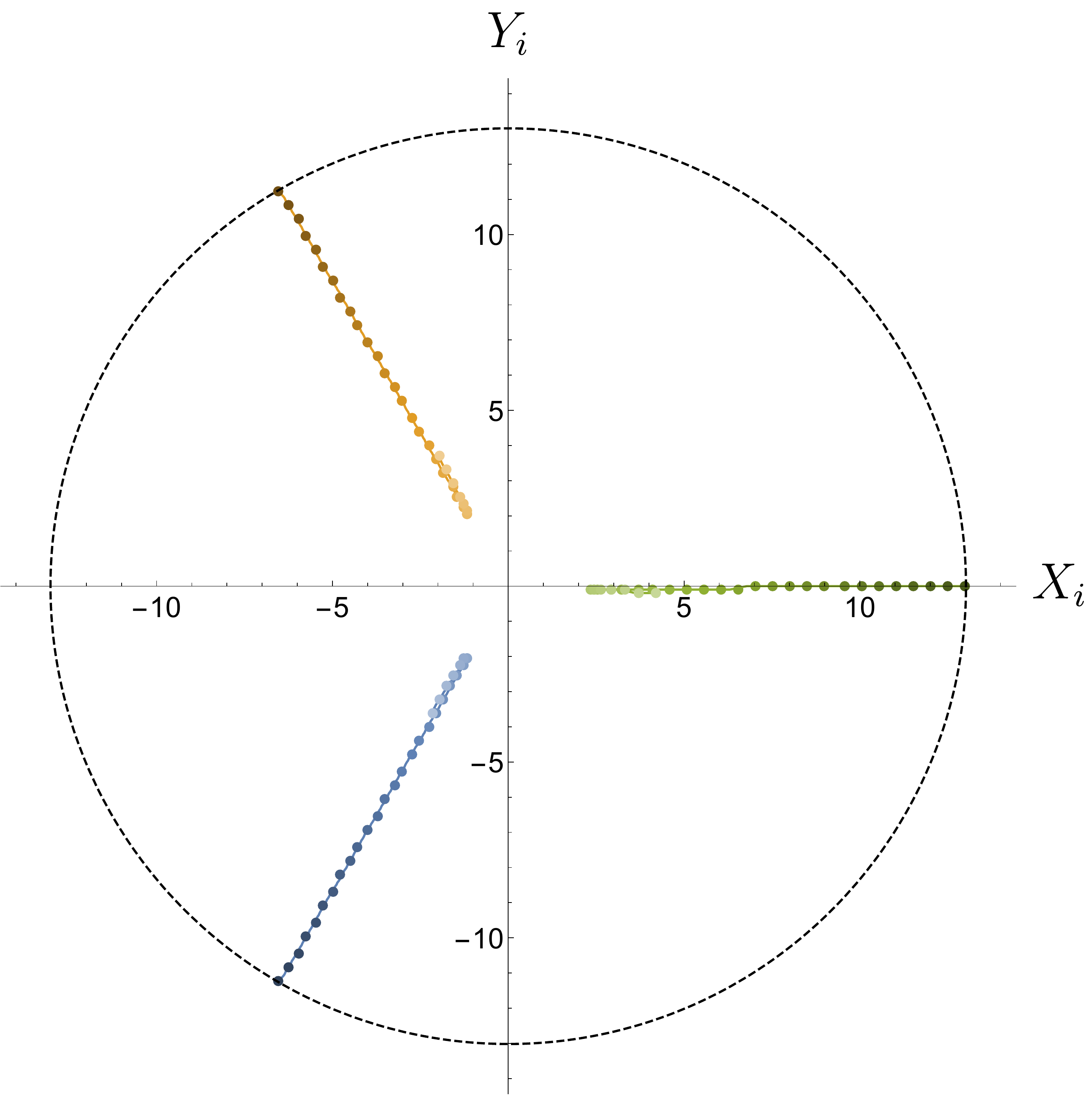} & \includegraphics[width=5cm]{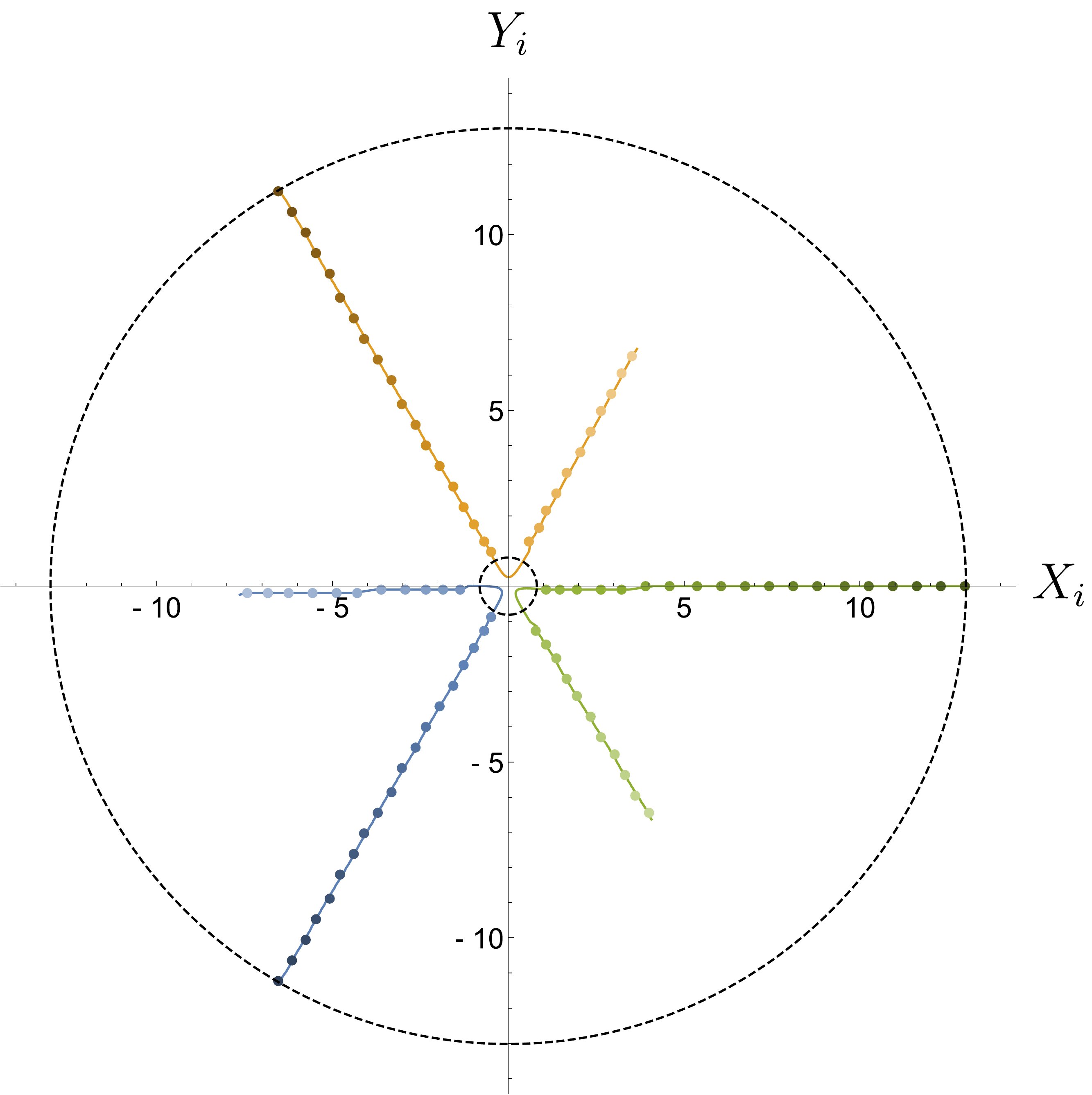}\\
(a) & (b)
\end{tabular}
\caption{Trajectories $(X_i(t), Y_i(t))$ of $n=3$ type II non-Abelian vortices with parallel internal orientations, scattering with initial velocity $u=0.5$ (a) and $u=0.7$ (b). The vortices start on the circle of diameter $a = 8L$ (outer dashed circle) moving first towards the origin and then scattering at away from it (points go from darker to lighter shades). The inner dashed circle in (b) has diameter $L/2$, and within it there is no way to distinguish individual vortices.}
\label{fig:XY_3vortex}
\end{center}
\end{figure}

In Fig.~\ref{fig:XY_3vortex} we show the trajectories of $n=3$ non-Abelian vortices of type II, scattering with initial velocities $u = 0.5$ (a) and $u = 0.7$ (b), with all vortices having parallel internal orientations. The inter-vortex forces being repulsive, we see that for low initial velocities the vortices bounce back before colliding, whereas at higher initial velocities they scatter at the origin. Note that because the three vortices are indistinguishable, the color association in the outer state of Fig.~\ref{fig:XY_3vortex}b is completely arbitrary and we could therefore also say \textit{e.g.} that the vortices pass through each other.

\begin{figure}[h]
\begin{center}
\includegraphics[width=15cm]{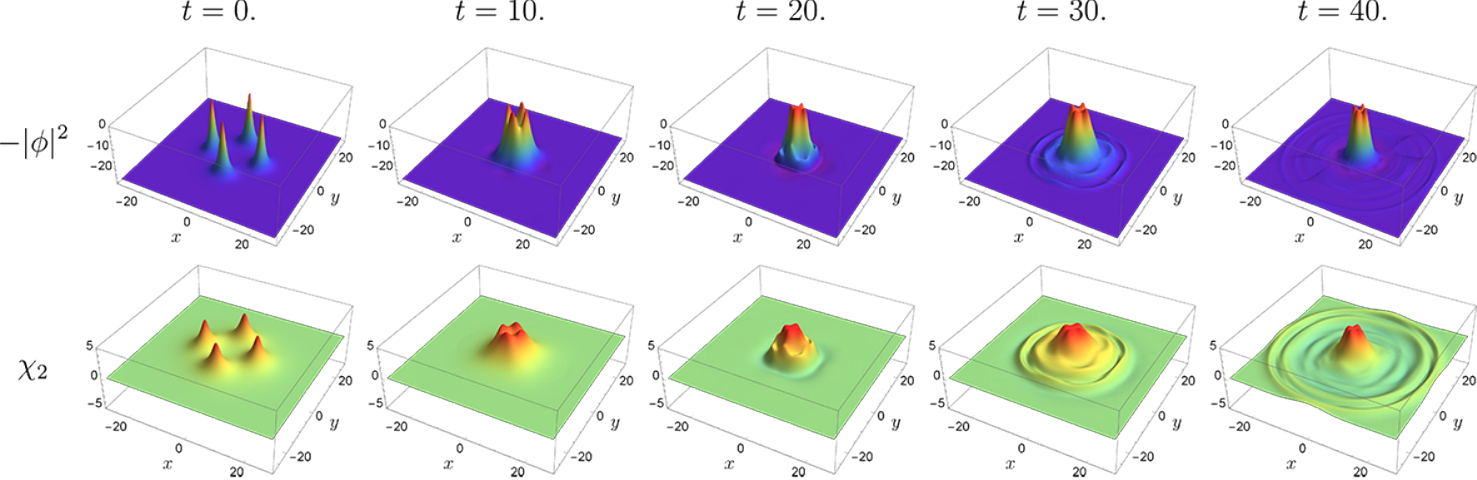}
\caption{Snapshots of $-|\phi|^2$ and $\chi_2$ for the scattering of $n = 4$ type I* non-Abelian vortices with parallel internal orientations and initial velocity $u=0.5$.}
\label{fig:3D_type1_4vortex_para_u5_01}
\end{center}
\end{figure}

A markedly different situation arises when the inter-vortex forces are attractive. Indeed, if the attraction is strong enough we can expect the formation of long-lived bound states of the $n$ vortices, see discussion around Fig.~\ref{fig:XYvsT_type1_para_01}. While we cannot guarantee this is the case based solely on numerical evidence, since our simulations necessarily involve a finite amount of time, this seems to be true \textit{e.g.} for $n = 4$ type I* non-Abelian vortices with parallel internal orientations prepared with initial velocity $u = 0.5$, as shown in Fig.~\ref{fig:3D_type1_4vortex_para_u5_01}.

\section{Summary and conclusion}
\label{concs}

In this paper we investigate in detail the dynamics, particularly in scattering problems, of non-Abelian vortices with internal $S^{N-1}$ orientational moduli on the two dimensional plane. The symmetry of the model under consideration is $U(1) \times SO(N)$, the $U(1)$ factor being spontaneously broken and giving rise to topologically
non-trivial vortices. We separately studied two cases: the first case corresponds to the $U(1)$ symmetry being a global symmetry, while in the second case it is a local symmetry.
The other factor $SO(N)$ is always global and it is unbroken in the vacuum. However, its spontaneous breaking
$SO(N) \to SO(N-1)$ locally occurs in the vicinity of the vortex center. This makes the vortices non-Abelian by accompanying the massless NG modes 
$S^{N-1} \simeq SO(N)/SO(N-1)$ (the internal orientation moduli). The presence of the internal orientation makes the dynamics of the vortices
quite rich and complicated. Clarifying this point was the main purpose of this paper.

In Sec.~\ref{sec:GNAV} we concentrated on the scattering problem of non-Abelian global vortices with $S^2$ ($N=3$) orientational moduli.
As usual Abelian global vortices, a long-range repulsive force mediated by the massless $U(1)$ NG mode in the bulk dominates the asymptotic inter-vortex forces.
Namely, irrespective of the internal orientation, the vortices asymptotically repel each other.
We numerically studied head-on scatterings for three different cases: the relative internal orientation being initially parallel, antiparallel, and orthogonal.
For each of these cases we considered various initial velocities $u$. Scatterings of the vortices with parallel orientations are qualitatively very similar to
well-known dynamics of Abelian global vortices. When $u$ is sufficiently small, the vortices bounce off backward.
On the other hand, when $u$ exceeds a critical velocity the vortices scatter at right angles.
In contrast, scatterings of the non-Abelian vortices with antiparallel orientations are different from those of the Abelian vortices.
The vortices always bounce back before their centers come into contact, at least for initial velocities up to 90 \% of speed of light.
The orthogonal case is the most interesting: 
when the initial velocity $u$ is small, the vortices bounce-back as in the case of parallel vortices, yet a difference can be found in the motion in internal space.
During the scatterings, in general the orientations are not fixed but change. Namely, energy transfer between real and internal space occurs.
When $u$ is sufficiently large, now the vortices scatter at right angles. When this happens, the orientations get closer as the vortices
approach in real space, and they become parallel at the moment of impact. After the collision, the orientations remain parallel.
As is expected, the presence of the internal orientations leads to rich dynamics.
We also briefly investigated scattering problems with non-zero impact parameter.

In Sec.~\ref{sec:static_int_lnA} and \ref{sec:scattering_lnA} we studied the local non-Abelian vortices with
$S^2$ orientation ($N=3$).
In Sec.~\ref{sec:static_int_lnA}, we first obtained the phase boundary in the parameter space $(m_\chi/m_\phi,m_\gamma/m_\phi)$
which separates two vortex types, corresponding to the Abelian ($SO(3)$ is unbroken) and non-Abelian ($SO(3)$ is broken to $SO(2)$) vortices.
Next, we studied the static inter-vortex forces between well-separated vortices. 
This was previously studied by two of the present authors \cite{Tallarita:2017opp} in a very similar model.
We carefully reexamined the asymptotic behavior of a single non-Abelian vortex, 
and found that a modification to the previous result is needed for the $\phi$ field (winding field). 
Reflecting the fact that the non-Abelian vortex exists only for  $2m_\chi \ll m_\phi$, the asymptotic interaction
associated with the $\phi$ field is modified.
This is essential to conclude that there are only two phases of the non-Abelian vortices,
the type II and type I*. Based on the analytic study of the asymptotic properties, 
we found that the former occurs for $m_\gamma < m_\chi$ and the latter for $m_\gamma > m_\chi$.
The asymptotic interaction between type II vortices is a magnetic repulsion, and is independent of the internal orientations. In comparison with the global vortices, 
the repulsive force is a short range force because it is mediated by the massive photon.
In contrast, the interaction between type I* vortices is sensitive to the internal orientations.
When they are parallel (antiparallel), the interaction is attractive (repulsive).
We numerically obtained the phase diagram of the vortex types, and verified that the analytic result based on the
asymptotic analysis is consistent with  the numerical results when $m_\chi/m_\phi$ and $m_\gamma/m_\phi$ 
are small. However, we found that the true type II region is slightly larger than the analytic estimation $m_\gamma < m_\chi$.

In Sec.~\ref{sec:scattering_lnA}, we performed numerical simulations of head-on scatterings of  type II and type I* vortices.  First, consistency between the numerical results and the analytic estimations provided a mutual check on the validity of both analysis.
We examined three different cases, with parallel, antiparallel and orthogonal internal orientations for both types of vortices.
We found that much richer phenomena occurs compared to the global vortices.
Firstly,  we examined type II vortices. The head-on collisions of parallel vortices are qualitatively similar to those
of non-Abelian global vortices with parallel orientations. The vortices bounce-back for a small initial velocity $u$,
and they scatter at right angles when $u$ is sufficiently large.
In contrast, the scatterings of antiparallel vortices show marked differences with respect to the global case. 
When $u$ is small, global and local vortices are qualitatively the same, namely they just bounce-back. 
But when $u$ is large, the local vortices scatter at right angles. At first glance, this is counterintuitive because
 right-angles scattering usually occurs for  two identical solitons, yet antiparallel vortices are not identical solitons.
To resolve the puzzle, we carefully observed the numerical simulations and found that the orientations scatter around 
the moment of impact, and indeed at this point the vortices have become identical. This does not happen for the global vortices because
they are protected by the stronger long range repulsive force. A further interesting phenomenon was observed for an
intermediate initial velocity between bounce-back scattering and right-angles scattering. The vortices bounce-back but
the orientations are exchanged between two vortices. 
The head-on scatterings of the orthogonal type II vortices proceed very similarly to those of the parallel cases if 
we only look at the spatial motion. However, the dynamics of the internal orientations are distinctive.
For small initial velocities, the vortices bounce-back and the initially orthogonal orientations non-trivially evolve in time: the relative angle is initially $\alpha_1-\alpha_2 = -\pi/2$, but immediately starts changing and continues to evolve throughout our simulations. For large $u$'s, the vortices scatter at right-angles and the orientations
become parallel after the collision.

Next we examined the scattering of type I* vortices. This case is the most different from the global case. 
The presence of the internal orientations is essential for separating Abelian and non-Abelian vortices.
However, it is washed out by the long range repulsive force (the massless NG boson) for the global vortices.
On the other hand, long range interactions are absent in the local model, and the type I* interaction depends crucially on the internal orientations.
Hence, it is the best playground to observe non-Abelian properties. The head-on scatterings of  parallel vortices are characteristic of the non-Abelian local vortices. The dominant inter-vortex force
is attractive, so the vortices never bounce-back but always scatter at right angles.
On the other hand, the antiparallel vortices feel the repulsion, they should bounce-back and indeed we observed this numerically.
For the orthogonal cases, we found that the scatterings in real space are almost the same as those of the parallel cases, but
motion of the internal orientations is different. The orientations of parallel vortices are fixed to be parallel throughout
the scatterings, but those of the orthogonal vortices approach each other and finally become parallel.
As we have seen, the scatterings of the non-Abelian vortices are quite rich and complicated. To further solidify our investigation,
we confirmed the asymptotic inter-vortex interactions. We found that it is indeed useful to have qualitative understandings
for various scatterings. We also included a more exotic section with multi-vortex scatterings.

One of the important discoveries of this work is to reveal a difference between two kinds of non-Abelian vortices,
those with $\mathbb{C}P^{N-1}$ internal orientations on one hand, and those with $S^{N-1}$ internal orientations on the other.
Head-on collisions of BPS non-Abelian vortices of the former type 
scatter at right angles both in real and internal spaces \cite{Eto:2006db}.
In contrast, collisions of the latter type become stuck in parallel orientations of the internal space after right-angles scattering. Scattering at right angles is related to reconnection probabilities of cosmic strings in $3+1$ dimensions. The reconnection probabilities of non-Abelian vortices with $\mathbb{C}P^{N-1}$ moduli are known to be unity \cite{Hashimoto:2005hi,Eto:2006db}. 
However, there exist no studies on the reconnection problem of non-Abelian cosmic strings with $S^{N-1}$ orientational moduli. 
This would be an interesting avenue of research, but is beyond the scope of this work, therefore we leave it for future work.

The setup used in this paper involves classical computations valid exclusively at small couplings of the fields. A relevant question is therefore what effects, if any, strong coupling has on the dynamics of vortices with orientational moduli. This is generically a hard question to answer. Initial studies in this direction were performed in \cite{Tallarita:2015mca,Tallarita:2019czh}, where holographic non-Abelian vortices were constructed, with and without backreaction on the gravitational sector. These models provide the first realizations of these kind of vortices in strongly coupled (albeit large N) field theories. An interesting avenue of future work would therefore involve the scattering of these vortices in a holographic setting. Presumably this strong coupling information is contained in gravitational waves, generated at the moment of collision. We leave this important question to a future study. 


\section*{Acknowledgements}
G. T. is funded by Fondecyt Grant 1200025. The research on this project has received funding from the European Research Council (ERC) under the European Union’s Horizon 2020 research and innovation program (QUASIFT grant agreement 677368).
The work of M. E. is supported in part by JSPS Grant-in-Aid for Scientific Research (KAKENHI Grant No. JP19K03839) and also by MEXT KAKENHI Grant-in-Aid for Scientific Research on Innovative Areas, Discrete Geometric Analysis for Materials Design, 
No. JP17H06462 from the MEXT of Japan. AP would like to thank Don Willcox and Ann Almgren for useful discussions on computational numerics.

\bibliographystyle{jhep}


\begin{thebibliography}{99}

\bibitem{abrikosov}
A. Abrikosov, Sov. Phys. JETP 32, 1442 (1957).

\bibitem{Nambu:1974zg}
Y.~Nambu,
Phys. Rev. D \textbf{10}, 4262 (1974)
doi:10.1103/PhysRevD.10.4262

\bibitem{thooft}
G. 't Hooft, Nucl. Phys. B 190, 455 (1981)
\bibitem{mandel}
S. Mandelstam, Phys. Rept. 23, 245 (1976).
\bibitem{Seiberg:1994rs} 
  N.~Seiberg and E.~Witten,
  Nucl.\ Phys.\ B {\bf 426}, 19 (1994)
  [Nucl.\ Phys.\ B {\bf 430}, 485 (1994)]
  [hep-th/9407087].

\bibitem{Vainshtein:2000hu}
A.~I.~Vainshtein and A.~Yung,
Nucl. Phys. B \textbf{614}, 3-25 (2001)
doi:10.1016/S0550-3213(01)00394-7
[arXiv:hep-th/0012250 [hep-th]].

\bibitem{shifman2}
A. Yung, M. Shifman: At the frontier of Particle Physics, vol 3* 1827-1857

\bibitem{Hanany:2003hp}
A.~Hanany and D.~Tong,
JHEP \textbf{07}, 037 (2003)
doi:10.1088/1126-6708/2003/07/037
[arXiv:hep-th/0306150 [hep-th]].

\bibitem{Auzzi:2003fs}
R.~Auzzi, S.~Bolognesi, J.~Evslin, K.~Konishi and A.~Yung,
Nucl. Phys. B \textbf{673}, 187-216 (2003)
doi:10.1016/j.nuclphysb.2003.09.029
[arXiv:hep-th/0307287 [hep-th]].

\bibitem{Carlino:2000uk}
G.~Carlino, K.~Konishi and H.~Murayama,
Nucl. Phys. B \textbf{590}, 37-122 (2000)
doi:10.1016/S0550-3213(00)00482-X
[arXiv:hep-th/0005076 [hep-th]].

\bibitem{Dorey:1999zk}
N.~Dorey, T.~J.~Hollowood and D.~Tong,
JHEP \textbf{05}, 006 (1999)
doi:10.1088/1126-6708/1999/05/006
[arXiv:hep-th/9902134 [hep-th]].

\bibitem{Hanany:2004ea}
A.~Hanany and D.~Tong,
JHEP \textbf{04}, 066 (2004)
doi:10.1088/1126-6708/2004/04/066
[arXiv:hep-th/0403158 [hep-th]].


\bibitem{Shifman:2004dr}
M.~Shifman and A.~Yung,
Phys. Rev. D \textbf{70}, 045004 (2004)
doi:10.1103/PhysRevD.70.045004
[arXiv:hep-th/0403149 [hep-th]].



\bibitem{Eto:2006dx}
M.~Eto, L.~Ferretti, K.~Konishi, G.~Marmorini, M.~Nitta, K.~Ohashi, W.~Vinci and N.~Yokoi,
Nucl. Phys. B \textbf{780}, 161-187 (2007)
doi:10.1016/j.nuclphysb.2007.03.040
[arXiv:hep-th/0611313 [hep-th]].




\bibitem{Tong:2005un}
D.~Tong,
[arXiv:hep-th/0509216 [hep-th]].

\bibitem{Eto:2006pg}
M.~Eto, Y.~Isozumi, M.~Nitta, K.~Ohashi and N.~Sakai,
J. Phys. A \textbf{39}, R315-R392 (2006)
doi:10.1088/0305-4470/39/26/R01
[arXiv:hep-th/0602170 [hep-th]].


\bibitem{Konishi:2008vj}
K.~Konishi,
Prog. Theor. Phys. Suppl. \textbf{177}, 83-98 (2009)
doi:10.1143/PTPS.177.83
[arXiv:0809.1370 [hep-th]].

\bibitem{shifbook1}
M.Shifman ``Advanced Topics in Quantum Field Theory: A lecture course". CUP
\bibitem{shifbook2}
M.Shifman and A. Yung ``Supersymmetric solitons" CUP 2009 


\bibitem{Vilenkin:2000jqa}
A.~Vilenkin and E.~P.~S.~Shellard,

\bibitem{Samols:1991ne}
T.~M.~Samols,
Commun. Math. Phys. \textbf{145}, 149-180 (1992)
doi:10.1007/BF02099284

\bibitem{Shellard:1988zx}
E.~P.~S.~Shellard and P.~J.~Ruback,
Phys. Lett. B \textbf{209}, 262-270 (1988)
doi:10.1016/0370-2693(88)90944-6



\bibitem{Speight:1996px}
J.~M.~Speight,
``Static intervortex forces,''
Phys. Rev. D \textbf{55}, 3830-3835 (1997)
doi:10.1103/PhysRevD.55.3830
[arXiv:hep-th/9603155 [hep-th]].

\bibitem{Nakano:2007dq}
E.~Nakano, M.~Nitta and T.~Matsuura,
``Interactions of non-Abelian global strings,''
Phys. Lett. B \textbf{672}, 61-64 (2009)
doi:10.1016/j.physletb.2008.11.049
[arXiv:0708.4092 [hep-ph]].

\bibitem{Auzzi:2007iv}
R.~Auzzi, M.~Eto and W.~Vinci,
``Type I non-Abelian superconductors in supersymmetric gauge theories,''
JHEP \textbf{11}, 090 (2007)
doi:10.1088/1126-6708/2007/11/090
[arXiv:0709.1910 [hep-th]].

\bibitem{Auzzi:2007wj}
R.~Auzzi, M.~Eto and W.~Vinci,
``Static Interactions of non-Abelian Vortices,''
JHEP \textbf{02}, 100 (2008)
doi:10.1088/1126-6708/2008/02/100
[arXiv:0711.0116 [hep-th]].

\bibitem{Tallarita:2017opp}
G.~Tallarita and A.~Peterson,
``Non-Abelian vortex lattices,''
Phys. Rev. D \textbf{97}, no.7, 076003 (2018)
doi:10.1103/PhysRevD.97.076003
[arXiv:1710.07806 [hep-th]].


\bibitem{Hashimoto:2005hi}
K.~Hashimoto and D.~Tong,
JCAP \textbf{09}, 004 (2005)
doi:10.1088/1475-7516/2005/09/004
[arXiv:hep-th/0506022 [hep-th]].

\bibitem{Eto:2006db}
M.~Eto, K.~Hashimoto, G.~Marmorini, M.~Nitta, K.~Ohashi and W.~Vinci,
Phys. Rev. Lett. \textbf{98}, 091602 (2007)
doi:10.1103/PhysRevLett.98.091602
[arXiv:hep-th/0609214 [hep-th]].

\bibitem{Eto:2011pj}
M.~Eto, T.~Fujimori, M.~Nitta, K.~Ohashi and N.~Sakai,
``Dynamics of Non-Abelian Vortices,''
Phys. Rev. D \textbf{84}, 125030 (2011)
doi:10.1103/PhysRevD.84.125030
[arXiv:1105.1547 [hep-th]].



\bibitem{Balachandran:2005ev}
A.~P.~Balachandran, S.~Digal and T.~Matsuura,
Phys. Rev. D \textbf{73}, 074009 (2006)
doi:10.1103/PhysRevD.73.074009
[arXiv:hep-ph/0509276 [hep-ph]].

\bibitem{Nakano:2007dr}
E.~Nakano, M.~Nitta and T.~Matsuura,
Phys. Rev. D \textbf{78}, 045002 (2008)
doi:10.1103/PhysRevD.78.045002
[arXiv:0708.4096 [hep-ph]].

\bibitem{Nakano:2008dc}
E.~Nakano, M.~Nitta and T.~Matsuura,
Prog. Theor. Phys. Suppl. \textbf{174}, 254-257 (2008)
doi:10.1143/PTPS.174.254
[arXiv:0805.4539 [hep-ph]].

\bibitem{Eto:2009kg}
M.~Eto and M.~Nitta,
Phys. Rev. D \textbf{80}, 125007 (2009)
doi:10.1103/PhysRevD.80.125007
[arXiv:0907.1278 [hep-ph]].

\bibitem{Eto:2009bh}
M.~Eto, E.~Nakano and M.~Nitta,
Phys. Rev. D \textbf{80}, 125011 (2009)
doi:10.1103/PhysRevD.80.125011
[arXiv:0908.4470 [hep-ph]].

\bibitem{Eto:2009tr}
M.~Eto, M.~Nitta and N.~Yamamoto,
Phys. Rev. Lett. \textbf{104}, 161601 (2010)
doi:10.1103/PhysRevLett.104.161601
[arXiv:0912.1352 [hep-ph]].

\bibitem{Gorsky:2011hd}
A.~Gorsky, M.~Shifman and A.~Yung,
Phys. Rev. D \textbf{83}, 085027 (2011)
doi:10.1103/PhysRevD.83.085027
[arXiv:1101.1120 [hep-ph]].

\bibitem{Eto:2011mk}
M.~Eto, M.~Nitta and N.~Yamamoto,
Phys. Rev. D \textbf{83}, 085005 (2011)
doi:10.1103/PhysRevD.83.085005
[arXiv:1101.2574 [hep-ph]].

\bibitem{Hirono:2010gq}
Y.~Hirono, T.~Kanazawa and M.~Nitta,
Phys. Rev. D \textbf{83}, 085018 (2011)
doi:10.1103/PhysRevD.83.085018
[arXiv:1012.6042 [hep-ph]].

\bibitem{Vinci:2012mc}
W.~Vinci, M.~Cipriani and M.~Nitta,
Phys. Rev. D \textbf{86}, 085018 (2012)
doi:10.1103/PhysRevD.86.085018
[arXiv:1206.3535 [hep-ph]].

\bibitem{Eto:2013hoa}
M.~Eto, Y.~Hirono, M.~Nitta and S.~Yasui,
PTEP \textbf{2014}, no.1, 012D01 (2014)
doi:10.1093/ptep/ptt095
[arXiv:1308.1535 [hep-ph]].

\bibitem{Chatterjee:2015lbf}
C.~Chatterjee and M.~Nitta,
Phys. Rev. D \textbf{93}, no.6, 065050 (2016)
doi:10.1103/PhysRevD.93.065050
[arXiv:1512.06603 [hep-ph]].

\bibitem{Alford:2016dco}
M.~G.~Alford, S.~K.~Mallavarapu, T.~Vachaspati and A.~Windisch,
Phys. Rev. C \textbf{93}, no.4, 045801 (2016)
doi:10.1103/PhysRevC.93.045801
[arXiv:1601.04656 [nucl-th]].

\bibitem{Chatterjee:2016ykq}
C.~Chatterjee, M.~Cipriani and M.~Nitta,
Phys. Rev. D \textbf{93}, no.6, 065046 (2016)
doi:10.1103/PhysRevD.93.065046
[arXiv:1602.01677 [hep-ph]].

\bibitem{Alford:2018mqj}
M.~G.~Alford, G.~Baym, K.~Fukushima, T.~Hatsuda and M.~Tachibana,
Phys. Rev. D \textbf{99}, no.3, 036004 (2019)
doi:10.1103/PhysRevD.99.036004
[arXiv:1803.05115 [hep-ph]].

\bibitem{Chatterjee:2018nxe}
C.~Chatterjee, M.~Nitta and S.~Yasui,
Phys. Rev. D \textbf{99}, no.3, 034001 (2019)
doi:10.1103/PhysRevD.99.034001
[arXiv:1806.09291 [hep-ph]].

\bibitem{Hirono:2018fjr}
Y.~Hirono and Y.~Tanizaki,
Phys. Rev. Lett. \textbf{122}, no.21, 212001 (2019)
doi:10.1103/PhysRevLett.122.212001
[arXiv:1811.10608 [hep-th]].

\bibitem{Chatterjee:2019tbz}
C.~Chatterjee, M.~Nitta and S.~Yasui,
JPS Conf. Proc. \textbf{26}, 024030 (2019)
doi:10.7566/JPSCP.26.024030
[arXiv:1902.00156 [hep-ph]].


\bibitem{Hidaka:2019jtv}
Y.~Hidaka, Y.~Hirono, M.~Nitta, Y.~Tanizaki and R.~Yokokura,
Phys. Rev. D \textbf{100}, no.12, 125016 (2019)
doi:10.1103/PhysRevD.100.125016
[arXiv:1903.06389 [hep-th]].


\bibitem{Lee:1973iz}
T.~D.~Lee,
Phys. Rev. D \textbf{8}, 1226-1239 (1973)
doi:10.1103/PhysRevD.8.1226

\bibitem{Dvali:1993sg}
G.~R.~Dvali and G.~Senjanovic,
Phys. Rev. Lett. \textbf{71}, 2376-2379 (1993)
doi:10.1103/PhysRevLett.71.2376
[arXiv:hep-ph/9305278 [hep-ph]].

\bibitem{Eto:2018hhg}
M.~Eto, M.~Kurachi and M.~Nitta,
Phys. Lett. B \textbf{785}, 447-453 (2018)
doi:10.1016/j.physletb.2018.09.002
[arXiv:1803.04662 [hep-ph]].

\bibitem{Eto:2018tnk}
M.~Eto, M.~Kurachi and M.~Nitta,
JHEP \textbf{08}, 195 (2018)
doi:10.1007/JHEP08(2018)195
[arXiv:1805.07015 [hep-ph]].


\bibitem{Eto:2019hhf}
M.~Eto, Y.~Hamada, M.~Kurachi and M.~Nitta,
Phys. Lett. B \textbf{802}, 135220 (2020)
doi:10.1016/j.physletb.2020.135220
[arXiv:1904.09269 [hep-ph]].

\bibitem{Eto:2020hjb}
M.~Eto, Y.~Hamada, M.~Kurachi and M.~Nitta,
JHEP \textbf{07}, 004 (2020)
doi:10.1007/JHEP07(2020)004
[arXiv:2003.08772 [hep-ph]].

\bibitem{Eto:2020opf}
M.~Eto, Y.~Hamada and M.~Nitta,
Phys. Rev. D \textbf{102}, no.10, 105018 (2020)
doi:10.1103/PhysRevD.102.105018
[arXiv:2007.15587 [hep-th]].

\bibitem{Abe:2020ure}
Y.~Abe, Y.~Hamada and K.~Yoshioka,
[arXiv:2010.02834 [hep-ph]].


\bibitem{Witten:1984eb}
E.~Witten,
Nucl. Phys. B \textbf{249}, 557-592 (1985)
doi:10.1016/0550-3213(85)90022-7

\bibitem{Shifman:2012vv}
M.~Shifman,
Phys. Rev. D \textbf{87}, no.2, 025025 (2013)
doi:10.1103/PhysRevD.87.025025
[arXiv:1212.4823 [hep-th]].


\bibitem{Peterson:2014nma}
A.~Peterson, M.~Shifman and G.~Tallarita,
Annals Phys. \textbf{353}, 48-63 (2014)
doi:10.1016/j.aop.2014.11.001
[arXiv:1409.1508 [hep-th]].


\bibitem{Peterson:2015tpa}
A.~J.~Peterson, M.~Shifman and G.~Tallarita,
Annals Phys. \textbf{363}, 515-532 (2015)
doi:10.1016/j.aop.2015.10.012
[arXiv:1508.01490 [hep-th]].

\bibitem{Shifman:2015ama}
M.~Shifman, G.~Tallarita and A.~Yung,
Phys. Rev. D \textbf{91}, no.10, 105026 (2015)
doi:10.1103/PhysRevD.91.105026
[arXiv:1503.08684 [hep-th]].

\bibitem{Canfora:2016spb}
F.~Canfora and G.~Tallarita,
Phys. Rev. D \textbf{94}, no.2, 025037 (2016)
doi:10.1103/PhysRevD.94.025037
[arXiv:1607.04140 [hep-th]].

\bibitem{Tallarita:2017xhh}
G.~Tallarita and A.~Peterson,
Phys. Rev. D \textbf{96}, no.11, 116017 (2017)
doi:10.1103/PhysRevD.96.116017
[arXiv:1711.06612 [hep-th]].

\bibitem{Tallarita:2015mca}
G.~Tallarita,
Phys. Rev. D \textbf{93}, no.6, 066011 (2016)
doi:10.1103/PhysRevD.93.066011
[arXiv:1510.06719 [hep-th]].

\bibitem{Tallarita:2019czh}
G.~Tallarita, R.~Auzzi and A.~Peterson,
JHEP \textbf{03}, 114 (2019)
doi:10.1007/JHEP03(2019)114
[arXiv:1901.05814 [hep-th]].


\bibitem{Hill:1987ye}
C.~T.~Hill, H.~M.~Hodges and M.~S.~Turner,
Phys. Rev. Lett. \textbf{59}, 2493 (1987)
doi:10.1103/PhysRevLett.59.2493

\bibitem{MacKenzie:1987ye}
R.~MacKenzie,
Phys. Lett. B \textbf{197}, 59 (1987)
[erratum: Phys. Lett. B \textbf{199}, 596 (1987)]
doi:10.1016/0370-2693(87)90342-X

\bibitem{Haws:1988ax}
D.~Haws, M.~Hindmarsh and N.~Turok,
Phys. Lett. B \textbf{209}, 255-261 (1988)
doi:10.1016/0370-2693(88)90943-4

\bibitem{Amsterdamski:1988zp}
P.~Amsterdamski and P.~Laguna-Castillo,
Phys. Rev. D \textbf{37}, 877-884 (1988)
doi:10.1103/PhysRevD.37.877

\bibitem{Babul:1987me}
A.~Babul, T.~Piran and D.~N.~Spergel,
Phys. Lett. B \textbf{202}, 307-314 (1988)
doi:10.1016/0370-2693(88)90476-5

\bibitem{Babul:1988qt}
A.~Babul, T.~Piran and D.~N.~Spergel,
Phys. Lett. B \textbf{209}, 477-484 (1988)
doi:10.1016/0370-2693(88)91177-X

\bibitem{Davis:1988jp}
R.~L.~Davis and E.~P.~S.~Shellard,
Phys. Lett. B \textbf{207}, 404-410 (1988)
doi:10.1016/0370-2693(88)90673-9

\bibitem{Davis:1988jq}
R.~L.~Davis and E.~P.~S.~Shellard,
Phys. Lett. B \textbf{209}, 485-490 (1988)
doi:10.1016/0370-2693(88)91178-1

\bibitem{Hill:1987qx}
C.~T.~Hill, H.~M.~Hodges and M.~S.~Turner,
Phys. Rev. D \textbf{37}, 263 (1988)
doi:10.1103/PhysRevD.37.263


\bibitem{Perivolaropoulos:1993uj}
L.~Perivolaropoulos,
Phys. Rev. D \textbf{48}, 5961-5962 (1993)
doi:10.1103/PhysRevD.48.5961
[arXiv:hep-ph/9310264 [hep-ph]].

\bibitem{Eto:2009wq}
M.~Eto, T.~Fujimori, T.~Nagashima, M.~Nitta, K.~Ohashi and N.~Sakai,
Phys. Lett. B \textbf{678}, 254-258 (2009)
doi:10.1016/j.physletb.2009.05.061
[arXiv:0903.1518 [hep-th]].


\end{thebibliography}
\end{document}